\documentclass[11pt,a4paper]{article}
\pdfoutput=1
\usepackage{jcappub}
\usepackage{float}
\usepackage{indentfirst}
\newcommand{\be}{\begin{equation}}
\newcommand{\ee}{\end{equation}}

\newcommand{\ba}{\begin{eqnarray}}
\newcommand{\ea}{\end{eqnarray}}

\usepackage{float}

\newcommand{\bege}{\begin{equation}}
\newcommand{\bpartial}{\mathop{\partial\kern -4pt\raisebox{.8pt}{$|$}}}
\newcommand{\enge}{\end{equation}}
\newcommand{\beq}{\begin{eqnarray}}
\newcommand{\benu}{\begin{enumerate}}

\newcommand{\enu}{\end{enumerate}}
\newcommand{\eeq}{\end{eqnarray}}

\begin{document}

\title{{Power-law and Logarithmic Entropy Corrected Holographic Dark Energy Models in Brans-Dicke Cosmology with Granda-Oliveros Cut-Off}}
\author[a]{Antonio Pasqua}
\affiliation[a]{Department of Physics,
University of Trieste, Via Valerio, 2 34127 Trieste, Italy.}
 \author[b]{Surajit Chattopadhyay}
\affiliation[b]{Pailan College of Management and Technology, Bengal Pailan Park, Kolkata-700 104, India.}
 \author[c]{Ratbay Myrzakulov}
\affiliation[c]{Eurasian International Center for Theoretical Physics, Eurasian National University, Astana 010008, Kazakhstan}

\emailAdd{toto.pasqua@gmail.com}
\emailAdd{surajcha@iucaa.ernet.in}
\emailAdd{rmyrzakulov@gmail.com}

\begin{abstract}{
In this paper, the cosmological implications of the Power Law Entropy Corrected Holographic Dark Energy (PLECHDE) and the Logarithmic Entropy Corrected Holographic Dark Energy (LECHDE) models in the context of Brans-Dicke (BD) cosmology for both non-interacting and interacting DE and Dark Matter (DM) are studied. As the system infrared cut-off, we choose the recently proposed Granda-Oliveros  cut-off, which contains a term proportional to the first time derivative of the Hubble parameter and one term proportional to $H^2$, i.e. the Hubble parameter squared. We obtain the expressions of three quantities, i.e. the Equation of State (EoS) parameter $\omega_D$, the deceleration parameter $q$ and the evolutionary form of the energy density parameter $\Omega'_D$ of the PLECHDE and LECHDE models in a non-flat Universe for non-interacting and interacting DE and DM as well. Moreover, we investigate  the limiting cases corresponding to: a) absence of entropy corrections; b) Einstein's gravity; and c) concomitantly absence of entropy corrections and Einstein's gravity. Furthermore, we consider the limiting case corresponding to the Ricci scale, which is recovered for some particular values of the parameters characterizing the Granda-Oliveros scale. We also study the statefinder diagnostic and the cosmographic parameters for both models considered in this work.
}
\end{abstract}

\flushbottom
\maketitle

\section{Introduction}
Cosmological and astrophysical data recently obtained  thanks to studies of Supernovae Ia  (SNeIa), the Large Scale Structures (LSS), the Cosmic Microwave Background Radiation (CMBR) anisotropies  and X-ray experiments provide strong evidences supporting a phase of accelerated expansion of the present Universe \cite{1,1a,1b,1c,1d,cmb3,planck,sds1,sds2,xray}.\\
In order to find a suitable model for our Universe, some possible reasons for  the accelerated expansion have been widely investigated. Three main classes of models are usually proposed to describe this phenomenon: a) a  Cosmological Constant $\Lambda$; b) Dark Energy (DE) models; c) modified theories for gravity.\\
The Cosmological Constant $\Lambda$, which has EoS parameter which is given by $\omega_{\Lambda} = -1$,  represents the simplest candidate proposed to have an explanation for the present expansion with accelerated rate of our Universe. However, $\Lambda$ is well known to be related to two main problems, i.e. the fine-tuning problem and the cosmic coincidence problem \cite{copeland-2006}. According to the fine tuning problem, the vacuum energy density is about $10^{123}$ times smaller than what is  observed. Instead, according to the cosmic coincidence problem, the vacuum energy and DM are almost equal nowadays, even if  they had an independent evolution and even if they evolved from different mass scales (which is a particular fortuity if no interactions exist between them). Many attempts have been made with the aim to propose a solution to the coincidence problem \cite{delcampo,delcampoa,delcampoc,delcampod,delcampoe,delcampoi,delcampol}.\\
The evidence of this cosmic expansion with accelerated rate we observe implies, if Einstein's theory of General Relativity (GR) is valid at  cosmological scales, that the Universe must be dominated by some still unknown type of missing component with some peculiar features. For instance, it must not be clustered on large scales and its pressure must be negative.
In relativistic cosmology, the observed cosmic accelerated expansion  can be described using a perfect fluid which can be characterized by a pressure $p$ and an energy density $\rho$ satisfying the relation $\rho + 3p < 0$. This kind of fluid with sufficiently high negative pressure is known as dark energy (DE). In other words, the relation  $\rho + 3p < 0$ implies that EoS parameter $\omega=  p/\rho$ for this fluid must obey the condition $\omega  <-1/3$, while from an observational point of view it is a difficult task to constrain its precise value. \\
The largest amount of the cosmic energy density $\rho$ is represented by the two dark sectors, i.e. DE and Dark Matter (DM) which represent, respectively, the 68.3$\%$ and the 26.8$\%$ of the energy density in the present day Universe \cite{twothirds}, while the observed ordinary baryonic matter contributes for only 4.9$\%$ of the cosmic energy density. Moreover, we can safely consider that the radiation term negligibly contributes to the cosmic energy density.

As an alternative to the cosmological constant $\Lambda$, different candidates have been studied in order to try to explain the nature of DE. Some of them include quintessence,  tachyon, quintom,  $k$-essence, phantom energy, Chaplygin gas and Agegraphic DE (ADE) model \cite{dil1,dil2,dil2-1,kess3,kess4,quint1,quint2,quint3,quint4,tac1,tac3,tac1-2,tac1-3,tac2-1,
pha1,pha2,pha5,qui2,qui3,qui4,cgas1,cgas3,cgas41,ade1,ade2,2a}.

One of the most studied candidate for the DE is the Holographic DE (HDE) model, which has its motivation from the holographic principle \cite{3,3b,4,5,5a}.  Cohen et al.\cite{7} have shown that, in the context of Quantum Field Theory (QFT), the ultraviolet (UV) cut-off $\Lambda_{UV}$ should be connected to the infrared (IR) cut-off $L$ by a limit set by the formation of a black hole. Indeed, if the vacuum energy density associated to the UV cut-off is given by $\rho_D = \Lambda_{UV}^4$, therefore the total energy per length $L$ should not be larger than the mass of the associated black hole:
\begin{eqnarray}
E_D \leq E_{BH}\;\; \Rightarrow\;\; L^3 \rho_D \leq M_p^2 L, \label{1}
\end{eqnarray}
where the quantity $M_p = \left(8\pi G_N  \right)^{-1/2} \approxeq 10^{18}$ GeV represents the Planck mass and $G_N = 6.67384\times 10^{-11} m^3 kg^{-1} s^{-2}$ denotes the Newton gravitational constant. If the largest possible cut-off $L$ is given by the one which saturates the inequality given in Eq. (\ref{1}), one obtains the energy density $\rho_D$ of the HDE  as follows:
\begin{eqnarray}
\rho_D = 3c^2 M_p^2 L^{-2}, \label{2}
\end{eqnarray}
where $c^2$ represents a dimensionless numerical constant which value $c$ is evinced by observational data:  for  a flat Universe (which means $k=0$) $c=0.818_{-0.097}^{+0.113}$  and in the case corresponding to a non-flat Universe (which means $k \neq 0$) we have  $c=0.815_{-0.139}^{+0.179}$ \cite{n2primo,n2secondo}.\\
According to a recent work made by Guberina et al. \cite{8}, the HDE model based on the entropy bound can be derived in an alternative way. In fact, in black hole thermodynamics \cite{9,9a} the maximum value of the entropy in a box with size $L$, known as the Bekenstein-Hawking entropy bound, is obtained using the relation:
\begin{eqnarray}
S_{BH} \approx M_p^2 L^2. \label{sbh}
\end{eqnarray}
We have that Eq. (\ref{sbh}) scales as the area of the box (which is given by$A \approx L^2$) rather than the volume of the box (which is given by the relation $\left(V \approx L^3\right)$). Moreover, for a macroscopic system with self-gravity effects which can not be ignored, we know that the Bekenstein entropy bound $S_B$ can be also obtained multiplying the energy $E = \rho_DL^3$ and the linear size of the system given by $L$. If we require that the Bekenstein entropy bound $S_B$ must be smaller than the Bekenstein-Hawking entropy (i.e. $S_B \leq S_{BH}$, which implies $EL \leq M_p^2 L^2$), it is possible to obtain the same result obtained from energy bound argument, i.e. $\rho_D \leq M_p^2L^{-2}$.

The HDE model has been widely investigated and studied in literature. Chen et al. \cite{10} used the HDE model in order to drive inflation in the early evolutionary phases of Universe. Jamil et al. \cite{11} studied the EoS parameter $\omega_D$ of HDE,  by considering a time-varying Newton's gravitational constant $G_N \equiv G_N\left( t \right)$. Moreover, they showed that $\omega_D$ can be significantly changed in the low redshift $z$ limit.
The HDE model was also studied in other works \cite{12,12a,12c,12d,12e,13,13c}  with different IR cut-offs, for example the particle, the Hubble and the future event horizons. Moreover, a correspondence between the HDE model and other scalar field models have been recently proposed \cite{14,14a,14b}. It was also demonstrated that the HDE model can precisely fit the cosmological data obtained from the SNe Ia and CMB \cite{16,16a,16b,16c}.

The cosmic acceleration has been also accurately studied by imposing the concept of modification of gravity  \cite{nojod}. This new model of gravity (predicted by string/M theory) gives a very natural gravitational alternative for exotic matter. The explanation of the phantom, non-phantom and quintom phases of the Universe can be well described using modified gravity without the necessity of the introduction of a negative kinetic term in DE models.
 The cosmic  acceleration is evinced by the straightforward fact that terms (like the $1/R$ term) might become fundamental at small curvatures. Furthermore, modified gravity models provide a natural way to join the late-time accelerated expansion (which our Universe is experiencing) and the early-time inflation (which is the first accelerated expansion our Universe experienced in past epochs). Such theories are also prime candidates for the explanation of the DE and DM, including for instance the anomalous galaxies rotation curves.  The effective DE dominance may be aided by a  modification of gravity. Therefore, the coincidence problem can be solved simply considering the fact that the Universe expands.
 Modified gravity theory is also expected and believed to be relevant and useful in the framework of High Energy physics,  explaining the hierarchy problem or the unification of the other forces with gravity \cite{nojod}.
Some of the most famous and known models of modified gravity are represented by braneworld models, $f\left(T\right)$ gravity (where $T$ indicates the torsion scalar),
$f \left(R\right)$ gravity (where the quantity $R$ indicates the Ricci scalar curvature), $f \left(G\right)$ gravity (where the quantity $G=R^2-4R_{\mu \nu}R^{\mu \nu} + R_{\mu \nu \lambda \sigma}R^{\mu \nu \lambda \sigma}$ is the Gauss-Bonnet invariant, with $R_{\mu \nu}$ representing the Ricci curvature tensor and $R_{\mu \nu \lambda \sigma}$ representing the Riemann curvature tensor), $f \left(R,T\right)$ gravity, DGP models, DBI models and Ho${\check{\rm r}}$ava-Lifshitz gravity  gravity \cite{15,15b,15c,15d,15g,15i,mio,miobd2,dgp1,miodbi,miohl,frt2,frt4,miofg1,fr1,fr2,miofr,fr8,fr9,fr10,fr12,fr14,fr15,
mioft1,mioft2,ft2,ft4,ft8,bra1,bra2}.\\
Since we have that the HDE energy density model is a dynamical model, we have that a  dynamical frame (like the Brans-Dicke (BD) theory) is required, in order to accommodate it instead of General Relativity (GR).  Moreover, considering the cut-off $L = H^{-1}$, where $H$ denotes the Hubble parameter, it is not possible to exactly determine the EoS parameter $\omega_D$ in the context of GR. By assuming that our Universe is entering the phantom phase, we have that modified gravity theories can provide a natural description to the passage from a non-phantom to a phantom phase, without being necessary the introduction of the idea of presence of exotic matter (like a scalar with wrong sign kinetic term or an ideal fluid with an EoS parameter which satisfy the relation $\omega_D<-1$). Moreover,  we have that the phantom phase in modified gravity theory is often transient and the Brans-Dicke scalar field is able to speed up the expansion rate of a dust matter dominated era reducing the deceleration and it can also slow down the expansion rate of a cosmological constant $\Lambda$ era reducing the acceleration.
By taking into account all the considerations done above, the study of HDE models in the context of Brans-Dicke theory results to be well motivated. For this reason, many HDE models have been well studied in the context of Brans-Dicke gravity \cite{23,23a,24a,24b,24c,24d,24e,25}. In these papers, many dynamical characteristics and features of the HDE models  have been studied for both a  flat and a non-flat Friedmann-Lema\^itre-Robertson-Walker (FLRW) background, for example the cosmic coincidence problem, the quintom behaviour, the phantom crossing at the present time, the EoS and the deceleration parameters $\omega$ and $q$.\\
It must be emphasized that the black hole entropy, indicated with $S$, assumes an important role during the derivation of the final expression of the HDE energy density. It is also well known that the derivation of the expression of the HDE energy density strongly relies on the entropy-area relation given, in General Relativity, by $S \approx A \approx L^2$, where the quantity $A$ indicates the area of the black hole horizon while the quantity $L$ indicates the characteristic length of the system. However, the definition of the entropy-area relation can be also modified by taking into account some quantum effects which have their motivation from the Loop Quantum Gravity (LQG). The relation $S=S\left(A\right)$, i.e. the entropy as function of the area $A$, has two interesting possible modifications/corrections, which are in particular the power-law \cite{das18,das18a,das18d} and the logarithmic  \cite{hei17} corrections.\\
In this paper, we decided to consider both the modifications of the HDE energy density due to the power-law and the logarithmic corrections to the entropy $S\left( A \right)$. \\
The power-law corrected form of the entropy is given by the following relation:
\begin{eqnarray}
    S\left( A \right)=c_0 \left( \frac{A}{a_1^2}  \right)\left[ 1+c_1f\left( A \right)  \right],\label{powerlawentropyold}
\end{eqnarray}
where:
\begin{eqnarray}
 f\left( A \right)  = \left( \frac{A}{a_1^2}  \right)^{-\nu},\label{powerlawentropyold}
\end{eqnarray}
with $c_0$ and $c_1$ being two constant parameters of the order of unity, $a_1$ being the ultraviolet cut-off at the horizon and $\nu$ being a fractional power which depends on the amount of mixing excited and ground states. In the limiting case corresponding to a large horizon area (i.e. when $A>>a_1^2$), the contribution of the term $f \left(A\right)$ is negligible and the mixed state entanglement entropy asymptotically approaches the ground state (Bekenstein-Hawking) entropy.\\
Another useful form of entanglement entropy is given by the following relation:
\begin{eqnarray}
  S\left ( A \right)= \frac{A}{4G_N}\left( 1-K_{\alpha}A^{1-\alpha /2}   \right), \label{3}
\end{eqnarray}
where  $A=4\pi R_h^2$ indicates the area of the horizon (with $R_h$ denoting the black hole event horizon), $\alpha$ represents a dimensionless constant while $K_{\alpha}$ represents a constant defined as follows:
\begin{eqnarray}
  K_{\alpha} = \frac{\alpha \left( 4\pi  \right)^{\alpha /2 -1}}{\left( 4-\alpha  \right)r_c^{2-\alpha}}, \label{4}
\end{eqnarray}
where $r_c$ indicates the cross-over scale. We have that the second term present in the expression of $  S\left ( A \right)$ given in Eq. (\ref{3}) represents the power law correction to the entropy-area law.\\
Instead, the logarithmic corrected expression of the entropy-area law has the following form:
\begin{eqnarray}
S\left( A \right)= \frac{A}{4G_N} + \hat{\alpha} \ln\left(\frac{A}{4G_N}\right) + \hat{\beta}, \label{3log}
\end{eqnarray}
where $\hat{\alpha}$ and $\hat{\beta}$ denote two dimensionless constant parameters.
We must also have $\hat{\alpha}>0$ in order the entropy defined in Eq. (\ref{3log}) is a well defined quantity.\\
Motivated by the entropy relations given in Eqs. (\ref{3}) and (\ref{3log}), new versions of HDE energy density, known respectively as Power Law Entropy Corrected HDE (PLECHDE) model and Logarithmic Entropy Corrected HDE (LECHDE) model, were recently proposed as follows:
\begin{eqnarray}
\rho_{Dpl} &=& 3c^2M_p^2L^{-2} - \lambda M_p^2 L^{-\delta}, \label{pl} \\
\rho_{Dlog} &=& 3c^2 M_p^2L^{-2}+ \varrho L^{-4} \log \left( M_p^2L^2 \right) + \epsilon L^{-4},\label{4log}
\end{eqnarray}
where $\delta$ denotes a positive power-law index while $\lambda$, $\varrho$ and $\epsilon$ represent three dimensionless constants.\\
In the limiting case corresponding to $\lambda =0$ and $\varrho = \epsilon = 0$ (i.e., in absence of entropy corrections), Eqs. (\ref{pl}) and (\ref{4log}) yield the expression of the HDE energy density  given in Eq. (\ref{2}).\\
The correction terms present in Eqs. (\ref{pl}) and (\ref{4log}) are of the same order of the first term only when the characteristic length of the system $L$ is a very small quantity. At the early evolutionary stages of the history of the Universe, when the Universe underwent an inflationary phase, the power-law and logarithmic correction terms present in the PLECHDE and LECHDE energy density models become more relevant. At the end of the inflationary era, we have that the Universe enters in the radiation era and then in the matter-dominated era: in these two eras, since the Universe  became much larger respect with the previous eras, we have that the power-law and the logarithmic entropy-corrected terms, i.e. the second term in Eq. (\ref{pl}) and the second and the third terms in Eq. (\ref{4log}), can be neglected since they do not give relevant contributions anymore. Therefore, the PLECHDE and the LECHDE models can be considered as models of entropic cosmology which are able to unify the early-time inflationary phase and the late-time cosmic accelerated expansion of the Universe.\\
In this paper, we decied to consider as the IR cut-off for the PLECHDE and the LECHDE energy density models a recently proposed cut-off, based on purely dimensional grounds,  and known as Granda-Oliveros (GO) scale, denoted by $L_{GO}$. The expression for $L_{GO}$ includes a term proportional to the first derivative of the Hubble parameter $H$ with respect to the cosmic time $t$ (i.e., $\dot{H}$) and one proportional to the Hubble parameter squared (i.e.,  $H^2$) and it is given by the following expression \cite{grandaoliveros,grandaoliverosa}:
\begin{equation}
L_{GO}=\left( \alpha H^{2}+\beta \dot{H}\right) ^{-1/2},  \label{5}
\end{equation}%
where the quantities $\alpha $ and $\beta $ indicate two positive constant parameters. In the limiting case corresponding to  $ \left\{\alpha, \beta\right\} = \left\{2,1\right\}$, we obtain that the GO scale $L_{GO}$ becomes proportional to the average radius of the Ricci scalar curvature (i.e., $L_{GO} \propto R^{-1/2}$) in the case the curvature parameter $k$ assumes the value of zero, corresponding to a  flat Universe.
Recently, Wang and Xu \cite{wangalfa} have constrained the new HDE model in the framework of non-flat Universe using observational data. They derived that the best fit values of the parameters  $\left(\alpha, \beta   \right)$ with their respective confidence levels are given by $\alpha  = 0.8824^{+0.2180}_{-0.1163}(1\sigma)\,^{+0.2213}_{-0.1378}(2\sigma)$ and $\beta = 0.5016^{+0.0973}_{-0.0871}(1\sigma)\,^{+0.1247}_{-0.1102}(2\sigma)$ for non flat Universe, while for flat Universe they found that are $\alpha  = 0.8502^{+0.0984}_{-0.0875}(1\sigma)\,^{+0.1299}_{-0.1064}(2\sigma)$ and $\beta = 0.4817^{+0.0842}_{-0.0773}(1\sigma)\,^{+0.1176}_{-0.0955}(2\sigma)$.\\
We decided to consider the GO scale $L_{GO}$ as infrared (IR) cut-off for some specific reasons. If the IR cut-off chosen is given by the particle horizon, it is obtained that the HDE model cannot produce an expansion of the Universe with accelerated rate. If we consider as cut-off of the system the future event horizon, the HDE model has a causality problem.
The DE models which consider the GO scale $L_{GO}$ depend only on local quantities, then it is possible to avoid the causality problem, moreover it is also possible to obtain the accelerated phase of the Universe.   \\
This paper is organized as follows. In Section 2, we study the PLECHDE and the LECHDE models in the context of Brans-Dicke cosmology for a non-flat Universe, deriving and discussing some of the physical features and characteristics of these models, like for example the EoS parameter $\omega_D$, the evolutionary form  of the dimensionless energy density parameter of DE $\Omega'_D$  and the deceleration parameter $q$ in absence and later on in the presence of an interaction between the two Dark Sectors, i.e. DE and DM. We will choose three different expressions for the interaction term $Q$ (which characterizes the interaction between the DM and DE): the expressions of $Q$ we will consider are functions of the Hubble parameter $H$ and of the energy densities of DE and DM. Furthermore, we calculate the limiting cases of the energy densities for the PLECHDE and LECHDE models corresponding to $\lambda =0$, which means no power law correction in the expression of the DE energy density, to $\varrho = \epsilon = 0$, which means no logarithmic corrections in the expression of the energy density, to the Einstein's gravity, to both $\lambda =0$ and Einstein gravity together and finally to both $\varrho = \epsilon = 0$ and Einstein gravity together. Moreover, we derive the expressions of the Hubble parameter $H$ and of the scale factor $a$ as functions of the cosmic time $t$.
In Section 3, we study the statefinder diagnostic (which is a useful tool studied in order to check deviations of the models considered from the $\Lambda$CDM one) for the models considered in this paper.
In Section 4, we derive the expressions of the cosmographic parameters for the models considered.
Finally, Section 5 is devoted to the Conclusions of this paper.

\section{PLECHDE and LECHDE Models in the Framework of Brans-Dicke Cosmology}
The main aim of this Section is the description of the most important cosmological features of the Brans-Dicke (BD) cosmology and the derivation of some important cosmological parameters, like the EoS parameter $\omega_D$, the deceleration parameter $q$ and the evolutionary form of the energy density parameter of DE $\Omega_D'$  for both non-interacting and interacting Dark Sectors and for both models considered.

Within the framework of the Friedmann-Lema\^itre-Robertson-Walker (FLRW) cosmology, the line element  can be expressed, in spherical coordinates, as follows:
\begin{eqnarray}
    ds^2=-dt^2+a^2\left(t\right)\left(\frac{dr^2}{1-kr^2} +r^2 d\Omega^2 \right).\label{6}
\end{eqnarray}
In Eq. (\ref{6}), $a\left(t\right)$ represents the scale factor (which gives information about the expansion of the Universe), $r$ indicates the radial component of the metric, $t$ is the cosmic time and $k$ indicates the curvature parameter, which can assume one of the values $-1$, $0$ or $+1$  yielding an open, a flat, or a closed FLRW Universe, respectively. Moreover, $d\Omega^2= r^2 \left(d\theta ^2 + \sin^2 \theta d\varphi ^2\right)$ denotes the solid angle element (squared). The two angles $\theta$ and $\varphi$ represent, respectively, the usual azimuthal and polar angles and their values are constrained in the ranges $0\leq \theta \leq \pi$ and $0\leq \varphi \leq 2\pi$. The four coordinates $\left(r , t, \varphi, \theta  \right)$ are also known as co-moving coordinates.\\
The action $S_{BD}$ of Brans-Dicke cosmology is given by the following expression:
\begin{eqnarray}
S_{BD}=\int d^4x\sqrt{-g}\left( -\varphi R+\frac{\omega}{\varphi} g^{\mu \nu}\partial_{\mu}\varphi \partial_{\nu}\varphi + L_m    \right). \label{7}
\end{eqnarray}
By defining $\varphi = \frac{\phi^2}{8\omega}$, the action $S_{BD}$ given in Eq. (\ref{7}) can be written in its canonical form \cite{26-1}:
\begin{eqnarray}
    S_{BD}=\int d^4x\sqrt{-g}\left( -\frac{1}{8\omega}\phi^2R+\frac{1}{2}g^{\mu \nu}\partial_{\mu}\phi \partial_{\nu}\phi+L_m    \right), \label{9}
\end{eqnarray}
where the quantities $g$, $\omega$, $\phi$, and $L_m$ represent, respectively, the determinant of the tensor metric $g^{\mu \nu}$, the BD parameter, the BD scalar field and the Lagrangian of the matter.  Moreover, $\partial_{\nu}$ indicates a covariant derivative.    We can straightforwardly obtain the expressions of Friedmann equations in the context  of Brans-Dicke cosmology making the variation of the expression of the action $S_{BD}$ given in Eq. (\ref{9}) with respect to the FLRW metric given in Eq. (\ref{6}), which yields to the following system of equations:
\begin{eqnarray}
\frac{3}{4\omega}\phi^2\left(H^2+\frac{k}{a^2}\right)-\frac{\dot{\phi}^2}{2}+\frac{3}{2\omega}H\dot{\phi}\phi    &=& \rho_D + \rho_m, \label{10} \\
-\frac{\phi^2}{4\omega}\left(\frac{2\ddot{a}}{a}+H^2 +\frac{k}{a^2}\right)-\frac{H\dot{\phi}\phi}{\omega}-\frac{\ddot{\phi}\phi}{2\omega}-\frac{\dot{\phi}^2}{2}\left(1+\frac{1}{\omega} \right) &=& p_D, \label{11}\\
    \ddot{\phi}+3H\dot{\phi}- \frac{3}{2\omega}\left(\frac{\ddot{a}}{a}+H^2+\frac{k}{a^2}  \right)\phi  &=& 0. \label{12}
\end{eqnarray}
The quantities $\rho_D$, $p_D$ and $\rho_m$ denote, respectively, the energy density of DE, the pressure of DE and the energy density of DM, which is considered to have pressure equal to zero, i.e. $p_m =0$. We are also considering the contribution of the radiation term $\rho_r$ negligible, since we have that $\rho_r \simeq 0$. Moreover, as usual, the symbols  $(\,\,\dot{}\,\,)$ and $(\,\,\ddot{}\,\,)$ denote the first and the second derivative with respect to the cosmic time $t$ of the relevant quantity, respectively.

We can assume that the BD field has a power-law dependance with the scale factor $a(t)$  given by
$\phi \propto a^{n}$, where $n$ represents a positive power-law index. It has been observed that small values of $n$  lead to consistent results \cite{28bane,29}. Particularly interesting is the case corresponding to small values of $n$ whereas $\omega$ assumes big values,  such that $n\omega \approx 1$. In fact, local astrophysical observations can impose  a lower bound on the value of $\omega$. For example the Cassini experiment sets  $\omega \gtrsim 10^4$ \cite{30,30-1}, and strong constraints on the value of $n$ can be determined.

The expression $\phi \propto a^{n}$ implies that the first and the second time derivatives of $\phi$ are given, respectively, by:
\begin{eqnarray}
\dot{\phi} &=&  \frac{d\phi}{dt} = n a^{n -1}\dot{a}= n H \phi, \label{13} \\
\ddot{\phi} &=& \frac{d^2\phi}{dt^2} =nH\dot{\phi} +  n \phi \dot{H} =     n ^2 H^2 \phi + n \phi \dot{H}, \label{14}
\end{eqnarray}
where we used the result of Eq. (\ref{13}) in Eq. (\ref{14}).\\
In the framework of BD theory, taking into account Eqs. (\ref{pl}) and (\ref{4log}), the energy densities of the PLECHDE  and LECHDE models, indicated with $\rho_{Dpl}$ and $\rho_{Dlog}$,  are given by the following relations:
\begin{eqnarray}
\rho_{Dpl} &=& \frac{3c^2\phi^2}{4\omega L^2}   -\frac{\lambda \phi^2}{4\omega L^{\delta}}, \label{15}\\
\rho_{Dlog} &=& \frac{3c^2\phi^2}{4\omega L^2}   + \varrho L^{-4}\ln \left(\frac{\phi^2 L^2}{4\omega}\right) + \epsilon L^{-4}, \label{15log}
\end{eqnarray}
which can be also rewritten in the following way:
\begin{eqnarray}
\rho_{Dpl} &=& \frac{3c^2\phi^2}{4\omega L^2}\gamma_{pl}, \label{16} \\
\rho_{Dlog} &=& \frac{3c^2\phi^2}{4\omega L^2}\gamma_{log}, \label{16log}
\end{eqnarray}
where $\gamma_{pl}$ and $\gamma_{log}$ are defined, respectively, as follows:
\begin{eqnarray}
\gamma_{pl}  &=& 1  -\frac{\lambda }{3c^2 L^{\delta -2}}, \label{17} \\
\gamma_{log} &=& 1   + \frac{4\omega \varrho }{3c^2\phi^2 L^2}\ln \left(\frac{\phi^2 L^2}{4\omega}\right) + \frac{4\omega \epsilon }{3c^2\phi^2 L^2}  \nonumber \\
&=&    1   + \frac{4\omega  }{3c^2\phi^2 L^2} \left[   \varrho\ln \left(\frac{\phi^2 L^2}{4\omega}\right) + \epsilon \right].  \label{17log}
\end{eqnarray}
Eqs. (\ref{15}) and (\ref{15log}) are motivated by the fact that, in BD cosmology, the scalar field plays the role of the Newton's constant $G_N$, i.e. $\phi^2 \propto 1/G_N$.

In the limiting cases corresponding to $\lambda = 0$ and $\varrho = \epsilon = 0$ (i.e. in absence of both power-law and logarithmic corrections), we have that $\gamma_{pl} =\gamma_{log} = 1$, therefore the energy densities of the PLECHDE and the LECHDE models $\rho_{pl}$ and $\rho_{log}$ defined, respectively, in  Eqs. (\ref{15}) and (\ref{15log}) reduce to:
\begin{eqnarray}
\rho_{Dpl}= \rho_{Dlog} = \frac{3c^2\phi^2}{4\omega L^2}, \label{18}
\end{eqnarray}
which is the expression of the HDE energy density in the framework of Brans-Dicke cosmology.\\
Now, by substituting $L$ with $L_{GO}$, we can write Eqs. (\ref{15}), (\ref{16}) and (\ref{17}), which describe the DE energy density for the PLECHDE model, as follows:
\begin{eqnarray}
\rho_{Dpl} &=& \frac{3c^2\phi^2}{4\omega L_{GO}^2}   -\frac{\lambda \phi^2}{4\omega L_{GO}^{\delta}}, \label{19}\\
\rho_{Dpl} &=& \left(\frac{3c^2\phi^2}{4\omega L_{GO}^2}\right)\gamma_{pl},  \label{20}\\
\gamma_{pl}  &=& 1  -\frac{\lambda }{3c^2 L_{GO}^{\delta -2}}.  \label{21}
\end{eqnarray}
Moreover, by substituting $L$ with $L_{GO}$, we can write Eqs. (\ref{15log}), (\ref{16log}) and (\ref{17log}), which describe the DE energy density for the LECHDE model, as follows:
\begin{eqnarray}
\rho_{Dlog} &=& \frac{3c^2\phi^2}{4\omega L_{GO}^2}   + \varrho L_{GO}^{-4}\ln \left(\frac{\phi^2 L_{GO}^2}{4\omega}\right) + \epsilon L_{GO}^{-4}, \label{19log}\\
\rho_{Dlog} &=& \left(\frac{3c^2\phi^2}{4\omega L_{GO}^2}\right)\gamma_{log},  \label{20log}\\
\gamma_{log}  &=& 1   + \frac{4\omega \varrho }{3c^2\phi^2 L_{GO}^2}\ln \left(\frac{\phi^2 L_{GO}^2}{4\omega}\right) + \frac{4\omega \epsilon }{3c^2\phi^2 L_{GO}^2}.  \label{21log}
\end{eqnarray}
In the limiting case corresponding to $\lambda = 0$ and $\varrho = \epsilon =0$ (which imply absence of both power-law and logarithmic corrections and $\gamma_{pl} = \gamma_{log} =1$), the energy densities for the PLECHDE and the LECHDE models defined in  Eqs. (\ref{19}), (\ref{20}), (\ref{19log}) and (\ref{20log}) reduce to:
\begin{eqnarray}
\rho_{Dpl} =   \rho_{Dlog}   = \frac{3c^2\phi^2}{4\omega L_{GO}^2}, \label{22}
\end{eqnarray}
which represents the expression of HDE energy density in BD cosmology with Granda-Oliveros cut-off $L_{GO}$, a model recently studied by Jamil et al. \cite{jgo}.

In Eqs. (\ref{15}), (\ref{16}), (\ref{18}),  (\ref{19}), (\ref{20}), (\ref{22}), we have  $\phi^2 = \frac{\omega}{2\pi G_{eff}}$, where $G_{eff}$ represents the effective gravitational constant. In the limiting case of Einstein's gravity, remembering that in this case we have $G_{eff}\rightarrow G$, we have $\phi$ and $M_p$ are related through the relation $\phi^2 = 4\omega M_p^2$ and Eqs. (\ref{19}) and (\ref{19log}) reduce, respectively, to the PLECHDE and to the LECHDE energy densities in Einstein's gravity which are given, respectively, by the following relations:
\begin{eqnarray}
\rho_{Dpl} &=& \frac{3c^2M_p^2}{ L_{GO}^2}   -\frac{\lambda M_p^2}{ L_{GO}^{\delta}}, \label{19bo}\\
\rho_{Dlog} &=& \frac{3c^2M_p^2}{L_{GO}^2}   + \varrho L_{GO}^{-4}\ln \left(M_p^2L_{GO}^2\right) + \epsilon L_{GO}^{-4}. \label{19logbo}
\end{eqnarray}
The critical energy density (i.e. the average energy density required for flatness) $\rho_{cr}$ and the energy density of the curvature parameter $\rho_k$ are defined, respectively, as follows:
\begin{eqnarray}
\rho_{cr} &=& \frac{3\phi^2 H^2}{4\omega}, \label{23}   \\
\rho_k &=& \frac{3k\phi^2}{4\omega a^2}.  \label{24}
\end{eqnarray}
Instead, the fractional energy densities for DM, curvature parameter $k$ and DE are defined, respectively, as follows:
\begin{eqnarray}
\Omega_m &=& \frac{\rho_m}{\rho_{cr}} = \frac{4\omega \rho_m}{3\phi^2H^2}, \label{25}\\
\Omega_k &=& \frac{\rho_k}{\rho_{cr}} = \frac{k}{a^2H^2},\label{26} \\
\Omega_{Dpl} &=& \frac{\rho_{Dpl}}{\rho_{cr}} =   \frac{4\omega \rho_{Dpl}}{3\phi^2 H^2} = \frac{c^2\gamma_{pl}}{L_{GO}^2H^2}, \label{27} \\
\Omega_{Dlog} &=& \frac{\rho_{Dlog}}{\rho_{cr}} = \frac{4\omega \rho_{Dlog}}{3\phi^2 H^2}   =  \frac{c^2\gamma_{log}}{L_{GO}^2H^2}, \label{27log}
\end{eqnarray}
where we used the definitions of $\rho_{pl}$ and $\rho_{log}$ given, respectively, in Eqs. (\ref{20}) and  (\ref{20log}).\\
Combining Eqs. (\ref{13}), (\ref{23}) and (\ref{24}) with the Friedmann equation given in Eq. (\ref{10}), we obtain the following expressions:
\begin{eqnarray}
\rho_{cr} + \rho_k &=& \rho_m + \rho_{Dpl} + \rho_{\phi}, \label{28}\\
\rho_{cr} + \rho_k &=& \rho_m + \rho_{Dlog} + \rho_{\phi}, \label{28log}
\end{eqnarray}
where $\rho_{\phi}$ is defined as follows:
\begin{eqnarray}
\rho_{\phi} = \frac{1}{2}nH^2\phi^2\left( n - \frac{3}{\omega}   \right),\label{29}
\end{eqnarray}
and it represents the energy density of the BD scalar field $\phi$. We can clearly see that Eqs. (\ref{28}) and (\ref{28log}) related all the energy densities considered.

By dividing Eqs. (\ref{28}) and (\ref{28log})  by the critical energy density $\rho_{cr}$ defined in Eq. (\ref{23}) and using the definitions of the fractional energy densities given in Eqs. (\ref{25}), (\ref{26}), (\ref{27}) and (\ref{27log}),   Eqs. (\ref{28}) and (\ref{28log}) can be rewritten as follows:
\begin{eqnarray}
1+\Omega_k &=& \Omega_{Dpl} + \Omega_m + \Omega_{\phi}, \label{30} \\
1+\Omega_k &=& \Omega_{Dlog} + \Omega_m + \Omega_{\phi}, \label{30log}
\end{eqnarray}
where $\Omega_{\phi}$ is given by:
\begin{eqnarray}
\Omega_{\phi}= 2n\left( \frac{n\omega}{3} -1 \right), \label{31}
\end{eqnarray}
and it represents the fractional energy density of the BD scalar field $\phi$. The main feature of Eqs. (\ref{30}) and (\ref{30log}) is that they relate the fractional energy densities of all the components defined in this work, i.e. DE, DM, curvature parameter $k$ and Brans-Dicke parameter $\phi$.\\
As stated before, we choose the GO scale $L_{GO}$ given in Eq. (\ref{5}) as IR cut-off of the system.
The first derivative with respect to the cosmic time $t$ of the expression of $L_{GO}$ defined in Eq. (\ref{5}) is given by the following relation:
\begin{equation}
\dot{L}_{GO}=-H^{3}L_{GO}^{3}\left[ \alpha \left(\frac{\dot{H}}{H^{2}}\right)+\beta \left( \frac{\ddot{H}}{2H^{3}}\right) \right],  \label{32}
\end{equation}
while the first time derivative of the energy densities of the PLECHDE and the LECHDE models $\rho_{Dpl}$ and $\rho_{Dlog}$ given, respectively, by Eqs. (\ref{19}) and (\ref{19log}), can be written as follows:
\begin{eqnarray}
\dot{\rho}_{Dpl} &=& 6H^3\left[ \alpha \left(\frac{\dot{H}}{H^{2}}\right)+\beta \left( \frac{\ddot{H}}{2H^{3}}\right) \right]_{pl} \frac{\phi^2 c^2}{4\omega}  \left( 1  - \frac{\lambda \delta  }{6 c^2L_{GO}^{\delta -2}} \right) + \frac{3c^2\phi^2}{2\omega L_{GO}^2}nH \left(1  -\frac{\lambda }{3c^2 L_{GO}^{\delta -2}}   \right)    \nonumber \\
&=& 6H^3\left[ \alpha \left(\frac{\dot{H}}{H^{2}}\right)+\beta \left( \frac{\ddot{H}}{2H^{3}}\right) \right]_{pl}    \frac{\phi^2 c^2}{4\omega}  \left( 1 - \frac{\lambda \delta }{6 c^2L_{GO}^{\delta -2}} \right) + \frac{3c^2\phi^2}{2\omega L_{GO}^2}nH \gamma_{pl}, \label{33} \\
\dot{\rho}_{Dlog} &=&  6H^3\left[ \alpha \left(\frac{\dot{H}}{H^{2}}\right)+\beta \left( \frac{\ddot{H}}{2H^{3}}\right) \right]_{log}\times \nonumber \\
 && \frac{\phi^2  c^2}{4\omega} \left\{ 1  +\frac{4\omega}{3\phi^2 c^2}L_{GO}^{-2}\left[ 2\varrho \ln \left( \frac{\phi^2 L_{GO}^2}{4\omega}\right) -\varrho + 2\epsilon  \right] \right\}   \nonumber \\
&&+ \frac{\phi^2 c^2 nH}{4\omega L_{GO}^2}\left( \frac{8\omega \varrho}{\phi^2 c^2 L_{GO}^2} + 6 \right), \label{33log}
\end{eqnarray}
where we used the expression of $\dot{L}_{GO}$ given in Eq. (\ref{32}).\\
By differentiating the Friedmann equation given in Eq. (\ref{10}) with respect to the cosmic time $t$ and using the results of Eqs. (\ref{33}) and (\ref{33log}), we can write the terms $\left[ \alpha \left(\frac{\dot{H}}{H^{2}}\right)+\beta \left( \frac{\ddot{H}}{2H^{3}}\right) \right]_{pl}$ and $\left[ \alpha \left(\frac{\dot{H}}{H^{2}}\right)+\beta \left( \frac{\ddot{H}}{2H^{3}}\right) \right]_{log}$ for the PLECHDE and the LECHDE models, respectively, as follows:
\begin{eqnarray}
\left(\alpha \frac{\dot{H}}{H^2} + \beta \frac{\ddot{H}}{2H^3}\right)_{pl}&=&
\left[-\frac{nc^2}{L_{GO}^2 H^2}\gamma_{pl} + \frac{\dot{H}}{H^2}\left(1 -\Omega_{\phi}\right)
\right. \nonumber \\
&&\left. +n\left( \Omega_{Dpl} +  \Omega_m \right) -\left(\Omega_{Dpl} + \Omega_{\phi} -1\right) + \frac{\Omega_m}{2}  \right]\times \nonumber \\
&& \left( c^2  - \frac{\lambda \delta  }{6 L_{GO}^{\delta -2}} \right)^{-1},   \label{34} \\
\left(\alpha \frac{\dot{H}}{H^2} + \beta \frac{\ddot{H}}{2H^3}\right)_{log} &=&
\left[-\frac{ nH}{L_{GO}^2}\left( \frac{8\omega \varrho}{\phi^2 L_{GO}^2} + 6c^2 \right) + \frac{\dot{H}}{H^2}\left(1 -\Omega_{\phi}\right) \right. \nonumber \\
&&\left.  +n\left( \Omega_{Dlog} + \Omega_m \right)  -\left(\Omega_{Dlog} + \Omega_{\phi}  -1\right) + \frac{\Omega_m}{2}  \right] \nonumber \\
&&\times \left\{ c^2  +\frac{4\omega}{3\phi^2}L_{GO}^{-2}\left[ 2\varrho \ln \left( \frac{\phi^2 L_{GO}^2}{4\omega}\right) -\varrho + 2\epsilon  \right] \right\}^{-1}. \label{34log}
\end{eqnarray}
Using the definition of $L_{GO}$ given in Eq. (\ref{5}) and the expression of the fractional energy densities of DE  given in Eqs. (\ref{27}) and (\ref{27log}), we obtain, after some calculations, the following expressions for the terms $\left(\frac{\dot{H}}{H^2}\right)_{pl}$ and $\left(\frac{\dot{H}}{H^2}\right)_{log}$:
\begin{eqnarray}
\left(\frac{\dot{H}}{H^2}\right)_{pl} &=& \frac{1}{\beta}\left(\frac{\Omega_{Dpl}}{c^2\gamma_{pl}}  -\alpha\right), \label{35} \\
\left(\frac{\dot{H}}{H^2}\right)_{log} &=& \frac{1}{\beta}\left(\frac{\Omega_{Dlog}}{c^2\gamma_{log}}  -\alpha\right). \label{35log}
\end{eqnarray}
Therefore, using the results obtained in Eqs. (\ref{34}), (\ref{34log}), (\ref{35}) and (\ref{35log}), Eqs. (\ref{33}) and (\ref{33log})  yield, respectively:
\begin{eqnarray}
\dot{\rho}_{Dpl} &=& \frac{3H^3\phi^2}{4 \omega}\left\{\frac{2}{\beta} \left[\left(\frac{\Omega_{Dpl}}{c^2 \gamma_{pl}}-\alpha\right)\left(1 -\Omega_{\phi}\right) +\beta \right] \right. \nonumber \\
 && \left. +\Omega_{Dpl} \left[u_{pl}\left( 2n+1\right) + 2\left( n-1\right)  \right] - 2 \Omega_{\phi}   \right\} \nonumber \\
&=& \frac{H\rho_{Dpl}}{\Omega_{Dpl}}\left\{\frac{2}{\beta} \left[\left(\frac{\Omega_{Dpl}}{c^2 \gamma_{pl}}-\alpha\right)\left(1 -\Omega_{\phi}\right) +\beta \right] \right. \nonumber \\
 && \left. +\Omega_{Dpl} \left[u_{pl}\left( 2n+1\right) + 2\left( n-1\right)  \right] - 2 \Omega_{\phi}   \right\}, \label{37}\\
\dot{\rho}_{Dlog} &=& \frac{3H^3\phi^2}{4 \omega}\left\{\frac{2}{\beta} \left[\left(\frac{\Omega_{Dlog}}{c^2 \gamma_{log}}-\alpha\right)\left(1 -\Omega_{\phi}\right) +\beta \right] \right. \nonumber \\
 && \left. + \Omega_{Dlog} \left[u_{log}\left( 2n+1\right) + 2\left( n-1\right)  \right] - 2 \Omega_{\phi}   \right\} \nonumber \\
 &=&\frac{H\rho_{Dlog}}{\Omega_{Dlog}}\left\{\frac{2}{\beta} \left[\left(\frac{\Omega_{Dlog}}{c^2 \gamma_{log}}-\alpha\right)\left(1 -\Omega_{\phi}\right) +\beta \right] \right. \nonumber \\
 && \left. + \Omega_{Dlog} \left[u_{log}\left( 2n+1\right) + 2\left( n-1\right)  \right] - 2 \Omega_{\phi}   \right\}.
 \label{37log}
\end{eqnarray}
In Eqs. (\ref{37}) and (\ref{37log}), we have used the definitions of $\Omega_{Dpl}$ and $\Omega_{Dlog}$ given in Eqs. (\ref{27}) and (\ref{27log}).\\
Eqs. (\ref{37}) and (\ref{37log}) also lead to the following expressions for the evolutionary forms of the energy densities of the PLECHDE and LECHDE models  $\rho'_{Dpl}$ and $\rho'_{Dlog}$:
\begin{eqnarray}
\rho'_{Dpl} = \frac{\dot{\rho}_{Dpl}}{H} &=&  \frac{\rho_{Dpl}}{\Omega_{Dpl}}\left\{\frac{2}{\beta} \left[\left(\frac{\Omega_{Dpl}}{c^2 \gamma_{pl}}-\alpha\right)\left(1 -\Omega_{\phi}\right) +\beta \right] \right. \nonumber \\
 && \left. +\Omega_{Dpl} \left[u_{pl}\left( 2n+1\right) + 2\left( n-1\right)  \right] - 2 \Omega_{\phi}   \right\} , \label{37ev}\\
\rho'_{Dlog} = \frac{\dot{\rho}_{Dlog}}{H} &=& \frac{\rho_{Dlog}}{\Omega_{Dlog}}\left\{\frac{2}{\beta} \left[\left(\frac{\Omega_{Dlog}}{c^2 \gamma_{log}}-\alpha\right)\left(1 -\Omega_{\phi}\right) +\beta \right] \right. \nonumber \\
 && \left. + \Omega_{Dlog} \left[u_{log}\left( 2n+1\right) + 2\left( n-1\right)  \right] - 2 \Omega_{\phi}   \right\}. \label{37logev}
\end{eqnarray}
In Eqs. (\ref{37}), (\ref{37log}), (\ref{37ev}) and (\ref{37logev}), the quantities  $u_{pl}$ and $u_{log}$ are given, respectively, by the following relations:
\begin{eqnarray}
u_{pl} &=& \frac{\rho_m}{\rho_{Dpl}} = \frac{\Omega_m}{\Omega_{Dpl}} = \frac{1+\Omega_k - \Omega_{\phi}}{\Omega_{Dpl}} -1, \label{38}\\
u_{log} &=& \frac{\rho_m}{\rho_{Dlog}} = \frac{\Omega_m}{\Omega_{Dlog}} = \frac{1+\Omega_k - \Omega_{\phi}}{\Omega_{Dlog}} -1, \label{38log}
\end{eqnarray}
where we used the relations between all fractional energy density obtained in Eqs. (\ref{30}) and (\ref{30log}).\\

\subsection{Non Interacting Case}
We now want to obtain some important physical quantities in the case DE and DM do not interact each other.\\
In order to satisfy the local energy-momentum conservation law, which is given by $\nabla_{\mu}T^{\mu \nu}=0$ (with the term $T^{\mu \nu}$ representing the energy-momentum tensor), the total energy density, defined as $\rho_{tot}= \rho_D + \rho_m$, must satisfy the continuity relation given by:
\begin{eqnarray}
    \dot{\rho}_{tot}+3H\left( 1+\omega_{tot} \right)\rho_{tot}=0,\label{39}
\end{eqnarray}
where $\omega_{tot} \equiv p_{tot}/\rho_{tot}$ represents the total EoS parameter, with $p_{tot}$  denoting the total pressure, which is equivalent to the DE pressure $p_D$ since we consider pressureless DM.\\
If there is not interaction between the two Dark Sectors, i.e. DE and DM, the two energy densities $\rho_D$ and $\rho_m$ for DE and DM  are conserved separately in the following way:
\begin{eqnarray}
\dot{\rho}_{Dpl}&+&3H\rho_{Dpl}\left(1+\omega_{Dpl}\right)=0, \label{40} \\
\dot{\rho}_{Dlog}&+&3H\rho_{Dlog}\left(1+\omega_{Dlog}\right)=0, \label{40log} \\
\dot{\rho}_m&+&3H\rho_m= 0,\label{41}
\end{eqnarray}
where $\omega_{Dpl}$ and $\omega_{Dlog}$  are the EoS parameters of the PLECHDE and LECHDE models, respectively.\\
Eqs. (\ref{40}), (\ref{40log}) and (\ref{41}) can be written in the following equivalent forms:
\begin{eqnarray}
\rho'_{Dpl}&+&3\rho_{Dpl}\left(1+\omega_{Dpl}\right)=0, \label{40ev} \\
\rho'_{Dlog}&+&3\rho_{Dlog}\left(1+\omega_{Dlog}\right)=0, \label{40logev} \\
\rho'_m&+&3\rho_m= 0.\label{41ev}
\end{eqnarray}
From the continuity equations obtained in Eqs. (\ref{40}), (\ref{40log}), (\ref{40ev}) and (\ref{40logev}), we can easily obtain the following expressions for the EoS parameters for the PLECHDE and LECHDE models  $\omega_{Dpl}$ and $\omega_{Dlog}$:
\begin{eqnarray}
\omega_{Dpl} &=& -1 - \frac{\dot{\rho}_{Dpl}}{3H\rho_{Dpl}} = -1 - \frac{\rho'_{Dpl}}{3\rho_{Dpl}}, \label{car1} \\
\omega_{Dlog} &=& -1 - \frac{\dot{\rho}_{Dlog}}{3H\rho_{Dlog}} = -1 - \frac{\rho'_{Dlog}}{3\rho_{Dlog}}. \label{car2}
\end{eqnarray}
By inserting in Eqs. (\ref{car1}) and (\ref{car2}) the expressions of $\dot{\rho}_{Dpl}$ and $\dot{\rho}_{Dlog}$ obtained in Eqs. (\ref{40}) and (\ref{40log}) (or equivalently the expressions of $\rho'_{Dpl}$ and $\rho'_{Dlog}$ obtained in Eqs. (\ref{40ev}) and (\ref{40logev})), the EoS parameters for the PLECHDE and the LECHDE models $\omega_{Dpl}$ and $\omega_{Dlog}$ can be written, respectively, as follows:
\begin{eqnarray}
\omega_{Dpl} &=& -1 -\frac{1}{3\Omega_{Dpl} }\left\{\frac{2}{\beta}\left[\left(\frac{\Omega_{Dpl} }{c^2 \gamma_{pl}}-\alpha\right)\left(1 -\Omega_{\phi}\right) +\beta \right] \right. \nonumber \\
 && \left. + \Omega_{Dpl} \left[u_{pl}\left( 2n+1\right) + 2\left( n-1\right)  \right] - 2 \Omega_{\phi}  \right\}, \label{42}\\
\omega_{Dlog} &=& -1 -\frac{1}{3\Omega_{Dlog}}\left\{\frac{2}{\beta}\left[\left(\frac{\Omega_{Dlog}}{c^2 \gamma_{log}}-\alpha\right)\left(1 -\Omega_{\phi}\right) +\beta \right] \right. \nonumber \\
 && \left. + \Omega_{Dlog} \left[u_{log}\left( 2n+1\right) + 2\left( n-1\right)  \right] - 2 \Omega_{\phi}   \right\}. \label{42log}
\end{eqnarray}
We can clearly observe that the expressions of $\omega_{Dpl}$ and $\omega_{Dlog}$ depend on most of the quantities characterizing the DE models and the Brans-Dicke theory.\\
We now want to derive, for both models we are considering, the expression of the deceleration parameter $q$, which is generally defined as follows:
\begin{eqnarray}
q =  -\frac{\ddot{a}a}{\dot{a}^2} =  -1 - \frac{\dot{H}}{H^2}. \label{43}
\end{eqnarray}
The expansion of the Universe results to be accelerated if the quantity $\ddot{a}$ assumes a positive value, as recent cosmological measurements suggest;  in this case, $q$ assumes a negative value. Positive values of $q$ indicate a non accelerating universe.\\
We can use two different approaches which allow to find the final expression for the deceleration parameter $q$.\\
In the first one, we can insert the results of  Eqs. (\ref{35}) and (\ref{35log}) in Eq. (\ref{43}), obtaining, respectively, the following expressions for $q_{pl}$ and $q_{log}$:
\begin{eqnarray}
q_{pl} &=&-1   -\frac{1}{\beta}\left( \frac{\Omega_{Dpl}}{c^2 \gamma_{pl}} - \alpha   \right) \nonumber \\
&=&   -\frac{1}{\beta}\left( \frac{\Omega_{Dpl}}{c^2 \gamma_{pl}} +\beta- \alpha   \right),  \label{44} \\
q_{log} &=& -1 - \frac{1}{\beta}\left( \frac{\Omega_{Dlog}}{c^2 \gamma_{log}} - \alpha   \right) \nonumber \\
&=& -\frac{1}{\beta}\left( \frac{\Omega_{Dpl}}{c^2 \gamma_{log}} +\beta - \alpha   \right). \label{44log}
\end{eqnarray}
In the second way, dividing Eq. (\ref{11}) by $H^2$ and by using the results of Eqs. (\ref{13}), (\ref{14}), (\ref{26}), (\ref{27}) and (\ref{27log}), the following relations for $q_{pl}$ and $q_{log}$ are obtained:
\begin{eqnarray}
q_{pl} &=& \frac{1}{2\left( n+1\right)}\left[ \left( 2n+1\right)^2 + 2n\left( n\omega - 1\right)+\Omega_k  + 3\Omega_{Dpl} \omega_{Dpl}   \right], \label{45} \\
q_{log} &=& \frac{1}{2\left( n+1\right)}\left[ \left( 2n+1\right)^2 + 2n\left( n\omega - 1\right)+\Omega_k  + 3\Omega_{Dlog} \omega_{Dlog}   \right]. \label{45log}
\end{eqnarray}
The final expression for the deceleration parameters $q_{pl}$ and $q_{log}$ can be  now derived by substituting in Eqs. (\ref{45}) and (\ref{45log}) the expression of the EoS parameters $\omega_{Dpl}$ and $\omega_{Dlog}$ derived, respectively, in Eqs. (\ref{42}) and (\ref{42log}). We can clearly realize that the final expressions of the deceleration parameters $q_{pl}$ and $q_{log}$ depend on the fractional energy densities of DE, on the EoS parameters of DE, on the fractional energy density of curvature $\Omega_k$, on the Brans-Dicke parameter $\omega$ and upon the index $n$. \\
We now want to study some limiting cases, in particular we will consider the equations obtained without entropy corrections (i.e. for $\lambda = 0$ for the power law correction and $\varrho = \varepsilon = 0$ for the logarithmic correction), in the limiting case of Einstein gravity (i.e., for $n=0$ and $\Omega_{\phi} = \rho_{\phi}=0$), without entropy corrections and in the limiting case of Einstein gravity together, for the Ricci scale (which is recovered when $\alpha =2$ and $\beta = 1$) and for the flat Dark Dominated Universe, which is recovered for  $\Omega_{Dpl}  =  \Omega_{Dlog}    = 1$, $\gamma_{pl}  =  \gamma_{log}   =  1$, $\Omega_m = \Omega_k = \Omega_{\phi} = 0$, $n=0$ and $u_{pl} =   u_{log}   = 0$.\\
In the limiting case corresponding to $\lambda =0$ (which implies $\gamma_{pl} =1$ and no power-law corrections), Eqs. (\ref{42}), (\ref{44}) and (\ref{45}) reduce, respectively, to:
\begin{eqnarray}
\omega_{Dpl_{\lambda=0}}\! &\!=\!&\! -1\! -\frac{1}{3\Omega_{Dpl_{\lambda=0}}}\left\{\frac{2}{\beta}\left[\left(\frac{\Omega_{Dpl_{\lambda=0}}}{c^2}-\alpha\right)\!\left(1 -\Omega_{\phi}\right) +\beta \right] +\right. \nonumber \\
&&\left. \Omega_{Dpl_{\lambda=0}}\left[u_{pl_{\lambda=0}}\left( 2n+1  \right) + 2\left( n-1  \right)   \right] - 2\Omega_{\phi}    \right\}, \label{42-1} \\
q_{pl,\lambda=0} &=& -\frac{1}{\beta}\left( \frac{\Omega_{Dpl_{\lambda=0}}}{c^2} + \beta - \alpha \right) \nonumber \\
 &=& -1-\frac{1}{\beta}\left( \frac{\Omega_{Dpl_{\lambda=0}}}{c^2}  - \alpha \right),  \label{44-1}\\
q_{pl,\lambda=0} &=& \frac{1}{2\left( n+1\right)}\left[ \left( 2n+1\right)^2 + 2n\left( n\omega - 1\right)+\Omega_k  + 3\Omega_{Dpl_{\lambda=0}} \omega_{Dpl_{\lambda=0}}   \right], \label{45-1-1}
\end{eqnarray}
while, in the limiting cases corresponding to $\varrho = \epsilon = 0 $  (which means $\gamma_{log} =1$, i.e. no logarithmic correction), Eqs. (\ref{42log}), (\ref{44log}) and (\ref{45log}) reduce, respectively, to:
\begin{eqnarray}
\omega_{Dlog_{\varrho = \epsilon =0}}\! &\!=\!&\! -1\! -\frac{1}{3\Omega_{Dlog_{\varrho = \epsilon =0}}}\left\{\frac{2}{\beta}\left[\left(\frac{\Omega_{Dlog_{\varrho = \epsilon =0}}}{c^2}-\alpha\right)\!\left(1 -\Omega_{\phi}\right)\! +\beta \right] \!+ \right. \nonumber \\
&&\left. \Omega_{Dlog_{\varrho = \epsilon =0}}\left[u_{log_{\varrho = \epsilon =0}}\left( 2n+1  \right) + 2\left( n-1  \right)   \right] - 2\Omega_{\phi}     \right\}, \label{42-1log} \\
q_{log,\varrho = \epsilon =0} &=& -\frac{1}{\beta}\left( \frac{\Omega_{Dlog_{\varrho = \epsilon =0}}}{c^2} + \beta - \alpha   \right) \nonumber \\
 &=& -1 -\frac{1}{\beta}\left( \frac{\Omega_{Dlog_{\varrho = \epsilon =0}}}{c^2} - \alpha   \right),  \label{44-1log}\\
q_{log,\varrho = \epsilon =0} &=& \frac{1}{2\left( n+1\right)}\left[ \left( 2n+1\right)^2 + 2n\left( n\omega - 1\right)+\Omega_k  + 3\Omega_{Dlog_{\varrho = \epsilon =0}} \omega_{Dlog_{\varrho = \epsilon =0}}   \right], \label{45-1-1log}
\end{eqnarray}
where, from Eqs. (\ref{27}), (\ref{27log}), (\ref{38}) and (\ref{38log}), we can derive, respectively, the following relations for $\Omega_{Dpl_{\lambda=0}}$, $\Omega_{Dlog_{\lambda=0}}$, $u_{pl,\lambda=0}$ and $u_{log,\lambda=0}$:
\begin{eqnarray}
\Omega_{Dpl_{\lambda=0}} &=& \Omega_{Dlog_{\varrho = \epsilon =0}} =  \frac{c^2}{L_{GO}^2H^2}  , \label{lambda01}\\
u_{pl,\lambda=0} &=& \frac{\Omega_m}{\Omega_{Dpl_{\lambda=0}}} =   \frac{1+\Omega_k - \Omega_{\phi}}{\Omega_{Dpl_{\lambda=0}}} -1 ,\label{lambda02} \\
u_{log,\varrho = \epsilon =0} &=& \frac{\Omega_m}{\Omega_{Dlog_{\varrho = \epsilon =0}}} =  \frac{1+\Omega_k - \Omega_{\phi}}{\Omega_{Dlog_{\varrho = \epsilon =0}}} -1.\label{lambda02-2}
\end{eqnarray}
Taking into account Eq. (\ref{lambda01}), we can conclude that Eqs. (\ref{lambda02}) and (\ref{lambda02-2}) are equivalent, i.e. $u_{pl,\lambda=0} = u_{log,\varrho = \epsilon =0} $.\\
In the limiting case corresponding to Einstein's gravity (i.e., for $n=0$ and $\Omega_{\phi} = \rho_{\phi}=0$), the EoS parameter $\omega_{Dpl}$ and the deceleration parameter $q_{pl}$ given in Eqs. (\ref{42}), (\ref{44})  and (\ref{45})  reduce, respectively, to:
\begin{eqnarray}
\omega_{Dpl,Ein} &=& -1 -\frac{1}{3\Omega_{Dpl,Ein}}\left\{\frac{2}{\beta}\left[\left(\frac{\Omega_{Dpl,Ein}}{c^2 \gamma_{pl,Ein}}-\alpha\right) +\beta \right] \right. \nonumber \\
&&\left.+ \Omega_{Dpl,Ein}\left[u_{pl,Ein} -2\right]  \right\} \nonumber \\
&=&  \frac{2}{3\beta \Omega_{Dpl,Ein}}\left(\frac{\Omega_{Dpl,Ein}}{c^2 \gamma_{pl,Ein}}-\alpha + \beta\right) - \frac{1+u_{pl,Ein}}{3}, \label{42-2}\\
q_{pl,Ein} &=&   -\frac{1}{\beta}\left( \frac{\Omega_{Dpl,Ein}}{c^2 \gamma_{pl,Ein}}+\beta - \alpha   \right)\nonumber \\ &=& -1-\frac{1}{\beta}\left( \frac{\Omega_{Dpl,Ein}}{c^2 \gamma_{pl,Ein}} - \alpha   \right) ,  \label{44-2} \\
q_{pl,Ein} &=& \frac{1}{2}\left[ 1 +\Omega_k  + 3\Omega_{Dpl,Ein} \omega_{Dpl,Ein}  \right], \label{45-1}
\end{eqnarray}
while the EoS parameter $\omega_{Dlog,Ein}$ and the deceleration parameter $q_{log,Ein}$ given in Eqs. (\ref{42log}), (\ref{44log})  and (\ref{45log})  reduce, respectively, to:
\begin{eqnarray}
\omega_{Dlog,Ein} &=& -1 -\frac{1}{3\Omega_{Dlog,Ein}}\left\{\frac{2}{\beta}\left[\left(\frac{\Omega_{Dlog,Ein}}{c^2 \gamma_{log,Ein}}-\alpha\right) +\beta \right]\right. \nonumber \\
&&\left. + \Omega_{Dlog,Ein}\left[u_{log,Ein} -2\right]  \right\}, \nonumber \\
&=& \frac{2}{3\beta \Omega_{Dlog,Ein}}\left(\frac{\Omega_{Dlog,Ein}}{c^2 \gamma_{log,Ein}}-\alpha + \beta\right) - \frac{1+u_{log,Ein}}{3}\label{42log-2}\\
q_{log,Ein} &=&   -\frac{1}{\beta}\left( \frac{\Omega_{Dlog,Ein}}{c^2 \gamma_{log,Ein}}+\beta - \alpha   \right) \nonumber \\ &=&  -1-\frac{1}{\beta}\left( \frac{\Omega_{Dlog,Ein}}{c^2 \gamma_{log,Ein}} - \alpha   \right)  \label{44log!!!}, \\
q_{log,Ein} &=& \frac{1}{2}\left[ 1 +\Omega_k  + 3\Omega_{Dlog,Ein} \omega_{Dlog,Ein}  \right]. \label{45log-1}
\end{eqnarray}
We have that the expressions of the EoS parameters obtained in Eqs. (\ref{42-2}) and (\ref{42log-2}) and the expressions of the deceleration parameters obtained in Eqs. (\ref{45-1}) and (\ref{45log-1})  are similar to the EoS and the deceleration parameter $q$ obtained in Khodam et al. \cite{kho8} in the limiting case of $c^2=1$.\\
Using the expression obtained in Eqs. (\ref{17}) and (\ref{17log}), we have that the quantities $\gamma_{pl,Ein}$ and $\gamma_{log,Ein}$ are given, respectively, by the following relations:
\begin{eqnarray}
\gamma_{pl,Ein}  &=&   \gamma_{pl}  = 1  -\frac{\lambda }{3c^2 L_{GO}^{\delta -2}}, \label{17ein} \\
\gamma_{log,Ein} &=& 1   + \frac{\varrho }{3c^2M_p^2 L_{GO}^2}\ln \left(M_p^2 L_{GO}^2\right) + \frac{\epsilon }{3c^2M_p^2 L_{GO}^2},  \label{17logein}
\end{eqnarray}
where we have considered the fact that, for Einstein gravity, the relation $\frac{\phi^2}{4\omega}=M_p^2$ is valid. Moreover, we have that the fractional energy densities of DE for the two models we are studying are given, respectively, by:
\begin{eqnarray}
\Omega_{Dpl,Ein} &=& \frac{c^2\gamma_{pl,Ein}}{L_{GO}^2H^2} =  \frac{c^2\gamma_{pl}}{L_{GO}^2H^2} , \label{27esoein} \\
\Omega_{Dlog,Ein} &=&  \frac{c^2\gamma_{log,Ein}}{L_{GO}^2H^2}. \label{27logesoein}
\end{eqnarray}
In the limiting case corresponding to $\lambda = 0$ (i.e. for $ \gamma_{pl} =1$) and to Einstein's gravity together, the EoS parameter $\omega_{Dpl}$ and the deceleration parameter $q_{pl}$ given Eqs. (\ref{42}), (\ref{44}) and (\ref{45}) lead, respectively, to:
\begin{eqnarray}
\omega_{Dpl,Ein_{\lambda=0}} &=& -1 -\frac{1}{3\Omega_{Dpl,Ein_{\lambda=0}}}\left\{\frac{2}{\beta}\left[\left(\frac{\Omega_{Dpl,Ein_{\lambda=0}}}{c^2 }-\alpha\right) +\beta \right] + \right. \nonumber \\
&&\left.\Omega_{Dpl,Ein_{\lambda=0}}u_{pl,Ein_{\lambda=0}}  \right\}, \label{42-3}\\
q_{pl,Ein_{\lambda=0}} &=& -\frac{1}{\beta}\left(\frac{\Omega_{Dpl,Ein_{\lambda=0}}}{c^2} +\beta - \alpha   \right) \nonumber \\ &=& -1-\frac{1}{\beta}\left(\frac{\Omega_{Dpl,Ein_{\lambda=0}}}{c^2} - \alpha   \right),  \label{44-2} \\
q_{pl,Ein_{\lambda=0}} &=& \frac{1}{2}\left( 1 +\Omega_k  + 3\Omega_{Dpl,Ein_{\lambda=0}} \omega_{Dpl,Ein_{\lambda=0}}   \right) \label{45-2}.
\end{eqnarray}
while, in the limiting case corresponding to $\varrho = \epsilon = 0 $ (i.e. for $ \gamma_{log} =1$) and to Einstein's gravity together, the EoS parameter $\omega_{Dlog}$ and the deceleration parameter $q_{log}$ given  Eqs. (\ref{42log}), (\ref{44log}) and (\ref{45log}) lead, respectively, to:
\begin{eqnarray}
\omega_{Dlog,Ein_{\varrho = \epsilon =0}} &=& -1 -\frac{1}{3\Omega_{Dlog,Ein_{\varrho = \epsilon =0}}}\left\{\frac{2}{\beta}\left[\left(\frac{\Omega_{Dlog,Ein_{\varrho = \epsilon =0}}}{c^2 }-\alpha\right) +\beta \right] +\right. \nonumber \\
&& \left. \Omega_{Dlog,Ein_{\varrho = \epsilon=0}}u_{log,Ein_{\varrho = \epsilon =0}}  \right\}, \label{42-3}\\
q_{log,Ein_{\varrho = \epsilon =0}} &=& -\frac{1}{\beta}\left(\frac{\Omega_{Dlog,Ein_{\varrho = \epsilon=0}}}{c^2} + \beta - \alpha   \right) \nonumber \\ &=& -1-\frac{1}{\beta}\left(\frac{\Omega_{Dlog,Ein_{\varrho = \epsilon=0}}}{c^2} - \alpha   \right),  \label{44-2} \\
q_{log,Ein_{\varrho = \epsilon =0}} &=& \frac{1}{2}\left( 1 +\Omega_k  + 3\Omega_{Dlog,Ein_{\varrho = \epsilon=0}} \omega_{Dlog,Ein_{\varrho = \epsilon=0}}   \right) \label{45-2},
\end{eqnarray}
where $\Omega_{Dpl_{\lambda=0}}$, $\Omega_{Dlog_{\varrho = \epsilon=0}}$, $u_{pl,\lambda=0}$ and $u_{log,\varrho = \epsilon=0}$ are defined, respectively, in Eqs. (\ref{lambda01}), (\ref{lambda02}) and (\ref{lambda02-2}).\\
For the Ricci scale, which is recovered when $\alpha =2$ and $\beta =1$,  we have that the EoS parameter $\omega_{Dpl}$ and the deceleration  parameter $q_{pl}$ for the power law case obtained Eqs. (\ref{42}), (\ref{44}) and (\ref{45}) reduce, respectively, to:
\begin{eqnarray}
\omega_{Dpl,Ricci} &=& -1 -\frac{1}{3\Omega_{Dpl,Ricci} }\left\{2\left[\left(\frac{\Omega_{Dpl,Ricci} }{c^2 \gamma_{pl,Ricci}}-2\right)\left(1 -\Omega_{\phi}\right) +1 \right] \right. \nonumber \\
 && \left. + \Omega_{Dpl,Ricci} \left[u_{pl,Ricci}\left( 2n+1\right) + 2\left( n-1\right)  \right] - 2 \Omega_{\phi}  \right\}, \label{42buby}\\
q_{pl,Ricci} &=&   -\left( \frac{\Omega_{Dpl,Ricci}}{c^2 \gamma_{pl,Ricci}} -1   \right),  \label{44buby} \\
q_{pl,Ricci} &=& \frac{1}{2\left( n+1\right)}\left[ \left( 2n+1\right)^2 + 2n\left( n\omega - 1\right)+\Omega_k  + 3\Omega_{Dpl,Ricci} \omega_{Dpl,Ricci}   \right], \label{45buby}
\end{eqnarray}
while the EoS parameter $\omega_{Dlog}$ and the deceleration parameter $q_{log}$ for the logarithmic case obtained in Eqs. (\ref{42log}), (\ref{44log}) and (\ref{45log}) reduce, respectively, to:
\begin{eqnarray}
\omega_{Dlog,Ricci} &=& -1 -\frac{1}{3\Omega_{Dlog,Ricci} }\left\{2\left[\left(\frac{\Omega_{Dlog,Ricci} }{c^2 \gamma_{log,Ricci}}-2\right)\left(1 -\Omega_{\phi}\right) +1 \right] \right. \nonumber \\
 && \left. + \Omega_{Dlog,Ricci} \left[u_{log,Ricci}\left( 2n+1\right) + 2\left( n-1\right)  \right] - 2 \Omega_{\phi}  \right\}, \label{42buby2}\\
q_{log,Ricci} &=&   -\left( \frac{\Omega_{Dlog,Ricci}}{c^2 \gamma_{log,Ricci}} - 1  \right),  \label{44buby2} \\
q_{log,Ricci} &=& \frac{1}{2\left( n+1\right)}\left[ \left( 2n+1\right)^2 + 2n\left( n\omega - 1\right)+\Omega_k  + 3\Omega_{Dlog,Ricci} \omega_{Dlog,Ricci}   \right]. \label{45buby2}
\end{eqnarray}
Finally, we have that, for the limiting case of a flat Dark Dominated Universe, the EoS parameter $\omega_{Dpl}$ and the deceleration  parameter $q_{pl}$ for the power law case obtained in Eqs. (\ref{42}), (\ref{42log}), (\ref{44}), (\ref{44log}), (\ref{45}) and (\ref{45log}) reduce, respectively, to the following quantities:
\begin{eqnarray}
\omega_{Dpl} &=& \omega_{Dlog} = -1 -\frac{2}{3\beta}\left(\frac{1}{c^2}-\alpha +\beta \right), \\
q_{pl} &=& q_{log} =   -\frac{1}{\beta}\left( \frac{1}{c^2}+\beta - \alpha   \right) \nonumber \\
 &=&-1 -\frac{1}{\beta}\left( \frac{1}{c^2} - \alpha   \right), \\
q_{pl} &=& \frac{1}{2}\left( 1   + 3 \omega_{Dpl}   \right), \\
 q_{log} &=& \frac{1}{2}\left( 1   + 3 \omega_{Dlog}   \right).
\end{eqnarray}
At Ricci scale, we obtain that the results for the flat Dark Dominated case lead to the following results:
\begin{eqnarray}
\omega_{Dpl} &=& \omega_{Dlog} = -1 -\frac{2}{3}\left(\frac{1}{c^2}-1 \right)  , \\
q_{pl} &=& q_{log} =   1 - \frac{1}{c^2}  , \\
q_{pl} &=& \frac{1}{2}\left( 1   + 3 \omega_{Dpl}   \right), \nonumber \\
q_{log} &=& \frac{1}{2}\left( 1   + 3 \omega_{Dlog}   \right).
\end{eqnarray}

We now want to obtain the final expression of the evolutionary form of the energy density parameter of DE $\Omega'_D$ for both the power law and the logarithmic corrections considered in this work.\\
Differentiating  Eqs. (\ref{27}) and (\ref{27log}) with respect to $x$ and using the relations $\dot{\Omega}_{Dpl}=H\Omega_{Dpl}'$ and $\dot{\Omega}_{Dlog}=H\Omega_{Dlog}'$, we derive that:
\begin{eqnarray}
\Omega_{Dpl}' &=& \Omega_{Dpl} \left\{  \frac{\rho_{Dpl}'}{\rho_{Dpl}}    -2\left[ \frac{\phi'}{\phi}+\left(\frac{\dot{H}}{H^2}\right)_{pl}  \right]  \right\},\label{cr7pl}\\
\Omega_{Dlog}' &=& \Omega_{Dlog} \left\{  \frac{\rho_{Dlog}'}{\rho_{Dlog}}    -2\left[ \frac{\phi'}{\phi}+\left(\frac{\dot{H}}{H^2}\right)_{log}  \right]  \right\},\label{cr7log}
\end{eqnarray}
where we used the fact that:
\begin{eqnarray}
\frac{\phi'}{\phi} = \frac{\dot{\phi}}{H\phi} = n.
\end{eqnarray}
Inserting in Eqs. (\ref{cr7pl}) and (\ref{cr7log}) the expressions of $\left(\frac{\dot{H}}{H^2}\right)_{pl}$, $\left(\frac{\dot{H}}{H^2}\right)_{log}$, $\rho_{Dpl}'$ and $\rho_{Dlog}'$, obtained, respectively, in Eqs. (\ref{35}), (\ref{35log}), (\ref{37ev}) and (\ref{37logev}), we obtain the following relations for $\Omega_{Dpl}'$ and $\Omega_{Dlog}'$:
\begin{eqnarray}
\Omega_{Dpl}' &=& \frac{2}{\beta} \left(  \frac{\Omega_{Dpl}}{c^2\gamma_{pl}} +\beta  - \alpha \right)\left(1-\Omega_{\phi} - \Omega_{Dpl}   \right)    + \Omega_{m}\left( 2n+1\right) \nonumber  \\
&=& \frac{2}{\beta} \left(  \frac{\Omega_{Dpl}}{c^2\gamma_{pl}} +\beta  - \alpha \right)\left(1-\Omega_{\phi} - \Omega_{Dpl}   \right)    + u_{pl}\Omega_{Dpl}\left( 2n+1\right), \label{52kissnat} \\
\Omega_{Dlog}' &=& \frac{2}{\beta} \left(  \frac{\Omega_{Dlog}}{c^2\gamma_{log}} +\beta  - \alpha \right)\left(1-\Omega_{\phi} - \Omega_{Dlog}  \right)    + \Omega_{m}\left( 2n+1\right) \nonumber \\
&=&\frac{2}{\beta} \left(  \frac{\Omega_{Dlog}}{c^2\gamma_{log}} +\beta  - \alpha \right)\left(1-\Omega_{\phi} - \Omega_{Dlog}  \right)    +u_{log}\Omega_{Dlog}\left( 2n+1\right), \label{52logkissnat}
\end{eqnarray}
where the prime $'$ denotes the first derivative with respect to the parameter $x =\ln a$. We must also underline that we used the relation $\frac{d}{dt} = H \frac{d}{dx}$.\\
We can now study some limiting cases of the evolutionary forms of the fractional energy densities obtained in Eqs. (\ref{52kissnat}) and (\ref{52logkissnat}), in particular we will consider the equations obtained without entropy corrections (i.e. for $\lambda = 0$ for the power law correction and $\varrho = \varepsilon = 0$ for the logarithmic correction), in the limiting case of Einstein gravity (i.e., for $n=0$ and $\Omega_{\phi} = \rho_{\phi}=0$), without entropy corrections and in the limiting case of Einstein gravity together, for the Ricci scale (which is recovered when $\alpha =2$ and $\beta = 1$) and for the flat Dark Dominated Universe, which is recovered for   $\Omega_{Dpl}  =  \Omega_{Dlog}    = 1$, $\gamma_{pl}  =  \gamma_{log}   =  1$, $\Omega_m = \Omega_k = \Omega_{\phi} = 0$, $n=0$ and $u_{pl} =   u_{log}   = 0$.\\
In the limiting cases corresponding to $\lambda =0$ and $\varrho = \epsilon =0$ (which imply that $\gamma_{pl} = \gamma_{log} = 1$), Eqs. (\ref{52kissnat}) and (\ref{52logkissnat}) lead, respectively, to:
\begin{eqnarray}
\Omega_{Dpl_{\lambda=0}}' &=& \frac{2}{\beta}\left( \frac{\Omega_{Dpl_{\lambda=0}}}{c^2 } +\beta -\alpha \right)\left(1-\Omega_{\phi} - \Omega_{Dpl_{\lambda=0}}   \right)     + \Omega_m \left( 2n+1   \right) \nonumber \\
&=& \frac{2}{\beta}\left( \frac{\Omega_{Dpl_{\lambda=0}}}{c^2 } +\beta -\alpha \right)\left(1-\Omega_{\phi} - \Omega_{Dpl_{\lambda=0}}   \right) \nonumber \\
&&+ u_{pl,\lambda=0}\Omega_{Dpl_{\lambda=0}}\left( 2n+1\right), \label{52-1catalina1}\\
\Omega_{Dlog_{\varrho = \epsilon=0}}' &=& \frac{2}{\beta}\left( \frac{\Omega_{Dlog_{\varrho = \epsilon=0}}}{c^2 } +\beta -\alpha \right)\left(1-\Omega_{\phi} - \Omega_{Dlog_{\varrho = \epsilon=0}}   \right)    + \Omega_m \left( 2n+1   \right) \nonumber  \\
&=& \frac{2}{\beta}\left( \frac{\Omega_{Dlog_{\varrho = \epsilon=0}}}{c^2 } +\beta -\alpha \right)\left(1-\Omega_{\phi} - \Omega_{Dlog_{\varrho = \epsilon=0}}   \right) \nonumber\\
&&+ u_{log,\varrho = \epsilon=0}\Omega_{Dlog_{\varrho = \epsilon=0}}\left( 2n+1\right). \label{52-1logcatalina2}
\end{eqnarray}
We can clearly observe, considering the expressions given in Eqs. (\ref{lambda01}), (\ref{lambda02}) and (\ref{lambda02-2}), that Eqs. (\ref{52-1catalina1}) and (\ref{52-1logcatalina2}) are identical.\\
Furthermore, in the limiting case corresponding to Einstein's gravity (i.e. for $n=0$ and $\Omega_{\phi} = \rho_{\phi}=0$), Eqs. (\ref{52kissnat}) and (\ref{52logkissnat}) reduce, respectively, to:
\begin{eqnarray}
\Omega_{Dpl,Ein}' &=& \frac{2}{\beta}\left( \frac{\Omega_{Dpl,Ein}}{c^2 \gamma_{pl,Ein}} +\beta -\alpha \right) \left(1 - \Omega_{Dpl,Ein}   \right)    + \Omega_m  \nonumber \\
&=& \frac{2}{\beta}\left( \frac{\Omega_{Dpl,Ein}}{c^2 \gamma_{pl,Ein}} +\beta -\alpha \right) \left(1 - \Omega_{Dpl,Ein}   \right)    + \Omega_{Dpl,Ein}u_{pl,Ein} , \label{52-2} \\
\Omega_{Dlog,Ein}' &=& \frac{2}{\beta}\left( \frac{\Omega_{Dlog,Ein}}{c^2 \gamma_{log,Ein}} +\beta -\alpha \right)\left(1 - \Omega_{Dlog,Ein}   \right)    + \Omega_m \nonumber \\
&=&\frac{2}{\beta}\left( \frac{\Omega_{Dlog,Ein}}{c^2 \gamma_{log,Ein}} +\beta -\alpha \right)\left(1 - \Omega_{Dlog,Ein}   \right)    + \Omega_{Dlog,Ein}u_{log,Ein}. \label{52-2log}
\end{eqnarray}
Finally, in the limiting case corresponding to $\lambda =0$, $\varrho = \epsilon=0$ (i.e. in the case of absence of entropy corrections)  and to Einstein's gravity together, Eqs. (\ref{52kissnat}) and (\ref{52logkissnat}) read, respectively, as follows:
\begin{eqnarray}
\Omega_{Dpl,Ein_{\lambda=0}}' &=&     \frac{2}{\beta}\left( \frac{\Omega_{Dpl,Ein_{\lambda=0}}}{c^2} +\beta -\alpha \right)\left(1 - \Omega_{Dpl,Ein_{\lambda=0}}   \right)    + \Omega_m \nonumber \\
 &=&     \frac{2}{\beta}\left( \frac{\Omega_{Dpl,Ein_{\lambda=0}}}{c^2} +\beta -\alpha \right)\left(1 - \Omega_{Dpl,Ein_{\lambda=0}}   \right)   \nonumber\\
 &&+ u_{pl,Ein_{\lambda=0}}\Omega_{Dpl,Ein_{\lambda=0}}, \label{52-3catalina22} \\
\Omega_{Dlog,Ein_{\varrho = \epsilon=0}}' &=&    \frac{2}{\beta}\left( \frac{\Omega_{Dlog,Ein_{\varrho = \epsilon=0}}}{c^2} +\beta -\alpha \right)\left(1- \Omega_{Dlog,Ein_{\varrho = \epsilon=0}}   \right)     + \Omega_m \nonumber \\
&=&    \frac{2}{\beta}\left( \frac{\Omega_{Dlog,Ein_{\varrho = \epsilon=0}}}{c^2} +\beta -\alpha \right)\left(1- \Omega_{Dlog,Ein_{\varrho = \epsilon=0}}   \right)   \nonumber\\
  &&+u_{log,Ein_{\varrho = \epsilon=0}} \Omega_{Dlog,Ein_{\varrho = \epsilon=0}} .\label{52-3logcatalina22}
\end{eqnarray}
Considering the results obtained in Eq. (\ref{lambda01}), (\ref{lambda02}) and (\ref{lambda02-2}), we derive that Eqs. (\ref{52-3catalina22}) and (\ref{52-3logcatalina22}) are equivalent.\\
For the Ricci scale (i.e. when $\alpha=2$ and $\beta =1$), we obtain that  Eqs. (\ref{52kissnat}) and (\ref{52logkissnat}) lead, respectively, to:
\begin{eqnarray}
\Omega_{Dpl,Ricci}' &=& 2 \left(  \frac{\Omega_{Dpl,Ricci}}{c^2\gamma_{pl,Ricci}} -1 \right)\left(1-\Omega_{\phi} - \Omega_{Dpl,Ricci}   \right)    + \Omega_m\left( 2n+1\right) \nonumber \\
&=&\left(  \frac{\Omega_{Dpl,Ricci}}{c^2\gamma_{pl,Ricci}} -1 \right)\left(1-\Omega_{\phi} - \Omega_{Dpl,Ricci}   \right)   \nonumber \\
&&+ u_{pl,Ricci}\Omega_{Dpl,Ricci}\left( 2n+1\right), \label{52ricciscale} \\
\Omega_{Dlog,Ricci}' &=& 2 \left(  \frac{\Omega_{Dlog,Ricci}}{c^2\gamma_{log,Ricci}} -1 \right)\left(1-\Omega_{\phi} - \Omega_{Dlog,Ricci}  \right)    + \Omega_m\left( 2n+1\right)\nonumber  \\
&=& 2 \left(  \frac{\Omega_{Dlog,Ricci}}{c^2\gamma_{log,Ricci}} -1 \right)\left(1-\Omega_{\phi} - \Omega_{Dlog,Ricci}  \right)    \nonumber\\
&&+u_{pl,Ricci} \Omega_{Dlog,Ricci}\left( 2n+1\right).\label{52logricciscale}
\end{eqnarray}
Finally, for the limiting case of a flat Dark Dominated Universe, we obtain that both expressions of $\Omega_{Dpl}' $ and $\Omega_{Dlog}'$ are equals to zero, as it must be since for a flat  Dark Dominated Universe we have that $\Omega_D = 1$.\\
It is also possible to find, after some algebraic calculation, the following relations between the EoS parameters $\omega_{Dpl}$ and $\omega_{Dlog}$ and the deceleration parameters $q_{pl}$ and $q_{log}$:
\begin{eqnarray}
\omega_{Dpl} &=& \frac{2q_{pl}}{3\Omega_{Dpl}}\left(1-\Omega_{\phi} \right) - 1 - \frac{1}{3}\left[ u_{pl}\left( 2n+1 \right) +2\left(  n-1 \right) \right] \nonumber \\
&=& \frac{2q_{pl}}{3\Omega_{Dpl}}\left(1-\Omega_{\phi} \right) - \frac{1}{3}\left[ \left(u_{pl}+1\right)\left( 2n+1 \right) \right], \label{carolinghia1} \\
\omega_{Dlog} &=& \frac{2q_{log}}{3\Omega_{Dlog}}\left(1-\Omega_{\phi} \right) - 1 - \frac{1}{3}\left[ u_{log}\left( 2n+1 \right) +2\left(  n-1 \right) \right] \nonumber \\
&=& \frac{2q_{log}}{3\Omega_{Dlog}}\left(1-\Omega_{\phi} \right) - \frac{1}{3}\left[ \left(u_{log}+1\right)\left( 2n+1 \right) \right]. \label{carolinghia2}
\end{eqnarray}
At present time, considering the present day values of the parameters involved, i.e. $\Omega_D =0.6911$, $\Omega_m =0.3089$, $u=0.446969$ and $n=10^{-4}$, we have that Eqs. (\ref{carolinghia1}) and (\ref{carolinghia2}) reduce to:
\begin{eqnarray}
\omega_{Dpl}  = \omega_{Dlog} \approx 0.964774q   -0.482419 .\label{calculon}
\end{eqnarray}
From Eq. (\ref{calculon}), we derive that our Universe exists in the accelerated phase (which is obtained when $q<0$) if we have $\omega < -0.482419$ while the phantom divide line (corresponding to $\omega_D=-1$) can be crossed provided that $q< -0.536478$.\\
In the limiting case of the Einstein's gravity, Eqs. (\ref{carolinghia1}) and (\ref{carolinghia2})  yield:
\begin{eqnarray}
\omega_{Dpl} &=& \frac{2q}{3\Omega_{Dpl}} - \frac{1+u_{pl}}{3}, \label{codam1} \\
\omega_{Dlog} &=& \frac{2q}{3\Omega_{Dlog}} - \frac{1+u_{log}}{3}, \label{codam2}
\end{eqnarray}
which has been already derived in the work of Khodam et al. \cite{kho8}. Moveover, in the limiting case of a flat Dark Dominated Universe (i.e. for  $\Omega_{Dpl}  =  \Omega_{Dlog}    = 1$, $\gamma_{pl}  =  \gamma_{log}   =  1$, $\Omega_m = \Omega_k = \Omega_{\phi} = 0$, $n=0$ and $u_{pl} =   u_{log}   = 0$), we have that Eqs. (\ref{codam1}) and  (\ref{codam2}) reduce, respectively, to:
\begin{eqnarray}
\omega_{Dpl} &=& \frac{2q_{pl}}{3} - \frac{1}{3}, \label{} \\
\omega_{Dlog} &=& \frac{2q_{log}}{3} - \frac{1}{3}. \label{}
\end{eqnarray}
Using the definition of fractional energy densities given in Eqs. (\ref{27}) and (\ref{27log}) along with the definition of the deceleration parameter $q$ given in Eq. (\ref{43}), we obtain the following relations:
\begin{eqnarray}
\frac{\Omega_{Dpl}}{c^2\gamma_{pl}} &=& L_{GO}^{-2}H^{-2} =  \alpha - \beta \left( 1+ q_{pl} \right), \label{risu1} \\
\frac{\Omega_{Dlog}}{c^2\gamma_{log}} &=& L_{GO}^{-2}H^{-2} =  \alpha - \beta \left( 1+ q_{log} \right).\label{risu2}
\end{eqnarray}
Using in Eqs. (\ref{risu1}) and (\ref{risu2}) the value of $q$ derived in order to have that the phantom divide is crossed, we obtain that:
\begin{eqnarray}
\frac{\Omega_{Dpl}}{c^2\gamma_{pl}} &=& \frac{\Omega_{Dlog}}{c^2\gamma_{log}} = L_{GO}^{-2}H^{-2} \approx \alpha - 0.463522\beta,
\end{eqnarray}
which leads, using the values of $\alpha$ and $\beta$ for a non flat Universe, i.e. $\alpha  = 0.8824$ and $\beta = 0.5016$, to:
\begin{eqnarray}
\frac{\Omega_{Dpl}}{c^2\gamma_{pl}} &=& \frac{\Omega_{Dlog}}{c^2\gamma_{log}} = L_{GO}^{-2}H^{-2} \approx 0.649898. \label{res1}
\end{eqnarray}
Instead, considering the values of $\alpha$ and $\beta$ in the case of flat Universe, case, i.e. for $\alpha  = 0.8502$ and $\beta = 0.4817$, we obtain:
\begin{eqnarray}
\frac{\Omega_{Dpl}}{c^2\gamma_{pl}} &=& \frac{\Omega_{Dlog}}{c^2\gamma_{log}} = L_{GO}^{-2}H^{-2} \approx 0.626922. \label{res1flat}
\end{eqnarray}
Moreover, considering the Ricci scale, recovered for $\alpha =2 $ and $\beta =1$,  we obtain:
\begin{eqnarray}
\frac{\Omega_{Dpl}}{c^2\gamma_{pl}} &=& \frac{\Omega_{Dlog}}{c^2\gamma_{log}} = L_{GO}^{-2}H^{-2} \approx 1.53648. \label{res2}
\end{eqnarray}

\subsection{Interacting Case}
We now extend the calculations accomplished in the previous Subsection to the case of interaction between the two dark sectors. \\
Recent observational evidences about the galaxy cluster Abell A586 convincingly support the presence of a kind of interaction between the two Dark Sectors, DE and DM \cite{q2,q2-2}.  However, the exact strength of the interaction between DE and DM is not precisely identified yet\cite{q3}. The presence of interaction between DE and DM can be detected during the formation of the Large Scale Structures (LSS). It has been pointed out that the dynamical equilibrium of the collapsed structures like clusters of galaxies (like for example Abell A586) would be affected by the presence of the coupling between the Dark Sectors, i.e. DE and DM \cite{1abd,10abd}. The main idea is that the
virial theorem results to be modified by the exchange of energy between DE and DM, which leads to a bias in the estimation of the virial masses of the clusters of galaxies when we employ the usual virial conditions. This gives a probe in the near Universe of the coupling of the Dark Sectors. We have that other observational signatures of the Dark Sectors mutual interaction can be also observed in the probes of the cosmic expansion history by
using the data from the Baryonic Acoustic Oscillations (BAO), the CMB shift data and the Supernovae Ia (SNeIa) \cite{A173,22abd}.\\
In presence of interaction, the energy densities of DE and DM $\rho_D$ and $\rho_m$ are separately conserved and the conservation equations have the following forms:
\begin{eqnarray}
\dot{\rho}_{Dpl}&+&3H\rho_{Dpl}\left(1+\omega_{Dpl}\right)=-Q, \label{46} \\
\dot{\rho}_{Dlog}&+&3H\rho_{Dlog}\left(1+\omega_{Dlog}\right)=-Q, \label{46log} \\
\dot{\rho}_m&+&3H\rho_m=Q .\label{47}
\end{eqnarray}
$Q$ indicates the interaction term which is in general an arbitrary function of the main cosmological parameters, like the Hubble parameter $H$ and the energy densities of DM and DE $\rho_m$ and $\rho_D$, i.e. $Q(H\rho_m,H\rho_D)$.
Many different candidates have been suggested in order to describe the behavior of the interaction term $Q$. In this paper, we have chosen to consider the following ones:
\begin{eqnarray}
Q_1 &=& 3b^2H\rho_m  ,\label{48} \\
Q_2 &=& 3b^2H\rho_D  ,\label{48due} \\
Q_3 &=& 3b^2H\left(\rho_m + \rho_D \right)  ,\label{48tre}
\end{eqnarray}
where $b^2$ represents a coupling parameter between DM and DE, also known as transfer strength \cite{q1,q1-4,q1-8,q1-9}.
Using the observational cosmological data given by the Gold SNeIa samples, the CMBR anisotropies data from the
WMAP and Planck satellites and the Baryonic Acoustic Oscillations (BAO) data obtained thanks to the Sloan Digital Sky Survey (SDSS),
the coupling parameter between DM and DE is estimated to assume a positive value which is supposed to be small (much smaller than unity and then closer to zero), fact which is able to satisfy the requirements for solving the cosmic coincidence problem and the constraints given by the second law of thermodynamics \cite{feng08}.
Observations of CMBR anisotropies and of clusters of galaxies show that $b^2 < 0.025$  \cite{q4}. This evidence is in agreement with the fact that $b^2$ must be taken in the range of values [0,1] \cite{zhang-02-2006}, with $b^2 = 0$ representing the non-interacting FLRW model. \\
We can now derive the expressions for $\dot{\rho}_{Dpl}$, $\rho'_{Dpl}$, $\dot{\rho}_{Dlog}$ and $\rho'_{Dlog}$ for the interacting case for the three interaction terms we have considered in this paper.\\
We start considering the expression of $Q_1$ given in Eq. (\ref{48}). Following the same procedure of the non interacting case, we obtain the following expressions for $\dot{\rho}_{Dpl}$ and $\dot{\rho}_{Dlog}$:
\begin{eqnarray}
\dot{\rho}_{Dpl} &=& \frac{3H^3\phi^2}{4 \omega}\left\{\frac{2}{\beta} \left[\left(\frac{\Omega_{Dpl}}{c^2 \gamma_{pl}}-\alpha\right)\left(1 -\Omega_{\phi}\right) +\beta \right] \right. \nonumber \\
 && \left. +\Omega_{Dpl}  \left[u_{pl} \left( 2n+1\right) + 2\left( n-1\right)  \right] - 2 \Omega_{\phi}- 3b^2\Omega_{m}  \right\} \nonumber \\
&=& \frac{H\rho_{Dpl}}{\Omega_{Dpl}}\left\{\frac{2}{\beta} \left[\left(\frac{\Omega_{Dpl}}{c^2 \gamma_{pl}}-\alpha\right)\left(1 -\Omega_{\phi}\right) +\beta \right] \right. \nonumber \\
 && \left. +\Omega_{Dpl}  \left[u_{pl} \left( 2n+1\right) + 2\left( n-1\right)  \right] - 2 \Omega_{\phi}  - 3b^2\Omega_{m} \right\}, \label{37intcataold}\\
\dot{\rho}_{Dlog} &=& \frac{3H^3\phi^2}{4 \omega}\left\{\frac{2}{\beta} \left[\left(\frac{\Omega_{Dlog}}{c^2 \gamma_{log}}-\alpha\right)\left(1 -\Omega_{\phi}\right) +\beta \right] \right. \nonumber \\
 && \left. + \Omega_{Dlog} \left[u_{log}\left( 2n+1\right) + 2\left( n-1\right)  \right] - 2 \Omega_{\phi}  - 3b^2\Omega_{m} \right\} \nonumber \\
 &=&\frac{H\rho_{Dlog}}{\Omega_{Dlog}}\left\{\frac{2}{\beta} \left[\left(\frac{\Omega_{Dlog}}{c^2 \gamma_{log}}-\alpha\right)\left(1 -\Omega_{\phi}\right) +\beta \right] \right. \nonumber \\
 && \left. + \Omega_{Dlog} \left[u_{log}\left( 2n+1\right) + 2\left( n-1\right)  \right] - 2 \Omega_{\phi} - 3b^2\Omega_{m}  \right\},  \label{37logintcataold}
\end{eqnarray}
which lead to the following expressions of the evolutionary forms of the energy density of DE $\rho_{Dpl}'$ and
$\rho_{Dlog}'$:
\begin{eqnarray}
\rho'_{Dpl} = \frac{\dot{\rho}_{Dpl}}{H} &=&  \frac{\rho_{Dpl}}{\Omega_{Dpl}}\left\{\frac{2}{\beta} \left[\left(\frac{\Omega_{Dpl}}{c^2 \gamma_{pl}}-\alpha\right)\left(1 -\Omega_{\phi}\right) +\beta \right] \right. \nonumber \\
 && \left. +\Omega_{Dpl}  \left[u_{pl} \left( 2n+1\right) + 2\left( n-1\right)  \right] - 2 \Omega_{\phi} - 3b^2\Omega_{m}  \right\} , \label{37evintcata}\\
\rho'_{Dlog} = \frac{\dot{\rho}_{Dlog}}{H} &=& \frac{\rho_{Dlog}}{\Omega_{Dlog}}\left\{\frac{2}{\beta} \left[\left(\frac{\Omega_{Dlog}}{c^2 \gamma_{log}}-\alpha\right)\left(1 -\Omega_{\phi}\right) +\beta \right] \right. \nonumber \\
 && \left. + \Omega_{Dlog}  \left[u_{log} \left( 2n+1\right) + 2\left( n-1\right)  \right] - 2 \Omega_{\phi} - 3b^2\Omega_{m}  \right\}. \label{37logevintcata}
\end{eqnarray}
From the continuity equations for DE obtained in Eqs. (\ref{46}) and (\ref{46log}), we derive the following expressions for the EoS parameters of the PLECHDE and LECHDE models $\omega_{Dpl}$ and $\omega_{Dlog}$:
\begin{eqnarray}
\omega_{Dpl} &=& -1 - \frac{\dot{\rho}_{Dpl}}{3H\rho_{Dpl}}  - \frac{Q}{3H\rho_{Dpl}}= -1 - \frac{\rho'_{Dpl}}{3\rho_{Dpl}}- \frac{Q}{3H\rho_{Dpl}}, \label{car1int} \\
\omega_{Dlog} &=& -1 - \frac{\dot{\rho}_{Dlog}}{3H\rho_{Dlog}}- \frac{Q}{3H\rho_{Dlog}} = -1 - \frac{\rho'_{Dlog}}{3\rho_{Dlog}}- \frac{Q}{3H\rho_{Dlog}}. \label{car2int}
\end{eqnarray}
We can now derive the final expression of the EoS parameters $\omega_{Dpl}$ and $\omega_{Dlog}$ for the first  interaction term we have chosen.\\
By inserting into Eqs. (\ref{car1int}) and (\ref{car2int}) the expressions of $\dot{\rho}_{Dpl}$ and $\dot{\rho}_{Dlog}$ obtained in Eqs. (\ref{37intcataold}) and (\ref{37logintcataold}) (or equivalently the expressions of $\rho'_{Dpl}$ and $\rho'_{Dlog}$ obtained in Eqs. (\ref{37evintcata}) and (\ref{37logevintcata})) along with the definition of $Q_1$ given in Eq. (\ref{48}), we have that the EoS parameters for the PLECHDE and the LECHDE models $\omega_{Dpl}$ and $\omega_{Dlog}$ can be written, respectively, as follows:
\begin{eqnarray}
\omega_{Dpl} &=& -1 -\frac{1}{3\Omega_{Dpl}}\left\{\frac{2}{\beta}\left[\left(\frac{\Omega_{Dpl}}{c^2 \gamma_{pl}}-\alpha\right)\left(1 -\Omega_{\phi}\right) +\beta \right]  \right. \nonumber \\
 && \left. + \Omega_{Dpl} \left[u_{pl}\left( 2n+1\right) + 2\left( n-1\right)  \right] - 2 \Omega_{\phi} \right\}      \label{49} \\
\omega_{Dlog} &=& -1 -\frac{1}{3\Omega_{Dlog}}\left\{\frac{2}{\beta}\left[\left(\frac{\Omega_{Dlog}}{c^2 \gamma_{log}}-\alpha\right)\left(1 -\Omega_{\phi}\right) +\beta \right] \right. \nonumber \\
 && \left. + \Omega_{Dlog} \left[u_{log}\left( 2n+1\right) + 2\left( n-1\right)  \right] - 2 \Omega_{\phi}   \right\}, \label{49log}
\end{eqnarray}
which are the same results obtained for non interacting case.\\
We now consider the second interaction term, i.e. $Q_2$, which has been defined in Eq. (\ref{48due}). Following the same procedure of the non interacting case, we obtain the following expressions for $\dot{\rho}_{Dpl}$ and $\dot{\rho}_{Dlog}$:
\begin{eqnarray}
\dot{\rho}_{Dpl} &=& \frac{3H^3\phi^2}{4 \omega}\left\{\frac{2}{\beta} \left[\left(\frac{\Omega_{Dpl}}{c^2 \gamma_{pl}}-\alpha\right)\left(1 -\Omega_{\phi}\right) +\beta \right] \right. \nonumber \\
 && \left. +\Omega_{Dpl} \left[u_{pl}\left( 2n+1\right) + 2\left( n-1\right)  \right] - 2 \Omega_{\phi}- 3b^2\Omega_{Dpl}  \right\} \nonumber \\
&=& \frac{H\rho_{Dpl}}{\Omega_{Dpl}}\left\{\frac{2}{\beta} \left[\left(\frac{\Omega_{Dpl}}{c^2 \gamma_{pl}}-\alpha\right)\left(1 -\Omega_{\phi}\right) +\beta \right] \right. \nonumber \\
 && \left. +\Omega_{Dpl}\left[u_{pl}\left( 2n+1\right) + 2\left( n-1\right)  \right] - 2 \Omega_{\phi}  - 3b^2\Omega_{Dpl} \right\}, \label{37intcata2old}\\
\dot{\rho}_{Dlog} &=& \frac{3H^3\phi^2}{4 \omega}\left\{\frac{2}{\beta} \left[\left(\frac{\Omega_{Dlog}}{c^2 \gamma_{log}}-\alpha\right)\left(1 -\Omega_{\phi}\right) +\beta \right] \right. \nonumber \\
 && \left. + \Omega_{Dlog} \left[u_{log}\left( 2n+1\right) + 2\left( n-1\right)  \right] - 2 \Omega_{\phi}  - 3b^2\Omega_{Dlog} \right\} \nonumber \\
 &=&\frac{H\rho_{Dlog}}{\Omega_{Dlog}}\left\{\frac{2}{\beta} \left[\left(\frac{\Omega_{Dlog}}{c^2 \gamma_{log}}-\alpha\right)\left(1 -\Omega_{\phi}\right) +\beta \right] \right. \nonumber \\
 && \left. + \Omega_{Dlog} \left[u_{log}\left( 2n+1\right) + 2\left( n-1\right)  \right] - 2 \Omega_{\phi} - 3b^2\Omega_{Dlog}  \right\},  \label{37logintcata2old}
\end{eqnarray}
which lead to the following expressions of the evolutionary form of the energy density of DE $\rho'_{Dpl} $ and $\rho'_{Dlog} $:
\begin{eqnarray}
\rho'_{Dpl} = \frac{\dot{\rho}_{Dpl}}{H} &=&  \frac{\rho_{Dpl}}{\Omega_{Dpl}}\left\{\frac{2}{\beta} \left[\left(\frac{\Omega_{Dpl}}{c^2 \gamma_{pl}}-\alpha\right)\left(1 -\Omega_{\phi}\right) +\beta \right] \right. \nonumber \\
 && \left. +\Omega_{Dpl} \left[u_{pl}\left( 2n+1\right) + 2\left( n-1\right)  \right] - 2 \Omega_{\phi} - 3b^2\Omega_{Dpl}  \right\} , \label{37evintcata2}\\
\rho'_{Dlog} = \frac{\dot{\rho}_{Dlog}}{H} &=& \frac{\rho_{Dlog}}{\Omega_{Dlog}}\left\{\frac{2}{\beta} \left[\left(\frac{\Omega_{Dlog}}{c^2 \gamma_{log}}-\alpha\right)\left(1 -\Omega_{\phi}\right) +\beta \right] \right. \nonumber \\
 && \left. + \Omega_{Dlog} \left[u_{log}\left( 2n+1\right) + 2\left( n-1\right)  \right] - 2 \Omega_{\phi} - 3b^2\Omega_{Dlog}  \right\}. \label{37logevintcata2}
\end{eqnarray}
By inserting into Eqs. (\ref{car1int}) and (\ref{car2int}) the expressions of $\dot{\rho}_{Dpl}$ and $\dot{\rho}_{Dlog}$ obtained in Eqs. (\ref{37intcata2old}) and (\ref{37logintcata2old}) (or equivalently the expressions of $\rho'_{Dpl}$ and $\rho'_{Dlog}$ obtained in Eqs. (\ref{37evintcata2}) and (\ref{37logevintcata2})) along with the definition of $Q_2$ given in Eq. (\ref{48due}), we have that the EoS parameters for the PLECHDE and the LECHDE models $\omega_{Dpl}$ and $\omega_{Dlog}$ can be written, respectively, as follows:
\begin{eqnarray}
\omega_{Dpl} &=&-1 -\frac{1}{3\Omega_{Dpl}}\left\{\frac{2}{\beta}\left[\left(\frac{\Omega_{Dpl}}{c^2 \gamma_{pl}}-\alpha\right)\left(1 -\Omega_{\phi}\right) +\beta \right]  \right. \nonumber \\
 && \left. + \Omega_{Dpl} \left[u_{pl}\left( 2n+1\right) + 2\left( n-1\right)  \right] - 2 \Omega_{\phi}\right\}, \label{492} \\
\omega_{Dlog}  &=& -1 -\frac{1}{3\Omega_{Dlog}}\left\{\frac{2}{\beta}\left[\left(\frac{\Omega_{Dlog}}{c^2 \gamma_{log}}-\alpha\right)\left(1 -\Omega_{\phi}\right) +\beta \right] \right. \nonumber \\
 && \left. + \Omega_{Dlog} \left[u_{log}\left( 2n+1\right) + 2\left( n-1\right)  \right] - 2 \Omega_{\phi}   \right\}, \label{49log2}
\end{eqnarray}
which are the same results obtained for non interacting case and for the interacting case with $Q_1$.\\
We finally consider the third and last interaction term considered in this work, i.e. $Q_3$, which has been defined in Eq. (\ref{48tre}). Following the same procedure of the non interacting case, we obtain the following expressions for $\dot{\rho}_{Dpl}$ and $\dot{\rho}_{Dlog}$:
\begin{eqnarray}
\dot{\rho}_{Dpl} &=& \frac{3H^3\phi^2}{4 \omega}\left\{\frac{2}{\beta} \left[\left(\frac{\Omega_{Dpl}}{c^2 \gamma_{pl}}-\alpha\right)\left(1 -\Omega_{\phi}\right) +\beta \right] \right. \nonumber \\
 && \left. +\Omega_{Dpl} \left[u_{pl}\left( 2n+1\right) + 2\left( n-1\right)  \right] - 2 \Omega_{\phi}- 3b^2\left(\Omega_m +  \Omega_{Dpl}\right)  \right\} \nonumber \\
&=& \frac{H\rho_{Dpl}}{\Omega_{Dpl}}\left\{\frac{2}{\beta} \left[\left(\frac{\Omega_{Dpl}}{c^2 \gamma_{pl}}-\alpha\right)\left(1 -\Omega_{\phi}\right) +\beta \right] \right. \nonumber \\
 && \left. +\Omega_{Dpl} \left[u_{pl}\left( 2n+1\right) + 2\left( n-1\right)  \right] - 2 \Omega_{\phi}  - 3b^2\left(\Omega_m  + \Omega_{Dpl}\right) \right\}, \label{37intold}\\
\dot{\rho}_{Dlog} &=& \frac{3H^3\phi^2}{4 \omega}\left\{\frac{2}{\beta} \left[\left(\frac{\Omega_{Dlog}}{c^2 \gamma_{log}}-\alpha\right)\left(1 -\Omega_{\phi}\right) +\beta \right] \right. \nonumber \\
 && \left. + \Omega_{Dlog}\left[u_{log}\left( 2n+1\right) + 2\left( n-1\right)  \right] - 2 \Omega_{\phi}  - 3b^2\left(\Omega_m  + \Omega_{Dlog}\right)\right\} \nonumber \\
 &=&\frac{H\rho_{Dlog}}{\Omega_{Dlog}}\left\{\frac{2}{\beta} \left[\left(\frac{\Omega_{Dlog}}{c^2 \gamma_{log}}-\alpha\right)\left(1 -\Omega_{\phi}\right) +\beta \right] \right. \nonumber \\
 && \left. + \Omega_{Dlog} \left[u_{log}\left( 2n+1\right) + 2\left( n-1\right)  \right] - 2 \Omega_{\phi} - 3b^2\left(\Omega_m  + \Omega_{Dlog}\right)  \right\},  \label{37logintold}
\end{eqnarray}
which lead to the following expressions of the evolutionary form of the energy density of DE $\rho'_{Dpl}$ and $\rho'_{Dlog}$:
\begin{eqnarray}
\rho'_{Dpl} = \frac{\dot{\rho}_{Dpl}}{H} &=&  \frac{\rho_{Dpl}}{\Omega_{Dpl}}\left\{\frac{2}{\beta} \left[\left(\frac{\Omega_{Dpl}}{c^2 \gamma_{pl}}-\alpha\right)\left(1 -\Omega_{\phi}\right) +\beta \right] \right. \nonumber \\
 && \left. +\Omega_{Dpl} \left[u_{pl}\left( 2n+1\right) + 2\left( n-1\right)  \right] \right. \nonumber \\
&& \left. - 2 \Omega_{\phi} - 3b^2\left(\Omega_m  + \Omega_{Dpl}\right)  \right\} , \label{37evint}\\
\rho'_{Dlog} = \frac{\dot{\rho}_{Dlog}}{H} &=& \frac{\rho_{Dlog}}{\Omega_{Dlog}}\left\{\frac{2}{\beta} \left[\left(\frac{\Omega_{Dlog}}{c^2 \gamma_{log}}-\alpha\right)\left(1 -\Omega_{\phi}\right) +\beta \right] \right. \nonumber \\
 && \left. + \Omega_{Dlog} \left[u_{log}\left( 2n+1\right) + 2\left( n-1\right)  \right]\right. \nonumber \\
  && \left.- 2 \Omega_{\phi} - 3b^2\left(\Omega_m  + \Omega_{Dlog}\right)  \right\}. \label{37logevint}
\end{eqnarray}
Finally, by inserting into Eqs. (\ref{car1int}) and (\ref{car2int}) the expressions of $\dot{\rho}_{Dpl}$ and $\dot{\rho}_{Dlog}$ obtained in Eqs. (\ref{37intold}) and (\ref{37logintold}) (or equivalently the expressions of $\rho'_{Dpl}$ and $\rho'_{Dlog}$ obtained in Eqs. (\ref{37evint}) and (\ref{37logevint})) along with the definition of $Q_3$ given in Eq. (\ref{48tre}), we have that the EoS parameters for the PLECHDE and the LECHDE models $\omega_{Dpl}$ and $\omega_{Dlog}$ can be written, respectively, as follows:
\begin{eqnarray}
\omega_{Dpl} &=& -1 -\frac{1}{3\Omega_{Dpl}}\left\{\frac{2}{\beta}\left[\left(\frac{\Omega_{Dpl}}{c^2 \gamma_{pl}}-\alpha\right)\left(1 -\Omega_{\phi}\right) +\beta \right]  \right. \nonumber \\
 && \left. + \Omega_{Dpl} \left[u_{pl}\left( 2n+1\right) + 2\left( n-1\right)  \right] - 2 \Omega_{\phi} \right\}, \label{493} \\
\omega_{Dlog} &=& -1 -\frac{1}{3\Omega_{Dlog}}\left\{\frac{2}{\beta}\left[\left(\frac{\Omega_{Dlog}}{c^2 \gamma_{log}}-\alpha\right)\left(1 -\Omega_{\phi}\right) +\beta \right] \right. \nonumber \\
 && \left. + \Omega_{Dlog} \left[u_{log}\left( 2n+1\right) + 2\left( n-1\right)  \right] - 2 \Omega_{\phi}   \right\}, \label{49log3}
\end{eqnarray}
which are the same results obtained for non interacting case and for the interacting cases with $Q_1$ and $Q_2$.\\
We now  derive an expression for the deceleration parameter $q$ in the case of interaction between DE and DM.\\
 Analogously to the previous Subsection, two different ways are possible in order to find the final expression for $q$.\\
 In the first one, we still consider the following expressions for $q_{pl}$ and $q_{log}$:
\begin{eqnarray}
q_{pl} &=&-1 -\frac{1}{\beta}\left( \frac{\Omega_{Dpl}}{c^2 \gamma_{pl}}  - \alpha   \right) \nonumber \\
 &=& -\frac{1}{\beta}\left( \frac{\Omega_{Dpl}}{c^2 \gamma_{pl}}+\beta - \alpha   \right), \label{50} \\
q_{log} &=& -1-\frac{1}{\beta}\left( \frac{\Omega_{Dlog}}{c^2 \gamma_{log}} - \alpha   \right)\nonumber \\
 &=&-\frac{1}{\beta}\left( \frac{\Omega_{Dlog}}{c^2 \gamma_{log}} +\beta - \alpha   \right). \label{50log}
\end{eqnarray}
Otherwise, Eqs. (\ref{45}) and (\ref{45log}) can be used in order to obtain the following relations  $q_{pl}$ and $q_{log}$:
\begin{eqnarray}
q_{pl} &=& \frac{1}{2\left( n+1\right)}\left[ \left( 2n+1\right)^2 + 2n\left( n\omega - 1\right)+\Omega_k  + 3\Omega_{Dpl} \omega_{Dpl}   \right], \label{51} \\
q_{log} &=& \frac{1}{2\left( n+1\right)}\left[ \left( 2n+1\right)^2 + 2n\left( n\omega - 1\right)+\Omega_k  + 3\Omega_{Dlog} \omega_{Dlog}   \right]. \label{51log}
\end{eqnarray}
The final expressions for the deceleration parameters $q_{pl}$ and $q_{log}$ can be now immediately  derived when in Eqs. (\ref{51}) and (\ref{51log}) the expressions of the relevant EoS parameters for the PLECHDE and the LECHDE models  are inserted.\\
We now  derive the expression of the evolutionary form of the energy density parameter of DE $\Omega'_D$ for both the power law and the logarithmic corrections considered in this work.\\
We still consider the general expressions of $\Omega_{Dpl}'$ and $\Omega_{Dlog}'$ given, respectively, in Eqs. (\ref{cr7pl}) and (\ref{cr7log}).\\
We start considering the case with the interaction term $Q_1$ defined in Eq. (\ref{48}).\\
Inserting in Eqs. (\ref{cr7pl}) and (\ref{cr7log}) the expressions of $\left(\frac{\dot{H}}{H^2}\right)_{pl}$, $\left(\frac{\dot{H}}{H^2}\right)_{log}$, $\rho_{Dpl}'$ and $\rho_{Dlog}'$, obtained, respectively, in Eqs. (\ref{35}), (\ref{35log}), (\ref{37evintcata}) and (\ref{37logevintcata}) , we derive that $\Omega_{Dpl}'$ and $\Omega_{Dlog}'$ can be expressed as follows:
\begin{eqnarray}
\Omega_{Dpl}' &=& \frac{2}{\beta} \left(  \frac{\Omega_{Dpl}}{c^2\gamma_{pl}} +\beta  - \alpha \right)\left(1-\Omega_{\phi} - \Omega_{Dpl}   \right)    + \Omega_m\left( 2n+1 -b^2\right) \nonumber \\
&=& \frac{2}{\beta} \left(  \frac{\Omega_{Dpl}}{c^2\gamma_{pl}} +\beta  - \alpha \right)\left(1-\Omega_{\phi} - \Omega_{Dpl}   \right)    + u_{pl}\Omega_{Dpl}\left(  2n+1-b^2\right), \label{52} \\
\Omega_{Dlog}' &=& \frac{2}{\beta} \left(  \frac{\Omega_{Dlog}}{c^2\gamma_{log}} +\beta  - \alpha \right)\left(1-\Omega_{\phi} - \Omega_{Dlog}  \right)    + \Omega_m\left( 2n+1 -b^2\right) \nonumber \\
&=& \frac{2}{\beta} \left(  \frac{\Omega_{Dlog}}{c^2\gamma_{log}} +\beta  - \alpha \right)\left(1-\Omega_{\phi} - \Omega_{Dlog}  \right)    + u_{log}\Omega_{Dlog}\left(  2n+1-b^2\right). \label{52log}
\end{eqnarray}
We can now study some limiting cases of the evolutionary forms of the fractional energy densities obtained in Eqs. (\ref{52}) and (\ref{52log}), in particular we will consider the equations obtained without entropy corrections (i.e. for $\lambda = 0$ for the power law correction and $\varrho = \varepsilon = 0$ for the logarithmic correction), in the limiting case of Einstein gravity (i.e., for $n=0$ and $\Omega_{\phi} = \rho_{\phi}=0$), without entropy corrections and in the limiting case of Einstein gravity together, for the Ricci scale (which is recovered when $\alpha =2$ and $\beta = 1$) and for the flat Dark Dominated Universe, which is recovered for   $\Omega_{Dpl}  =  \Omega_{Dlog}    = 1$, $\gamma_{pl}  =  \gamma_{log}   =  1$, $\Omega_m = \Omega_k = \Omega_{\phi} = 0$, $n=0$ and $u_{pl} =   u_{log}   = 0$.\\
In the limiting cases corresponding to $\lambda =0$ and $\varrho = \epsilon =0$ (which imply that $\gamma_{pl} = \gamma_{log} = 1$), Eqs. (\ref{52}) and (\ref{52log}) lead, respectively, to:
\begin{eqnarray}
\Omega_{Dpl_{\lambda=0}}' &=& \frac{2}{\beta}\left( \frac{\Omega_{Dpl_{\lambda=0}}}{c^2 } +\beta -\alpha \right)\left(1-\Omega_{\phi} - \Omega_{Dpl_{\lambda=0}}   \right)   \nonumber \\
&&+ \Omega_m \left( 2n+1  -b^2 \right) \nonumber\\
&=& \frac{2}{\beta}\left( \frac{\Omega_{Dpl_{\lambda=0}}}{c^2 } +\beta -\alpha \right)\left(1-\Omega_{\phi} - \Omega_{Dpl_{\lambda=0}}   \right) \nonumber\\
&&+ u_{pl,\lambda=0}\Omega_{Dpl_{\lambda=0}}\left( 2n+1-b^2\right), \label{52-1na}\\
\Omega_{Dlog_{\varrho = \epsilon=0}}' &=& \frac{2}{\beta}\left( \frac{\Omega_{Dlog_{\varrho = \epsilon=0}}}{c^2 } +\beta -\alpha \right)\left(1-\Omega_{\phi} - \Omega_{Dlog_{\varrho = \epsilon=0}}   \right) \nonumber\\
&&+ \Omega_m \left( 2n+1 -b^2  \right) \nonumber \\
&=& \frac{2}{\beta}\left( \frac{\Omega_{Dlog_{\varrho = \epsilon=0}}}{c^2 } +\beta -\alpha \right)\left(1-\Omega_{\phi} - \Omega_{Dlog_{\varrho = \epsilon=0}}   \right) \nonumber \\
&&+ u_{log,\varrho = \epsilon=0}\Omega_{Dlog_{\varrho = \epsilon=0}}\left( 2n+1-b^2\right). \label{52-1logna}
\end{eqnarray}
We can clearly observe, considering the expressions given in Eq. (\ref{lambda01}), (\ref{lambda02}) and (\ref{lambda02-2}), that Eqs. (\ref{52-1na}) and (\ref{52-1logna}) are identical.\\
Furthermore, in the limiting case corresponding to Einstein's gravity (i.e. for $n=0$ and $\Omega_{\phi} = \rho_{\phi}=0$), Eqs. (\ref{52}) and (\ref{52log}) reduce, respectively, to:
\begin{eqnarray}
\Omega_{Dpl,Ein}' &=& \frac{2}{\beta}\left( \frac{\Omega_{Dpl,Ein}}{c^2 \gamma_{pl,Ein}} +\beta -\alpha \right) \left(1 - \Omega_{Dpl,Ein}   \right)    + \Omega_m\left(  1-b^2 \right)  \nonumber \\
&=& \frac{2}{\beta}\left( \frac{\Omega_{Dpl,Ein}}{c^2 \gamma_{pl,Ein}} +\beta -\alpha \right) \left(1 - \Omega_{Dpl,Ein}   \right) \nonumber\\
&&   + u_{pl,Ein}\Omega_{Dpl,Ein}\left(  1-b^2 \right)  , \label{52-2} \\
\Omega_{Dlog,Ein}' &=& \frac{2}{\beta}\left( \frac{\Omega_{Dlog,Ein}}{c^2 \gamma_{log,Ein}} +\beta -\alpha \right)\left(1 - \Omega_{Dlog,Ein}   \right)    + \Omega_m\left(  1-b^2 \right)\nonumber \\
&=&\frac{2}{\beta}\left( \frac{\Omega_{Dlog,Ein}}{c^2 \gamma_{log,Ein}} +\beta -\alpha \right)\left(1 - \Omega_{Dlog,Ein}   \right)    \nonumber \\
&&+ u_{log,Ein}\Omega_{Dlog,Ein}\left(  1-b^2 \right). \label{52-2log}
\end{eqnarray}
Finally, in the limiting case corresponding to $\lambda =0$, $\varrho = \epsilon=0$  and to Einstein's gravity together, Eqs. (\ref{52}) and (\ref{52log}) read, respectively:
\begin{eqnarray}
\Omega_{Dpl,Ein_{\lambda=0}}' &=&     \frac{2}{\beta}\left( \frac{\Omega_{Dpl,Ein_{\lambda=0}}}{c^2} +\beta -\alpha \right)\left(1 - \Omega_{Dpl,Ein_{\lambda=0}}   \right)  \nonumber \\
&&  + \Omega_m\left(  1-b^2 \right) \nonumber \\
 &=&     \frac{2}{\beta}\left( \frac{\Omega_{Dpl,Ein_{\lambda=0}}}{c^2} +\beta -\alpha \right)\left(1 - \Omega_{Dpl,Ein_{\lambda=0}}   \right)  \nonumber \\
&&+ u_{pl,Ein_{\lambda=0}}\Omega_{Dpl,Ein_{\lambda=0}}\left(  1-b^2 \right), \label{52-3nalol} \\
\Omega_{Dlog,Ein_{\varrho = \epsilon=0}}' &=&    \frac{2}{\beta}\left( \frac{\Omega_{Dlog,Ein_{\varrho = \epsilon=0}}}{c^2} +\beta -\alpha \right)\left(1- \Omega_{Dlog,Ein_{\varrho = \epsilon=0}}   \right) \nonumber \\
    &&  + \Omega_m\left( 1-b^2\right) \nonumber \\
&=&    \frac{2}{\beta}\left( \frac{\Omega_{Dlog,Ein_{\varrho = \epsilon=0}}}{c^2} +\beta -\alpha \right)\left(1- \Omega_{Dlog,Ein_{\varrho = \epsilon=0}}   \right)   \nonumber \\
  &&+u_{log,Ein_{\varrho = \epsilon=0}}\Omega_{Dlog,Ein_{\varrho = \epsilon=0}} \left( 1-b^2\right). \label{52-3lognalol}
\end{eqnarray}
Considering the expression given in Eq. (\ref{lambda01}), (\ref{lambda02}) and (\ref{lambda02-2}), we derive that Eqs. (\ref{52-3nalol}) and (\ref{52-3lognalol}) are equivalent.\\
For the Ricci scale (i.e., for $\alpha=2$ and $\beta =1$), we obtain that  Eqs. (\ref{52}) and (\ref{52log}) lead, respectively, to:
\begin{eqnarray}
\Omega_{Dpl,Ricci}' &=& 2 \left(  \frac{\Omega_{Dpl,Ricci}}{c^2\gamma_{pl,Ricci}} -1 \right)\left(1-\Omega_{\phi} - \Omega_{Dpl,Ricci}   \right) \nonumber \\
&&   + \Omega_m\left( 2n+1-b^2\right) \nonumber \\
&=&\left(  \frac{\Omega_{Dpl,Ricci}}{c^2\gamma_{pl,Ricci}} -1 \right)\left(1-\Omega_{\phi} - \Omega_{Dpl,Ricci}   \right)  \nonumber \\
  &&+ u_{pl,Ricci}\Omega_{Dpl,Ricci}\left( 2n+1-b^2\right), \label{52ricciscale} \\
\Omega_{Dlog,Ricci}' &=& 2 \left(  \frac{\Omega_{Dlog,Ricci}}{c^2\gamma_{log,Ricci}} -1 \right)\left(1-\Omega_{\phi} - \Omega_{Dlog,Ricci}  \right)\nonumber \\
&&    + \Omega_m\left( 2n+1-b^2\right) \nonumber  \\
&=& 2 \left(  \frac{\Omega_{Dlog,Ricci}}{c^2\gamma_{log,Ricci}} -1 \right)\left(1-\Omega_{\phi} - \Omega_{Dlog,Ricci}  \right) \nonumber \\
&&+ u_{pl,Ricci}\Omega_{Dlog,Ricci}\left( 2n+1-b^2\right).\label{52logricciscale}
\end{eqnarray}
Finally, for a flat Dark Dominated Universe, we obtain that both expressions of $\Omega_{Dpl}' $ and $\Omega_{Dlog}'$ are equals to zero, as it must be since for a flat Dark Dominated Universe we have that $\Omega_D = 1$.\\
We now consider the case with the interaction term $Q_2$ defined in Eq. (\ref{48due}).\\
Inserting in Eqs. (\ref{cr7pl}) and (\ref{cr7log}) the expressions of $\left(\frac{\dot{H}}{H^2}\right)_{pl}$, $\left(\frac{\dot{H}}{H^2}\right)_{log}$, $\rho_{Dpl}'$ and $\rho_{Dlog}'$, obtained, respectively, in Eqs. (\ref{35}), (\ref{35log}), (\ref{37evintcata2}) and (\ref{37logevintcata2}) , we derive that $\Omega_{Dpl}'$ and $\Omega_{Dlog}'$ can be expressed as follows:
\begin{eqnarray}
\Omega_{Dpl}' &=& \frac{2}{\beta} \left(  \frac{\Omega_{Dpl}}{c^2\gamma_{pl}} +\beta  - \alpha \right)\left(1-\Omega_{\phi} - \Omega_{Dpl}   \right)    + \Omega_m\left( 2n+1\right) -b^2\Omega_{Dpl} \nonumber \\
&=& \frac{2}{\beta} \left(  \frac{\Omega_{Dpl}}{c^2\gamma_{pl}} +\beta  - \alpha \right)\left(1-\Omega_{\phi} - \Omega_{Dpl}   \right)    + \Omega_{Dpl}\left[u_{pl}\left(  2n+1\right) -b^2\right], \label{52caro} \\
\Omega_{Dlog}' &=& \frac{2}{\beta} \left(  \frac{\Omega_{Dlog}}{c^2\gamma_{log}} +\beta  - \alpha \right)\left(1-\Omega_{\phi} - \Omega_{Dlog}  \right)    + \Omega_m\left( 2n+1\right)-b^2\Omega_{Dlog} \nonumber \\
&=& \frac{2}{\beta} \left(  \frac{\Omega_{Dlog}}{c^2\gamma_{log}} +\beta  - \alpha \right)\left(1-\Omega_{\phi} - \Omega_{Dlog}  \right)    + \Omega_{Dlog}\left[ u_{log}\left(  2n+1\right) - b^2\right]. \label{52logcaro}
\end{eqnarray}
We can now study some limiting cases of the evolutionary forms of the fractional energy densities obtained in Eqs. (\ref{52caro}) and (\ref{52logcaro}), in particular we will consider the equations obtained without entropy corrections (i.e. for $\lambda = 0$ for the power law correction and $\varrho = \varepsilon = 0$ for the logarithmic correction), in the limiting case of Einstein gravity (i.e., for $n=0$ and $\Omega_{\phi} = \rho_{\phi}=0$), without entropy corrections and in the limiting case of Einstein gravity together, for the Ricci scale (which is recovered when $\alpha =2$ and $\beta = 1$) and for the flat Dark Dominated Universe, which is recovered for   $\Omega_{Dpl}  =  \Omega_{Dlog}    = 1$, $\gamma_{pl}  =  \gamma_{log}   =  1$, $\Omega_m = \Omega_k = \Omega_{\phi} = 0$, $n=0$ and $u_{pl} =   u_{log}   = 0$.\\
In the limiting cases corresponding to $\lambda =0$ and $\varrho = \epsilon =0$ (which imply that $\gamma_{pl} = \gamma_{log} = 1$), Eqs. (\ref{52caro}) and (\ref{52logcaro}) lead, respectively, to:
\begin{eqnarray}
\Omega_{Dpl_{\lambda=0}}' &=& \frac{2}{\beta}\left( \frac{\Omega_{Dpl_{\lambda=0}}}{c^2 } +\beta -\alpha \right)\left(1-\Omega_{\phi} - \Omega_{Dpl_{\lambda=0}}   \right)  \nonumber\\
&&+ \Omega_m \left( 2n+1   \right) - b^2\Omega_{Dpl_{\lambda=0}}\nonumber \\
&=& \frac{2}{\beta}\left( \frac{\Omega_{Dpl_{\lambda=0}}}{c^2 } +\beta -\alpha \right)\left(1-\Omega_{\phi} - \Omega_{Dpl_{\lambda=0}}   \right) \nonumber \\
&&+ \Omega_{Dpl_{\lambda=0}}\left[u_{pl_{\lambda=0}}\left(  2n+1\right) -b^2\right], \label{52-1cata}\\
\Omega_{Dlog_{\varrho = \epsilon=0}}' &=& \frac{2}{\beta}\left( \frac{\Omega_{Dlog_{\varrho = \epsilon=0}}}{c^2 } +\beta -\alpha \right)\left(1-\Omega_{\phi} - \Omega_{Dlog_{\varrho = \epsilon=0}}   \right)  \nonumber\\
&& + \Omega_m \left( 2n+1   \right)  -b^2\Omega_{Dlog_{\varrho = \epsilon=0}} \nonumber \\
&=& \frac{2}{\beta}\left( \frac{\Omega_{Dlog_{\varrho = \epsilon=0}}}{c^2} +\beta -\alpha \right)\left(1-\Omega_{\phi} - \Omega_{Dlog_{\varrho = \epsilon=0}}   \right)\nonumber \\
 &&+ \Omega_{Dlog_{\varrho = \epsilon=0}}\left[u_{log_{\varrho = \epsilon=0}}\left(  2n+1\right) -b^2\right]. \label{52-1logcata}
\end{eqnarray}
We can clearly observe, considering the expressions given in Eq. (\ref{lambda01}),(\ref{lambda02}) and (\ref{lambda02-2}),  that Eqs. (\ref{52-1cata}) and (\ref{52-1logcata}) are identical.\\
Furthermore, in the limiting case corresponding to Einstein's gravity (i.e. for $n=0$ and $\Omega_{\phi} = \rho_{\phi}=0$), Eqs. (\ref{52}) and (\ref{52log}) reduce, respectively, to:
\begin{eqnarray}
\Omega_{Dpl,Ein}' &=& \frac{2}{\beta}\left( \frac{\Omega_{Dpl,Ein}}{c^2 \gamma_{pl,Ein}} +\beta -\alpha \right) \left(1 - \Omega_{Dpl,Ein}   \right)    + \Omega_m - b^2\Omega_{Dpl,Ein}\nonumber  \\
&=& \frac{2}{\beta}\left( \frac{\Omega_{Dpl,Ein}}{c^2 \gamma_{pl,Ein}} +\beta -\alpha \right) \left(1 - \Omega_{Dpl,Ein}   \right)    + \Omega_{Dpl,Ein}\left(u_{pl,Ein} -b^2\right) , \label{52-2} \\
\Omega_{Dlog,Ein}' &=& \frac{2}{\beta}\left( \frac{\Omega_{Dlog,Ein}}{c^2 \gamma_{log,Ein}} +\beta -\alpha \right)\left(1 - \Omega_{Dlog,Ein}   \right)    + \Omega_m- b^2\Omega_{Dlog,Ein}\nonumber \\
&=&\frac{2}{\beta}\left( \frac{\Omega_{Dlog,Ein}}{c^2 \gamma_{log,Ein}} +\beta -\alpha \right)\left(1 - \Omega_{Dlog,Ein}   \right)  \nonumber \\
&&  +  \Omega_{Dlog,Ein}\left(u_{log,Ein} -b^2\right) . \label{52-2log}
\end{eqnarray}
Finally, in the limiting case corresponding to $\lambda =0$, $\varrho = \epsilon=0$  and to Einstein's gravity together, Eqs. (\ref{52}) and (\ref{52log}) read, respectively:
\begin{eqnarray}
\Omega_{Dpl,Ein_{\lambda=0}}' &=&     \frac{2}{\beta}\left( \frac{\Omega_{Dpl,Ein_{\lambda=0}}}{c^2} +\beta -\alpha \right)\left(1 - \Omega_{Dpl,Ein_{\lambda=0}}   \right) \nonumber \\
   &&+ \Omega_m -b^2\Omega_{Dpl,Ein_{\lambda=0}}\nonumber \\
 &=&     \frac{2}{\beta}\left( \frac{\Omega_{Dpl,Ein_{\lambda=0}}}{c^2} +\beta -\alpha \right)\left(1 - \Omega_{Dpl,Ein_{\lambda=0}}   \right)    \nonumber \\
&&+ \Omega_{Dpl,Ein_{\lambda=0}}\left(u_{pl,Ein_{\lambda=0}}-b^2\right), \label{cicci2} \\
\Omega_{Dlog,Ein_{\varrho = \epsilon=0}}' &=&    \frac{2}{\beta}\left( \frac{\Omega_{Dlog,Ein_{\varrho = \epsilon=0}}}{c^2} +\beta -\alpha \right)\left(1- \Omega_{Dlog,Ein_{\varrho = \epsilon=0}}   \right)  \nonumber\\
&& + \Omega_m -b^2\Omega_{Dlog,Ein_{\varrho = \epsilon=0}} \nonumber \\
&=&    \frac{2}{\beta}\left( \frac{\Omega_{Dlog,Ein_{\varrho = \epsilon=0}}}{c^2} +\beta -\alpha \right)\left(1- \Omega_{Dlog,Ein_{\varrho = \epsilon=0}}   \right)    \nonumber \\
 &&+\Omega_{Dlog,Ein_{\varrho = \epsilon=0}}\left(u_{log,Ein_{\varrho = \epsilon=0}} -b^2\right). \label{cicci1}
\end{eqnarray}
Considering the expression given in Eqs. (\ref{lambda01}), (\ref{lambda02}) and (\ref{lambda02-2})  we derive that Eqs. (\ref{cicci2}) and (\ref{cicci1}) are equivalent.\\
For the Ricci scale (i.e, for. $\alpha=2$ and $\beta =1$), we obtain that  Eqs. (\ref{52caro}) and (\ref{52logcaro}) lead, respectively, to:
\begin{eqnarray}
\Omega_{Dpl,Ricci}' &=& 2 \left(  \frac{\Omega_{Dpl,Ricci}}{c^2\gamma_{pl,Ricci}} -1 \right)\left(1-\Omega_{\phi} - \Omega_{Dpl,Ricci}   \right)   \nonumber\\
 &&+ \Omega_m\left( 2n+1\right) -b^2\Omega_{Dpl,Ricci}\nonumber \\
&=&\left(  \frac{\Omega_{Dpl,Ricci}}{c^2\gamma_{pl,Ricci}} -1\right)\left(1-\Omega_{\phi} - \Omega_{Dpl,Ricci}   \right)   \nonumber \\
 &&+ \Omega_{Dpl,Ricci}\left[u_{pl,Ricci}\left( 2n+1\right) -b^2  \right], \label{52ricciscale} \\
\Omega_{Dlog,Ricci}' &=& 2 \left(  \frac{\Omega_{Dlog,Ricci}}{c^2\gamma_{log,Ricci}} -1 \right)\left(1-\Omega_{\phi} - \Omega_{Dlog,Ricci}  \right) \nonumber\\
   &&+ \Omega_m\left( 2n+1\right)-b^2\Omega_{Dlog,Ricci}  \nonumber \\
&=& 2 \left(  \frac{\Omega_{Dlog,Ricci}}{c^2\gamma_{log,Ricci}} -1 \right)\left(1-\Omega_{\phi} - \Omega_{Dlog,Ricci}  \right)  \nonumber\\
  &&+ \Omega_{Dlog,Ricci}\left[u_{pl,Ricci}\left( 2n+1\right)-b^2\right].\label{52logricciscale}
\end{eqnarray}
Finally, for a flat Dark Dominated Universe, we obtain that both expressions of $\Omega_{Dpl}' $ and $\Omega_{Dlog}'$ are equals to zero, as it must be since for Dark Dominated Universe we have that $\Omega_D = 1$.\\
We finally consider the case with the interaction term $Q_3$ defined in Eq. (\ref{48tre}).\\
Inserting in Eqs. (\ref{cr7pl}) and (\ref{cr7log}) the expressions of $\left(\frac{\dot{H}}{H^2}\right)_{pl}$, $\left(\frac{\dot{H}}{H^2}\right)_{log}$, $\rho_{Dpl}'$ and $\rho_{Dlog}'$, obtained, respectively, in Eqs. (\ref{35}), (\ref{35log}), (\ref{37evint}) and (\ref{37logevint}) , we derive that $\Omega_{Dpl}'$ and $\Omega_{Dlog}'$ can be expressed as follows:
\begin{eqnarray}
\Omega_{Dpl}' &=& \frac{2}{\beta} \left(  \frac{\Omega_{Dpl}}{c^2\gamma_{pl}} +\beta  - \alpha \right)\left(1-\Omega_{\phi} - \Omega_{Dpl}   \right) \nonumber \\
&&   + \Omega_{pl}\left[u_{pl}\left(2n+1 -b^2  \right)  -b^2 \right],\label{52caro2} \\
\Omega_{Dlog}' &=& \frac{2}{\beta} \left(  \frac{\Omega_{Dlog}}{c^2\gamma_{log}} +\beta  - \alpha \right)\left(1-\Omega_{\phi} - \Omega_{Dlog}  \right)  \nonumber \\
&&+ \Omega_{log}\left[u_{log}\left(2n+1 -b^2  \right)  -b^2 \right]. \label{52logcaro2}
\end{eqnarray}
We can now study some limiting cases of the evolutionary forms of the fractional energy densities obtained in Eqs. (\ref{52caro2}) and (\ref{52logcaro2}), in particular we will consider the equations obtained without entropy corrections (i.e. for $\lambda = 0$ for the power law correction and $\varrho = \varepsilon = 0$ for the logarithmic correction), in the limiting case of Einstein gravity (i.e., for $n=0$ and $\Omega_{\phi} = \rho_{\phi}=0$), without entropy corrections and in the limiting case of Einstein gravity together, for the Ricci scale (which is recovered when $\alpha =2$ and $\beta = 1$) and for the flat Dark Dominated Universe, which is recovered for   $\Omega_{Dpl}  =  \Omega_{Dlog}    = 1$, $\gamma_{pl}  =  \gamma_{log}   =  1$, $\Omega_m = \Omega_k = \Omega_{\phi} = 0$, $n=0$ and $u_{pl} =   u_{log}   = 0$.\\
In the limiting cases corresponding to $\lambda =0$ and $\varrho = \epsilon =0$ (which imply that $\gamma_{pl} = \gamma_{log} = 1$), Eqs. (\ref{52caro2}) and (\ref{52logcaro2}) lead, respectively, to:
\begin{eqnarray}
\Omega_{Dpl_{\lambda=0}}' &=& \frac{2}{\beta}\left( \frac{\Omega_{Dpl_{\lambda=0}}}{c^2 } +\beta -\alpha \right)\left(1-\Omega_{\phi} - \Omega_{Dpl_{\lambda=0}}   \right)  \nonumber\\
 &&+ \Omega_m \left( 2n+1   \right) -b^2\left(\Omega_m + \Omega_{Dpl_{\lambda=0}}\right)\nonumber \\
&=& \frac{2}{\beta}\left( \frac{\Omega_{Dpl_{\lambda=0}}}{c^2 } +\beta -\alpha \right)\left(1-\Omega_{\phi} - \Omega_{Dpl_{\lambda=0}}   \right)\nonumber\\
&&+\Omega_{Dpl_{\lambda=0}}\left[ u_{pl,\lambda=0}\left( 2n+1-b^2\right)   -b^2\right], \label{52-1}\\
\Omega_{Dlog_{\varrho = \epsilon=0}}' &=& \frac{2}{\beta}\left( \frac{\Omega_{Dlog_{\varrho = \epsilon=0}}}{c^2 } +\beta -\alpha \right)\left(1-\Omega_{\phi} - \Omega_{Dlog_{\varrho = \epsilon=0}}   \right)\nonumber\\
&&+ \Omega_m \left( 2n+1   \right)  -b^2\left(\Omega_m + \Omega_{Dlog_{\lambda=0}}\right)\nonumber \\
&=& \frac{2}{\beta}\left( \frac{\Omega_{Dlog_{\varrho = \epsilon=0}}}{c^2 } +\beta -\alpha \right)\left(1-\Omega_{\phi} - \Omega_{Dlog_{\varrho = \epsilon=0}}   \right) \nonumber\\
&&+ \Omega_{Dlog_{\varrho = \epsilon=0}}\left[u_{log,\varrho = \epsilon=0} \left( 2n+1-b^2\right)-b^2\right]. \label{52-1log}
\end{eqnarray}
We can clearly observe, considering the expression given in Eqs. (\ref{lambda01}), (\ref{lambda02}) and (\ref{lambda02-2}) that Eqs. (\ref{52-1}) and (\ref{52-1log}) are identical.\\
Furthermore, in the limiting case corresponding to Einstein's gravity (i.e., for $n=0$ and $\Omega_{\phi} = \rho_{\phi}=0$), Eqs. (\ref{52}) and (\ref{52log}) reduce, respectively, to:
\begin{eqnarray}
\Omega_{Dpl,Ein}' &=& \frac{2}{\beta}\left( \frac{\Omega_{Dpl,Ein}}{c^2 \gamma_{pl,Ein}} +\beta -\alpha \right) \left(1 - \Omega_{Dpl,Ein}   \right)  \nonumber \\
&&  + \Omega_m -b^2\left(\Omega_m + \Omega_{Dpl,Ein} \right)\nonumber \\
&=& \frac{2}{\beta}\left( \frac{\Omega_{Dpl,Ein}}{c^2 \gamma_{pl,Ein}} +\beta -\alpha \right) \left(1 - \Omega_{Dpl,Ein}   \right) \nonumber \\
&&+\Omega_{Dpl,Ein}\left[ u_{pl,Ein}\left(1-b^2\right)   -b^2\right] , \label{52-2} \\
\Omega_{Dlog,Ein}' &=& \frac{2}{\beta}\left( \frac{\Omega_{Dlog,Ein}}{c^2 \gamma_{log,Ein}} +\beta -\alpha \right)\left(1 - \Omega_{Dlog,Ein}   \right)  \nonumber \\
&&  + \Omega_m -b^2\left(\Omega_m + \Omega_{Dlog,Ein} \right)\nonumber \\
&=&\frac{2}{\beta}\left( \frac{\Omega_{Dlog,Ein}}{c^2 \gamma_{log,Ein}} +\beta -\alpha \right)\left(1 - \Omega_{Dlog,Ein}   \right) \nonumber \\
&&+ \Omega_{Dlog,Ein}\left[ u_{log,Ein}\left(1-b^2\right)   -b^2\right]. \label{52-2log}
\end{eqnarray}
Finally, in the limiting case corresponding to $\lambda =0$, $\varrho = \epsilon=0$  and to Einstein's gravity together, Eqs. (\ref{52}) and (\ref{52log}) read, respectively:
\begin{eqnarray}
\Omega_{Dpl,Ein_{\lambda=0}}' &=&     \frac{2}{\beta}\left( \frac{\Omega_{Dpl,Ein_{\lambda=0}}}{c^2} +\beta -\alpha \right)\left(1 - \Omega_{Dpl,Ein_{\lambda=0}}   \right)    \nonumber\\
&&+ \Omega_m - b^2\left( \Omega_m + \Omega_{Dpl,Ein_{\lambda=0}}  \right)\nonumber \\
 &=&     \frac{2}{\beta}\left( \frac{\Omega_{Dpl,Ein_{\lambda=0}}}{c^2} +\beta -\alpha \right)\left(1 - \Omega_{Dpl,Ein_{\lambda=0}}   \right) \nonumber \\
 &&+ \Omega_{Dpl,Ein_{\lambda=0}}\left[ u_{pl,Ein_{\lambda=0}}\left(1-b^2\right)   -b^2\right] , \label{52-3rita} \\
\Omega_{Dlog,Ein_{\varrho = \epsilon=0}}' &=&    \frac{2}{\beta}\left( \frac{\Omega_{Dlog,Ein_{\varrho = \epsilon=0}}}{c^2} +\beta -\alpha \right)\left(1- \Omega_{Dlog,Ein_{\varrho = \epsilon=0}}   \right)\nonumber\\
 &&+ \Omega_m - b^2\left( \Omega_m + \Omega_{Dlog,Ein_{\varrho = \epsilon =0}}  \right)\nonumber \\
&=&    \frac{2}{\beta}\left( \frac{\Omega_{Dlog,Ein_{\varrho = \epsilon=0}}}{c^2} +\beta -\alpha \right)\left(1- \Omega_{Dlog,Ein_{\varrho = \epsilon=0}}   \right) \nonumber\\
 &&+\Omega_{Dlog,Ein_{\varrho = \epsilon=0}}\left[ u_{log,Ein_{\varrho = \epsilon=0}}\left(1-b^2\right)   -b^2\right]  . \label{52-3logrita}
\end{eqnarray}
Considering the expression given in Eq. (\ref{lambda01}) (\ref{lambda02}) and (\ref{lambda02-2}), we derive that Eqs. (\ref{52-3rita}) and (\ref{52-3logrita}) are equivalent.\\
For the Ricci scale (i.e., for  $\alpha=2$ and $\beta =1$), we obtain that  Eqs. (\ref{52}) and (\ref{52log}) lead, respectively, to:
\begin{eqnarray}
\Omega_{Dpl,Ricci}' &=& 2 \left(  \frac{\Omega_{Dpl,Ricci}}{c^2\gamma_{pl,Ricci}} -1 \right)\left(1-\Omega_{\phi} - \Omega_{Dpl,Ricci}   \right) \nonumber\\
&& + \Omega_m\left( 2n+1\right) -b^2\left(  \Omega_m + \Omega_{Dpl,Ricci} \right) \nonumber \\
&=&\left(  \frac{\Omega_{Dpl,Ricci}}{c^2\gamma_{pl,Ricci}} -1 \right)\left(1-\Omega_{\phi} - \Omega_{Dpl,Ricci}   \right)  \nonumber\\
 &&+ \Omega_{Dpl,Ricci}\left[u_{pl,Ricci}\left( 2n+1-b^2\right)  -b^2 \right], \label{52ricciscale} \\
\Omega_{Dlog,Ricci}' &=& 2 \left(  \frac{\Omega_{Dlog,Ricci}}{c^2\gamma_{log,Ricci}} -1 \right)\left(1-\Omega_{\phi} - \Omega_{Dlog,Ricci}  \right)\nonumber\\
&& + \Omega_m\left( 2n+1\right)-b^2\left(  \Omega_m + \Omega_{Dlog,Ricci} \right) \nonumber \\
&=& 2 \left(  \frac{\Omega_{Dlog,Ricci}}{c^2\gamma_{log,Ricci}} -1 \right)\left(1-\Omega_{\phi} - \Omega_{Dlog,Ricci}  \right) \nonumber\\
 &&+ \Omega_{Dlog,Ricci}\left[u_{pl,Ricci}\left( 2n+1-b^2\right)  -b^2 \right].\label{52logricciscale}
\end{eqnarray}
Finally, for a flat Dark Dominated Universe, we obtain that both expressions of $\Omega_{Dpl}' $ and $\Omega_{Dlog}'$ are equals to zero, as it must be since for Dark Dominated Universe we have that $\Omega_D = 1$.\\

\section{Statefinder Diagnostic}
The investigation of cosmological parameters (like the Hubble parameter $H$, the EoS parameter $\omega_D$ and the deceleration parameter $q$)  have attracted a lot of attention in modern cosmology. Since different DE models lead to a positive Hubble parameter and a negative deceleration parameter (i.e. $H > 0$ and $q < 0$) at the present epoch, the Hubble and the deceleration parameters $H$ and $q$ can not effectively discriminate the various DE models. A higher order of time derivatives of the scale factor $a\left( t \right)$ is then required in order to have a better comprehension of the model taken into account. Sahni et al. \citep{sah} and Alam et al. \citep{alam}, using the third time derivative of scale factor $a(t)$, introduced the statefinder pair, indicated with $\left\{r,s\right\}$, with the main purpose to eliminate the known degeneracy of the Hubble and the deceleration parameters $H$ and $q$ at the
present epoch. The general expressions of the two statefinder parameters $r$ and $s$ are given, respectively, by the following relations:
\begin{eqnarray}
r &=& \frac{\dddot{a}}{aH^3}, \label{r1}\\
s &=&   \frac{r -1}{3\left(q-1/2\right)}, \label{s1}
\end{eqnarray}
where $q$ represents the deceleration parameter, which has been already studied in the previous Section.\\
The statefinder parameters can be also written as functions of the energy densities and the pressure of the model considered as follows:
\begin{eqnarray}
r &=& 1 + \frac{9}{2}\left(\frac{\rho + p}{\rho}\right)\frac{\dot{p}}{\dot{\rho}} \nonumber \\
 &=& 1 + \frac{9}{2}\left(\frac{\rho + p}{\rho}\right)\frac{p'}{\rho'} , \label{34p2}\\
s&=& \left(\frac{\rho + p}{p}\right)\frac{\dot{p}}{\dot{\rho}} \nonumber \\
&=&  \left(\frac{\rho + p}{p}\right)\frac{p'}{\rho'}. \label{35p2}
\end{eqnarray}
Considering that the total pressure $p$ and the total energy density $\rho$ are given, respectively, by $p = p_D$ and $\rho = \rho_D + \rho_m$, we can write the expressions of $r$ and $s$ as follows:
\begin{eqnarray}
r &=& 1 + \frac{9}{2}\left(\frac{\rho_D + \rho_m + p_D}{\rho_D + \rho_m}\right)\frac{\dot{p}_D}{\dot{\rho}_D+ \dot{\rho}_m} \nonumber \\
 &=& 1 + \frac{9}{2}\left(\frac{\rho_D +\rho_m+ p_D}{\rho_D + \rho_m}\right)\frac{p_D'}{\rho_D' + \rho_m'} , \label{34p2}\\
s&=& \left(\frac{\rho_D + \rho_m + p_D}{p_D}\right)\frac{\dot{p}_D}{\dot{\rho}_D+ \dot{\rho}_m} \nonumber \\
 &=&  \left(\frac{\rho_D + \rho_m + p_D}{p_D}\right)\frac{p_D'}{\rho_D' + \rho_m'}. \label{35p2}
\end{eqnarray}
An alternative way to write the statefinder parameters $r$ and $s$ is the following one:
\begin{eqnarray}
r&=& 1 + 3\left(\frac{\dot{H}}{H^2}\right)+ \frac{\ddot{H}}{H^3}, \label{r2}\\
s&=& -\frac{3H\dot{H}+\ddot{H}}{3H\left( 2\dot{H}+3H^2  \right)}\nonumber \\
&=&  -\frac{3\dot{H}+\ddot{H}/H}{3\left( 2\dot{H}+3H^2  \right)}. \label{s2}
\end{eqnarray}
The most important property of the statefinder parameters is that the point with coordinates corresponding to $\left\{r, s\right\} = \left\{1, 0\right\}$ indicates the fixed point in the $r-s$ plane  which corresponds to the flat $\Lambda$CDM model \cite{huang}. Departures of other DE models from the point corresponding to the $\Lambda$CDM model are good ways to establish the distance of these models from the flat $\Lambda$CDM model. \\
Moreover, we must underline here that, in the $r-s$ plane, a positive value of $s$ (i.e. $s > 0$) indicates a quintessence-like model of DE, instead we have that a negative value of $s$ (i.e. $s < 0$) leads to a phantom-like model of DE. Furthermore, an evolution from a phantom-like to a quintessence-like model (or the inverse) is obtained by crossing of the point $\left\{r, s\right\} = \left\{1, 0\right\}$ in the $r-s$ plane \cite{wu1}.\\
Braneworld, cosmological constant, Chaplygin gas and quintessence models were investigated by Alam et al. \cite{alam} using the statefinder diagnostic: they observed that the statefinder pair could differentiate between these different models. An investigation on statefinder parameters for differentiating between DE and modified gravity was carried out in Wang et al. \cite{wang}. Statefinder diagnostics for the $f\left(T\right)$ modified gravity model has been considered in the paper of Wu $\&$ Yu \cite{wu1}. \\
Other authors have been studied the properties of various DE models from the viewpoint of statefinder diagnostic \cite{kho1,kho5,kho7}.\\

\subsection{Non Interacting Case}
We now  study the statefinder pair for the model considered in this paper, for this reason we need to derive the quantities useful in order to obtain the final expression of the pair $\left\{ r,s\right\}$. We underline here that for $r$ we will use the expression given in Eq. (\ref{r2}) since it will be easier to calculate the terms involved: in fact, we already calculated $\frac{\dot{H}}{H^2}$ and using its expression we can easily derive one for $\frac{\ddot{H}}{H^3}$.\\
We start considering the non interacting case.\\
We have already obtained the expression of $\left(\frac{\dot{H}}{H^2}\right)_{pl}$ and $\left(\frac{\dot{H}}{H^2}\right)_{log}$ in Eqs. (\ref{35}) and (\ref{35log}). \\
Differentiating the expressions of $\left(\frac{\dot{H}}{H^2}\right)_{pl}$ and $\left(\frac{\dot{H}}{H^2}\right)_{log}$ obtained in Eqs. (\ref{35}) and (\ref{35log}), we can easily find the following expressions for $\left(\frac{\ddot{H}}{H^3}\right)_{pl} $ and $\left(\frac{\ddot{H}}{H^3}\right)_{log} $:
\begin{eqnarray}
\left(\frac{\ddot{H}}{H^3}\right)_{pl} &=&    2\left( \frac{\dot{H}}{H^2}  \right)^2_{pl} + \left( \frac{\dot{H}}{H^2}  \right)_{pl}',  \label{carolinghia8}\\
\left(\frac{\ddot{H}}{H^3}\right)_{log} &=&    2\left( \frac{\dot{H}}{H^2}  \right)^2_{log} + \left( \frac{\dot{H}}{H^2}  \right)_{log}'.  \label{carolinghia9}
\end{eqnarray}
We than have that the expressions of the statefinder parameters  $r_{pl} $ and $r_{log} $ can be also written as follows:
\begin{eqnarray}
r_{pl} &=& 1 + 3\left( \frac{\dot{H}}{H^2}  \right)_{pl} + 2\left( \frac{\dot{H}}{H^2}  \right)_{pl}^2  + \left( \frac{\dot{H}}{H^2}  \right)_{pl}',\label{carolinghia10} \\
r_{log} &=& 1 + 3\left( \frac{\dot{H}}{H^2}  \right)_{log} + 2\left( \frac{\dot{H}}{H^2}  \right)_{log}^2   + \left( \frac{\dot{H}}{H^2}  \right)_{log}'.  \label{carolinghia11}
\end{eqnarray}
Considering the expressions of $\left( \frac{\dot{H}}{H^2}  \right)_{pl}$ and $\left( \frac{\dot{H}}{H^2}  \right)_{log}$ given, respectively, in Eqs. (\ref{35}) and (\ref{35log}), we obtain the following expressions:
\begin{eqnarray}
\left( \frac{\dot{H}}{H^2}  \right)'_{pl} &=& \frac{1}{\beta}\left( \frac{\Omega'_{Dpl}}{c^2\gamma_{pl}} - \frac{\Omega_{Dpl}\gamma'_{pl}}{c^2\gamma^2_{pl}} \right), \label{carolinghia12}\\
\left( \frac{\dot{H}}{H^2}  \right)'_{log} &=& \frac{1}{\beta}\left( \frac{\Omega'_{Dlog}}{c^2\gamma_{log}} - \frac{\Omega_{Dlog}\gamma'_{log}}{c^2\gamma^2_{log}} \right). \label{carolinghia13}
\end{eqnarray}
We now need to make some considerations about the expressions of $\gamma'_{pl}$  and $\gamma'_{log}$.\\
Using the definition of $\gamma_{pl}$ and $\gamma_{log}$ given, respectively, in Eqs. (\ref{17}) and (\ref{17log}), we obtain that $\gamma_{pl}'$ and $\gamma_{log}'$ are given by the following relations:
\begin{eqnarray}
\gamma_{pl}' &=& \frac{\dot{\gamma}_{pl}}{H} = \frac{\left( \delta -2  \right)\lambda}{3c^2 L_{GO}^{\delta -2}}\left(\frac{L_{GO}'}{L_{GO}}\right)_{pl} \nonumber \\
&=& \left( \delta -2  \right) \left(1-\gamma_{pl}\right) \left(\frac{L_{GO}'}{L_{GO}}\right)_{pl},   \\
\gamma_{log}' &=&   \frac{\dot{\gamma}_{log}}{H} = \frac{ 8\omega}{3c^2\phi^2 L_{GO}^2}\left\{ \varrho\left[  1-\log \left(\frac{\phi^2L_{GO}^2}{4\omega}\right) -\epsilon \right]  \right\}\left[\left(\frac{L_{GO}'}{L_{GO}}\right)_{log}   +n   \right],
\end{eqnarray}
where
\begin{eqnarray}
\left(\frac{L_{GO}'}{L_{GO}}\right)_{pl} &=& \left(\frac{\dot{L}_{GO}}{HL_{GO}}\right)_{pl},  \\
\left(\frac{L_{GO}'}{L_{GO}}\right)_{log} &=& \left(\frac{\dot{L}_{GO}}{HL_{GO}}\right)_{log}.
\end{eqnarray}
Using the definition of the Granda-Oliveros cut off $L_{GO}$ given in Eq. (\ref{5})  along with the definition of its time derivative previously derived in Eq. (\ref{32}), we obtain the following relations for $\left(\frac{L_{GO}'}{L_{GO}}\right)_{pl}$ and $\left(\frac{L_{GO}'}{L_{GO}}\right)_{log}$:
\begin{eqnarray}
\left(\frac{L_{GO}'}{L_{GO}}\right)_{pl} &=& -H^2L^2 \left(  \alpha\frac{\dot{H}}{H^2} + \beta \frac{\ddot{H}}{2H^3}  \right)_{pl}\nonumber \\
&=& -\frac{c^2\gamma_{pl}}{\Omega_{Dpl}}\left( \alpha\frac{\dot{H}}{H^2} + \beta \frac{\ddot{H}}{2H^3}  \right)_{pl} \nonumber \\
&=&  \frac{3c^2\gamma_{pl}}{2}  \left(1+\omega_{Dpl} - \frac{2n}{3}\right)\left( c^2  - \frac{\lambda \delta  }{6 L_{GO}^{\delta -2}} \right)^{-1}, \label{carolinghia14}\nonumber \\
\left(\frac{L_{GO}'}{L_{GO}}\right)_{log} &=& -H^2L^2 \left(  \alpha\frac{\dot{H}}{H^2} + \beta \frac{\ddot{H}}{2H^3}  \right)_{log} \nonumber \\
&=& -\frac{c^2\gamma_{log}}{\Omega_{log}}\left(  \alpha\frac{\dot{H}}{H^2} + \beta \frac{\ddot{H}}{2H^3}   \right)_{log} \nonumber \\
&=&  \frac{3c^2\gamma_{log}}{2}  \left(1+\omega_{Dlog} - \frac{2n}{3}\right)\times\nonumber \\
&&\left\{ c^2  +\frac{4\omega}{3\phi^2}L_{GO}^{-2}\left[ 2\varrho \ln \left( \frac{\phi^2 L_{GO}^2}{4\omega}\right) -\varrho + 2\epsilon  \right] \right\}^{-1}. \label{carolinghia15}
\end{eqnarray}
In order to study the present day behavior of $\gamma_{pl0}$, $\gamma_{pl0}'$, $\gamma_{log0}$ and $\gamma_{log0}'$ we need to make some preliminary considerations. We know that the power-law correction and the logarithmic corrections to the entropy gives reasonable contributions only at early stages of the Universe history while with the passing of the time their contributions become less important. For this reason, we can tell that, for $\gamma_{pl0}$, $\gamma_{pl0}'$, $\gamma_{log0}$ and $\gamma_{log0}'$, the power-law and the logarithmic corrections can be considered practically negligible. We then have:
\begin{eqnarray}
\gamma_{pl0} &=&  \gamma_{log0} \approx 1 , \label{lim1}\\
\gamma_{pl0}' &=& \gamma_{log0}'  \approx 0.\label{lim2}
\end{eqnarray}
In all the following calculations, we will take into account the consideration made in Eqs. (\ref{lim1}) and (\ref{lim2}), for this reason we will neglect all the derivatives of  $\gamma_{pl}$  and $\gamma_{log}$.\\
Using the relations for $r_{pl}$ and $r_{log}$ obtained in Eqs. (\ref{carolinghia10}) and (\ref{carolinghia11})  along with the results obtained in Eqs. (\ref{carolinghia12}) and  (\ref{carolinghia13}), we finally derived the following relations for $r_{pl}$ and $r_{log}$:
\begin{eqnarray}
r_{pl} &=& 1 +  \frac{3}{\beta}\left(\frac{\Omega_{Dpl}}{c^2\gamma_{pl}}  -\alpha\right) +\frac{2}{\beta^2}\left[\left(\frac{\Omega_{Dpl}}{c^2\gamma_{pl}}  -\alpha\right)\right]^2+\frac{1}{\beta}\left( \frac{\Omega'_{Dpl}}{c^2\gamma_{pl}}  \right),\label{carolinghia16} \\
r_{log} &=& 1 +  \frac{3}{\beta}\left(\frac{\Omega_{Dlog}}{c^2\gamma_{log}}  -\alpha\right) + \frac{2}{\beta^2}\left[\left(\frac{\Omega_{Dlog}}{c^2\gamma_{log}}  -\alpha\right)\right]^2 +\frac{1}{\beta}\left( \frac{\Omega'_{Dlog}}{c^2\gamma_{log}}  \right). \label{carolinghia17}
\end{eqnarray}
We start considering the non interacting case.\\
Substituting in Eqs. (\ref{carolinghia16}) and (\ref{carolinghia17}) the expressions of $\Omega'_{Dpl}$ and $\Omega'_{Dlog}$ for the non interacting case, we obtain that the present day values of $r_{pl}$ and $r_{log}$ are given, respectively, by:
\begin{eqnarray}
r_{pl,0} &=& 1 +  \frac{3}{\beta}\left(\frac{\Omega_{Dpl,0}}{c^2\gamma_{pl,0}}  -\alpha\right) +\frac{2}{\beta^2}\left(\frac{\Omega_{Dpl,0}}{c^2\gamma_{pl,0}}  -\alpha\right)^2   \nonumber \\
&&\,+\frac{1}{\beta}\left[\frac{2}{\beta c^2\gamma_{pl,0}} \left(  \frac{\Omega_{Dpl,0}}{c^2\gamma_{pl,0}} +\beta  - \alpha \right)\left(1-\Omega_{\phi} - \Omega_{Dpl,0}   \right) \right.  \nonumber \\
 &&\left. + \left(\frac{\Omega_{Dpl,0}}{c^2\gamma_{pl,0}}\right)u_{pl,0}\left( 2n+1\right)\right],   \label{carolinghia20} \\
r_{log,0} &=& 1 +  \frac{3}{\beta}\left(\frac{\Omega_{Dlog,0}}{c^2\gamma_{log,0}}  -\alpha\right) + \frac{2}{\beta^2}\left[\left(\frac{\Omega_{Dlog,0}}{c^2\gamma_{log,0}}  -\alpha\right)\right]^2 \nonumber \\
&&\, +\frac{1}{\beta} \left[\frac{2}{\beta c^2\gamma_{log,0}} \left(  \frac{\Omega_{Dlog,0}}{c^2\gamma_{log,0}} +\beta  - \alpha \right)\left(1-\Omega_{\phi} - \Omega_{Dlog,0}   \right)  \right.   \nonumber \\
&&\left. +\left(\frac{\Omega_{Dlog,0}}{c^2\gamma_{log,0}}\right)u_{log,0}\left( 2n+1\right)\right].  \label{carolinghia21}
\end{eqnarray}
Considering the values of $\alpha$ and $\beta$ defined in the Introduction for the non flat Universe (i.e.  $\alpha  = 0.8824$ and $\beta = 0.5016$) along with the values of $\frac{\Omega_{Dpl,0}}{c^2\gamma_{pl,0}}$ and $\frac{\Omega_{Dlog,0}}{c^2\gamma_{log,0}}$ defined in Eq. (\ref{res1}) and the value of $n$ $n=10^{-4}$, we obtain that:
\begin{eqnarray}
r_{pl,0} = r_{log,0} \approx 1.11598 \label{rcaso1}.
\end{eqnarray}
Using the results of Eq. (\ref{rcaso1}) along with the value of $q$ obtained in order to have the crossing of the phantom divide line, we have that the statefinder parameters $s_{pl,0}$ and $s_{log,0}$ assume the following value:
\begin{eqnarray}
s_{pl,0} = s_{log,0} \approx -0.0372979. \label{scaso1}
\end{eqnarray}
Considering the values of $\alpha$ and $\beta$ defined in the Introduction for the  flat Universe (i.e.  $\alpha  = 0.8502$ and $\beta = 0.4817$) along with the values of $\frac{\Omega_{Dpl,0}}{c^2\gamma_{pl,0}}$ and $\frac{\Omega_{Dlog,0}}{c^2\gamma_{log,0}}$ defined in Eq. (\ref{res1}) and the value of $n$ $n=10^{-4}$, we obtain that:
\begin{eqnarray}
r_{pl,0} = r_{log,0} \approx 1.13534 \label{rcaso1flat}.
\end{eqnarray}
Using the results of Eq. (\ref{rcaso1flat}) along with the value of $q$ obtained in order to have the crossing of the phantom divide line, we have that the statefinder parameters $s_{pl,0}$ and $s_{log,0}$ assume the following value:
\begin{eqnarray}
s_{pl,0} = s_{log,0} \approx -0.0435269. \label{scaso1flat}
\end{eqnarray}
Instead,  considering the values of $\alpha$ and $\beta$ corresponding to the Ricci scale (i.e.  $\alpha  = 2$ and $\beta = 1$) along with the values of $\frac{\Omega_{Dpl,0}}{c^2\gamma_{pl,0}}$ and $\frac{\Omega_{Dlog,0}}{c^2\gamma_{log,0}}$ defined in Eq. (\ref{res2}) and the value of $n$ $n=10^{-4}$, we obtain that:
\begin{eqnarray}
r_{pl,0} = r_{log,0} \approx 0.975634\label{rcaso2}.
\end{eqnarray}
Using the results of Eq. (\ref{rcaso2}) along with the value of $q$ obtained in order to have the crossing of the phantom divide line, we have that the statefinder parameters $s_{pl,0}$ and $s_{log,0}$ assume the following value:
\begin{eqnarray}
s_{pl,0} = s_{log,0} \approx 0.00783629. \label{scaso2}
\end{eqnarray}
We have, then, that for the non flat case, i.e. for  $\alpha  = 0.8824$ and $\beta = 0.5016$, the models we are studying lead to point $\left\{r,s\right\} = \left\{1.11598,  -0.0372979 \right\}$ in the $r-s$ plane. Instead, for the flat case, i.e. for  $\alpha  = 0.8502$ and $\beta = 0.4817$, the models we are studying lead to point $\left\{r,s\right\} = \left\{1.13534,  -0.0435269 \right\}$ in the $r-s$ plane. Therefore, we obtain two points which are slightly far from the point $\left\{r,s\right\} = \left\{1, 0 \right\}$, with the non flat case closer to the point corresponding to the $\Lambda$CDM model. Moreover, since we have $s<0$, we can conclude that the models considered have a phantom-like behavior in the case of absence of interacting between Dark Sectors for both flat and non flat cases.\\
Instead, for the case corresponding to the Ricci scale, i.e. for $\alpha =2$ and $\beta =1$, the models we are studying lead to point $\left\{r,s\right\} = \left\{0.975634, 0.00783629 \right\}$ in the $r-s$ plane, i.e. to a point closer to $\left\{r,s\right\} = \left\{1, 0 \right\}$ if compared with the other two models. However, since for the Ricci scale we obtained $s>0$, it means that for this limiting case we deal with a quintessence-like model. 

\subsection{Interacting Case}
We now consider  the interacting case for all the three different interacting case taken into account. \\
Substituting in Eqs. (\ref{carolinghia16}) and (\ref{carolinghia17}) the expressions of $\Omega'_{Dpl}$ and $\Omega'_{Dlog}$ for the first  interacting cases derived in Eqs. (\ref{52}) and (\ref{52log}), we obtain that the following present day values of $r_{pl}$ and $r_{log}$ are given, respectively, by:
\begin{eqnarray}
r_{pl,0} &=& 1 +  \frac{3}{\beta}\left(\frac{\Omega_{Dpl,0}}{c^2\gamma_{pl,0}}  -\alpha\right) +\frac{2}{\beta^2}\left[\left(\frac{\Omega_{Dpl,0}}{c^2\gamma_{pl,0}}  -\alpha\right)\right]^2    \nonumber \\
&&\,+\frac{1}{\beta} \left[\frac{2}{\beta c^2\gamma_{pl,0}} \left(  \frac{\Omega_{Dpl,0}}{c^2\gamma_{pl,0}} +\beta  - \alpha \right)\left(1-\Omega_{\phi} - \Omega_{Dpl,0}   \right) \right.  \nonumber \\
&& \left. + \left(\frac{\Omega_{Dpl,0}}{c^2\gamma_{pl,0}}\right)u_{pl,0}\left( 2n+1-b^2\right)\right] ,\label{} \\
r_{log,0} &=& 1 +  \frac{3}{\beta}\left(\frac{\Omega_{Dlog,0}}{c^2\gamma_{log,0}}  -\alpha\right) + \frac{2}{\beta^2}\left[\left(\frac{\Omega_{Dlog,0}}{c^2\gamma_{log,0}}  -\alpha\right)\right]^2 \nonumber \\
&&\, +\frac{1}{\beta}\left[\frac{2}{\beta c^2\gamma_{log,0}} \left(  \frac{\Omega_{Dlog,0}}{c^2\gamma_{log,0}} +\beta  - \alpha \right)\left(1-\Omega_{\phi} - \Omega_{Dlog,0}   \right) \right.  \nonumber \\
&& \left.+ \left(\frac{\Omega_{Dlog,0}}{c^2\gamma_{log,0}}\right)u_{log,0}\left( 2n+1-b^2\right)\right]. \label{}
\end{eqnarray}
Considering the values of $\alpha$ and $\beta$ defined in the Introduction for the non flat Universe (i.e.  $\alpha  = 0.8824$ and $\beta = 0.5016$) along with the values of $\frac{\Omega_{Dpl,0}}{c^2\gamma_{pl,0}}$ and $\frac{\Omega_{Dlog,0}}{c^2\gamma_{log,0}}$ defined in Eq. (\ref{res1}) and the value of $n$ $n=10^{-4}$, we obtain that:
\begin{eqnarray}
r_{pl,0} = r_{log,0} \approx 1.11598 - 0.579114b^2. \label{rcasoint1}
\end{eqnarray}
Using the results of Eq. (\ref{rcasoint1}) along with the value of $q$ obtained in order to have the crossing of the phantom divide line, we have that the statefinder parameters $s_{pl,0}$ and $s_{log,0}$ assume the following expression:
\begin{eqnarray}
s_{pl,0} = s_{log,0} \approx -0.0372979+ 0.186244b^2. \label{scasoint1}
\end{eqnarray}
Considering the values of $\alpha$ and $\beta$ defined in the Introduction for the  flat Universe (i.e.  $\alpha  = 0.8502$ and $\beta = 0.4817$) along with the values of $\frac{\Omega_{Dpl,0}}{c^2\gamma_{pl,0}}$ and $\frac{\Omega_{Dlog,0}}{c^2\gamma_{log,0}}$ defined in Eq. (\ref{res1}) and the value of $n$ $n=10^{-4}$, we obtain that:
\begin{eqnarray}
r_{pl,0} = r_{log,0} \approx 1.13534 - 0.58172b^2. \label{rcasoint1flat}
\end{eqnarray}
Using the results of Eq. (\ref{rcasoint1flat}) along with the value of $q$ obtained in order to have the crossing of the phantom divide line, we have that the statefinder parameters $s_{pl,0}$ and $s_{log,0}$ assume the following expression:
\begin{eqnarray}
s_{pl,0} = s_{log,0} \approx -0.0435269 + 0.187082b^2. \label{scasoint1flat}
\end{eqnarray}
Instead,  considering the values of $\alpha$ and $\beta$ corresponding to the Ricci scale (i.e., for  $\alpha  = 2$ and $\beta = 1$) along with the values of $\frac{\Omega_{Dpl,0}}{c^2\gamma_{pl,0}}$ and $\frac{\Omega_{Dlog,0}}{c^2\gamma_{log,0}}$ defined in Eq. (\ref{res2}) and the value of $n$ $n=10^{-4}$, we obtain that:
\begin{eqnarray}
r_{pl,0} = r_{log,0} \approx 0.975634 -0.686758 b^2 \label{rcaso2int1}.
\end{eqnarray}
Using the results of Eq. (\ref{rcaso2int1}) along with the value of $q$ obtained in order to have the crossing of the phantom divide line, we have that the statefinder parameters $s_{pl,0}$ and $s_{log,0}$ assume the following value:
\begin{eqnarray}
s_{pl,0} = s_{log,0} \approx  0.00783629 + 0.220862b^2. \label{scaso2int1}
\end{eqnarray}

We now consider the second interacting case, corresponding to $Q_2$. \\
Substituting in Eqs. (\ref{carolinghia16}) and (\ref{carolinghia17}) the expressions of $\Omega'_{Dpl}$ and $\Omega'_{Dlog}$ for the second  interacting cases derived in Eqs. (\ref{52caro}) and (\ref{52logcaro}), we obtain that  the present day values of $r_{pl}$ and $r_{log}$ are given, respectively, by:
\begin{eqnarray}
r_{pl,0} &=& 1 +  \frac{3}{\beta}\left(\frac{\Omega_{Dpl,0}}{c^2\gamma_{pl,0}}  -\alpha\right) +\frac{2}{\beta^2}\left[\left(\frac{\Omega_{Dpl,0}}{c^2\gamma_{pl,0}}  -\alpha\right)\right]^2    \nonumber \\
&&\,+\frac{1}{\beta}\left\{\frac{2}{\beta c^2\gamma_{pl,0}} \left(  \frac{\Omega_{Dpl,0}}{c^2\gamma_{pl,0}} +\beta  - \alpha \right)\left(1-\Omega_{\phi} - \Omega_{Dpl,0}   \right)\right.  \nonumber \\
&& \left. + \left(\frac{\Omega_{Dpl,0}}{c^2\gamma_{pl,0}}\right)\left[u_{pl,0}\left(  2n+1\right) -b^2\right]     \right\}  ,\label{} \\
r_{log,0} &=& 1 +  \frac{3}{\beta}\left(\frac{\Omega_{Dlog,0}}{c^2\gamma_{log,0}}  -\alpha\right) + \frac{2}{\beta^2}\left[\left(\frac{\Omega_{Dlog,0}}{c^2\gamma_{log,0}}  -\alpha\right)\right]^2 \nonumber \\
&&\, +\frac{1}{\beta}\left\{\frac{2}{\beta c^2\gamma_{log,0}} \left(  \frac{\Omega_{Dlog,0}}{c^2\gamma_{log,0}} +\beta  - \alpha \right)\left(1-\Omega_{\phi} - \Omega_{Dlog,0}   \right) \right.  \nonumber \\
&& \left.+ \left(\frac{\Omega_{Dlog,0}}{c^2\gamma_{log,0}}\right)\left[u_{log,0}\left(  2n+1\right) -b^2\right]     \right\} . \label{}
\end{eqnarray}
Considering the values of $\alpha$ and $\beta$ defined in the Introduction for the non flat Universe (i.e.  $\alpha  = 0.8824$ and $\beta = 0.5016$) along with the values of $\frac{\Omega_{Dpl,0}}{c^2\gamma_{pl,0}}$ and $\frac{\Omega_{Dlog,0}}{c^2\gamma_{log,0}}$ defined in Eq. (\ref{res1}) and the value of $n$ $n=10^{-4}$, we obtain that:
\begin{eqnarray}
r_{pl,0} = r_{log,0} \approx  1.11598 - 1.29565b^2   .\label{rcasoint2}
\end{eqnarray}
Using the results of Eq. (\ref{rcasoint2}) along with the value of $q$ obtained in order to have the phantom divide crossing, we have that the statefinder parameter $s_{pl,0}$ and $s_{log,0}$ assume the following expression:
\begin{eqnarray}
s_{pl,0} = s_{log,0} \approx  -0.0372979+ 0.416683b^2.\label{scasoint2}
\end{eqnarray}
Considering the values of $\alpha$ and $\beta$ defined in the Introduction for the  flat Universe (i.e.  $\alpha  = 0.8502$ and $\beta = 0.4817$) along with the values of $\frac{\Omega_{Dpl,0}}{c^2\gamma_{pl,0}}$ and $\frac{\Omega_{Dlog,0}}{c^2\gamma_{log,0}}$ defined in Eq. (\ref{res1}) and the value of $n$ $n=10^{-4}$, we obtain that:
\begin{eqnarray}
r_{pl,0} = r_{log,0} \approx 1.13534 - 1.30148b^2  . \label{rcasoint2flat}
\end{eqnarray}
Using the results of Eq. (\ref{rcasoint2flat}) along with the value of $q$ obtained in order to have the phantom divide crossing, we have that the statefinder parameters $s_{pl,0}$ and $s_{log,0}$ assume the following expression:
\begin{eqnarray}
s_{pl,0} = s_{log,0} \approx  -0.0435269 + 0.418557b^2. \label{scasoint2flat}
\end{eqnarray}
Instead,  considering the values of $\alpha$ and $\beta$ corresponding to the Ricci scale (i.e., for  $\alpha  = 2$ and $\beta = 1$) along with the values of $\frac{\Omega_{Dpl,0}}{c^2\gamma_{pl,0}}$ and $\frac{\Omega_{Dlog,0}}{c^2\gamma_{log,0}}$ defined in Eq. (\ref{res2}) and the value of $n$ $n=10^{-4}$, we obtain that:
\begin{eqnarray}
r_{pl,0} = r_{log,0} \approx 0.975634 -1.53648b^2   \label{rcaso2int2}.
\end{eqnarray}
Using the results of Eq. (\ref{rcaso2int2}) along with the value of $q$ obtained in order to have the phantom divide crossing, we have that the statefinder parameters $s_{pl,0}$ and $s_{log,0}$ assume the following value:
\begin{eqnarray}
s_{pl,0} = s_{log,0} \approx 0.00783629 + 0.49134b^2. \label{scaso2int2}
\end{eqnarray}

We finally consider the third interacting case corresponding to $Q_3$.\\
Substituting in Eqs. (\ref{carolinghia16}) and (\ref{carolinghia17}) the expressions of $\Omega'_{Dpl}$ and $\Omega'_{Dlog}$ for the third  interacting cases derived in Eqs. (\ref{52caro2}) and (\ref{52logcaro2}), we obtain that the present day values:
\begin{eqnarray}
r_{pl,0} &=& 1 +  \frac{3}{\beta}\left(\frac{\Omega_{Dpl,0}}{c^2\gamma_{pl,0}}  -\alpha\right) +\frac{2}{\beta^2}\left[\left(\frac{\Omega_{Dpl,0}}{c^2\gamma_{pl,0}}  -\alpha\right)\right]^2    \nonumber \\
&&\,+\frac{1}{\beta} \left\{\frac{2}{\beta c^2\gamma_{pl,0}} \left(  \frac{\Omega_{Dpl,0}}{c^2\gamma_{pl,0}} +\beta  - \alpha \right)\left(1-\Omega_{\phi} - \Omega_{Dpl,0}   \right) \right.  \nonumber \\
 && \left.  + \left(\frac{\Omega_{Dpl,0}}{c^2\gamma_{pl,0}}\right)\left[u_{pl,0}\left(  2n+1-b^2\right) -b^2\right]     \right\}  ,\label{} \\
r_{log,0} &=& 1 +  \frac{3}{\beta}\left(\frac{\Omega_{Dlog,0}}{c^2\gamma_{log,0}}  -\alpha\right) + \frac{2}{\beta^2}\left[\left(\frac{\Omega_{Dlog,0}}{c^2\gamma_{log,0}}  -\alpha\right)\right]^2 \nonumber \\
&&\, +\frac{1}{\beta}\left\{\frac{2}{\beta c^2\gamma_{log,0}} \left(  \frac{\Omega_{Dlog,0}}{c^2\gamma_{log,0}} +\beta  - \alpha \right)\left(1-\Omega_{\phi} - \Omega_{Dlog,0}   \right) \right.  \nonumber \\
 && \left.  + \left(\frac{\Omega_{Dlog,0}}{c^2\gamma_{log,0}}\right)\left[u_{log,0}\left(  2n+1-b^2\right) -b^2\right]     \right\} . \label{}
\end{eqnarray}
Considering the values of $\alpha$ and $\beta$ defined in the Introduction for the non flat Universe (i.e.  $\alpha  = 0.8824$ and $\beta = 0.5016$) along with the values of $\frac{\Omega_{Dpl,0}}{c^2\gamma_{pl,0}}$ and $\frac{\Omega_{Dlog,0}}{c^2\gamma_{log,0}}$ defined in Eq. (\ref{res1}) and the value of $n$ $n=10^{-4}$, we obtain that:
\begin{eqnarray}
r_{pl,0} = r_{log,0} \approx 1.11598 - 1.87476b^2   .\label{rcasoint3}
\end{eqnarray}
Using the results of Eq. (\ref{rcasoint3}) along with the value of $q$ obtained in order to have the phantom divide crossing, we have that the statefinder parameters $s_{pl,0}$ and $s_{log,0}$ assume the following expression:
\begin{eqnarray}
s_{pl,0} = s_{log,0} \approx   -0.0372979+ 0.602917b^2.\label{scaso4int3}
\end{eqnarray}
Considering the values of $\alpha$ and $\beta$ defined in the Introduction for the  flat Universe (i.e.  $\alpha  = 0.8502$ and $\beta = 0.4817$) along with the values of $\frac{\Omega_{Dpl,0}}{c^2\gamma_{pl,0}}$ and $\frac{\Omega_{Dlog,0}}{c^2\gamma_{log,0}}$ defined in Eq. (\ref{res2}) and the value of $n$ $n=10^{-4}$, we obtain that:
\begin{eqnarray}
r_{pl,0} = r_{log,0} \approx 1.13534 - 1.8832b^2  .\label{rcasoint3flat}
\end{eqnarray}
Using the results of Eq. (\ref{rcasoint3flat}) along with the value of $q$ obtained in order to have the phantom divide crossing, we have that the statefinder parameters $s_{pl,0}$ and $s_{log,0}$ assume the following expression:
\begin{eqnarray}
s_{pl,0} = s_{log,0} \approx   -0.0435269 + 0.60564b^2.\label{scaso4int3}
\end{eqnarray}
Instead,  considering the values of $\alpha$ and $\beta$ corresponding to the Ricci scale (i.e.  $\alpha  = 2$ and $\beta = 1$) along with the values of $\frac{\Omega_{Dpl,0}}{c^2\gamma_{pl,0}}$ and $\frac{\Omega_{Dlog,0}}{c^2\gamma_{log,0}}$ defined in Eq. (\ref{res2}) and the value of $n$ $n=10^{-4}$, we obtain that:
\begin{eqnarray}
r_{pl,0} = r_{log,0} \approx 0.975634 - 2.22324b^2\label{rcaso2int3}.
\end{eqnarray}
Using the results of Eq. (\ref{rcaso2int3}) along with the value of $q$ obtained in order to have the phantom divide crossing, we have that the statefinder parameters $s_{pl,0}$ and $s_{log,0}$ assume the following value:
\begin{eqnarray}
s_{pl,0} = s_{log,0} \approx 0.00783629 +0.714997 b^2. \label{scaso2int3}
\end{eqnarray}
We can observe that, in the limiting case of $b^2=0$, i.e. in absence of interaction, we recover the same results of the non interacting case for all the three interacting cases and for all the three limiting cases considered.\\
Considering $b^2$ in the range $[0,0.025]$ for the cases corresponding to $\alpha  = 0.8824$ and $\beta = 0.5016$ and to $\alpha  = 0.8502$ and $\beta = 0.4817$, we can observe that, for all the three interacting cases considered, we find a point in the $r-s$ plane which is closer to the one corresponding to the $\Lambda$CDM model if compared with the non interacting case. This fact is more evident for the third interacting case, while it is slightly less for the second and the first interacting cases. Moreover, for all the three cases considered, we have $s<0$, which indicates a phantom-like model.\\
For the limiting case corresponding to the Ricci scale, we obtain that for all the three interacting case, there is a departure from the point $\left\{ r,s  \right\} =\left\{ 1,0  \right\}$, corresponding to the $\Lambda$CDM model.

\section{Cosmographic Parameters}
In this Section, we obtain some important cosmological information about the model we are considering using the properties of the cosmographic parameters.\\
Standard candles (like SNe Ia) represent powerful instruments in present day cosmology since they can be used in order to reconstruct the Hubble diagram, i.e. the redshift-distance relation up to high redshifts $z$. It is common to constrain a parameterized model against the data in order to check its validity and to constraint its free parameters. However, this type of approach is highly model-dependent, so there are still some debate in scientific community on the validity of the constraints on the derived cosmological quantities.\\
In order to avoid this kind of problem, it is possible to use cosmography, i.e. expanding the scale factor $a\left( t \right)$ in Taylor series with respect to the cosmic time $t$. This type of expansion leads to a distance-redshift relation which it is fully model independent since it is independent on the particular form of the solution of cosmic equations. Cosmography can be considered as a milestone in the study of the main properties of Universe dynamics, which any theoretical model considered has to consider and also to satisfy. It is useful to introduce the following quantities \cite{cosmo1,cosmo2}:
\begin{eqnarray}
q &=& -\frac{\ddot{a}}{a}H^{-2}= -\frac{\ddot{a}a}{\dot{a}^2} = -\frac{a^{\left(2\right)}a}{\dot{a}^2}, \label{par1} \\
j &=& \frac{1}{a}\frac{d^3a}{dt^3}H^{-3} = \frac{a^{\left(3\right)}a^2}{\dot{a}^3}, \label{par2} \\
s &=& \frac{1}{a}\frac{d^4a}{dt^4}H^{-4}= -\frac{a^{\left(4\right)}a^3}{\dot{a}^4} \label{par3} .
\end{eqnarray}
In general, we have that the $i$-th parameter $x^i$ can be obtained thanks to the following expression:
\begin{eqnarray}
x^{i} &=& \left( -1  \right)^{i+1}\frac{1}{H^{i}}\frac{a^{\left(i\right)}}{a}= \left( -1  \right)^{i+1} \frac{a^{\left(i\right)}a^{i-1}}{\dot{a}^{i+1}},
\end{eqnarray}
where the index $i$, when in parenthesis, indicates the order of the derivative with respect to the cosmic time $t$ while, when not in parenthesis, indicates the power law index of the corresponding quantity.\\
The quantities given in Eqs. (\ref{par1}), (\ref{par2}) and (\ref{par3}) are known, respectively, as deceleration, jerk and snap parameters.\\
Since we have that one of the statefinder parameters studied in the previous Section is indicated with $s$, we will indicate the snap parameter with $s_{cosmo}$ in order to avoid confusion.\\
The present-day values of these parameters, denoted with the subscript 0 (which indicates the value of the parameter for $z=0$ or equivalently for $t=0$) can be used in order to characterize the evolutionary status of the Universe. For example, a negative value of  $q_0$ indicates an accelerated expansion of the Universe, while the value of $j_0$ allows us to discriminate among different accelerating models. \\
Using the definitions given in Eqs. (\ref{par1}), (\ref{par2}) and (\ref{par3}), we can easily obtain the fourth order Taylor expansion of the scale factor $a\left( t \right)$ as follows:
\begin{eqnarray}
\frac{a\left(t\right)}{a\left(t_0\right)} &=& 1+H_0 \left( t -t_0 \right) - \frac{q_0}{2}H_0^2 \left( t -t_0 \right)^2 +  \frac{j_0}{3!}H_0^3 \left( t -t_0 \right)^3  \nonumber \\
 &&+\frac{s_0}{4!}H_0^4 \left( t -t_0 \right)^4   +  O \left[ \left( t-t_0  \right)^5 \right], \label{exp}
\end{eqnarray}
where $t_0 \approx 1/H_0$ gives the present day age of the Universe. It must be here underlined that Eq. (\ref{exp}) is also the fourth order expansion of $\left(1 + z\right)^{-1}$, since, from the definition of redshift $z$,  we obtain that:
\begin{eqnarray}
z = \frac{a\left( t_0 \right)}{a\left( t \right)} -1. \label{}
\end{eqnarray}
The deceleration parameter $q$ has been already introduced and studied in Section 2. We have seen that, in order to obtain the crossing divide line, $q$ must assume a negative value, which is in agreement with the present theories.\\
The jerk parameter $l$ is also another way to call the statefinder parameter $r$  and it represents a natural next step beyond the Hubble and the deceleration parameters $H$ and $q$ \cite{alam,arab}.\\
The snap parameter $s_{cosmo}$, which depends on the fourth derivative with respect to the cosmic time of the scale factor $a\left(t \right)$, is also known with the name of kerk parameter and it has been well studied and discussed in the recent papers written by Dabrowski \cite{dabro}, Dunajski $\&$ Gibbons \cite{duna} and Arabsalmania $\&$ Sahni \cite{arab}.
Another useful expression which can be used in order to find the expression of $s_{cosmo}$ involves the deceleration and the jerk parameters and it is given by:
\begin{eqnarray}
s_{cosmo} = \frac{\dot{j}}{h}-j\left( 2+3q  \right) = j' -j\left( 2+3q  \right) . \label{spara}
\end{eqnarray}
Some constraints about the value of $s$ have been recently obtained. For example, Capozziello $\&$ Izzo \cite{values0} have found that $s_0 = 8.32 \pm 12.16$, while John \cite{values0-1} has derived that  $s_0 = 36.5 \pm 52.9$. The errors of the values derived in these two works are also of the order of 200$\%$, for future more precise comparisons between cosmological constraints of $s$ and values obtained from theoretical models, it will be useful to have better constraints.

\subsection{Non Interacting Case}
We now  derive an expression for the snap parameter $s$ for the models we are dealing with. We start with the non interacting case.\\
Using the definition given in Eq. (\ref{spara}), we obtain the following expressions for $s_{pl,cosmo}$ and $s_{log,cosmo}$:
\begin{eqnarray}
s_{pl,cosmo} &=& \frac{1}{\beta}\frac{\Omega''_{Dpl}}{c^2\gamma_{pl}} + \frac{\Omega'_{Dpl}}{c^2\gamma_{pl}}\left[\frac{4}{\beta} +  \frac{7}{\beta^2}\left(\frac{\Omega_{Dpl}}{c^2\gamma_{pl}}  -\alpha\right) \right] \nonumber \\
&&+\left\{1 +  \frac{3}{\beta}\left(\frac{\Omega_{Dpl}}{c^2\gamma_{pl}}  -\alpha\right) +\frac{2}{\beta^2}\left[\left(\frac{\Omega_{Dpl}}{c^2\gamma_{pl}}  -\alpha\right)\right]^2 \right\} \times \nonumber \\
&& \left[ 1 +\frac{3}{\beta}\left( \frac{\Omega_{Dpl}}{c^2 \gamma_{pl}} - \alpha   \right) \right],  \\
s_{log,cosmo} &=& \frac{1}{\beta}\frac{\Omega''_{Dlog}}{c^2\gamma_{log}} + \frac{\Omega'_{Dlog}}{c^2\gamma_{log}}\left[\frac{4}{\beta} +  \frac{7}{\beta^2}\left(\frac{\Omega_{Dlog}}{c^2\gamma_{log}}  -\alpha\right) \right] \nonumber \\
&&+\left\{1 +  \frac{3}{\beta}\left(\frac{\Omega_{Dlog}}{c^2\gamma_{log}}  -\alpha\right) +\frac{2}{\beta^2}\left[\left(\frac{\Omega_{Dlog}}{c^2\gamma_{log}}  -\alpha\right)\right]^2 \right\} \times \nonumber \\
 &&\left[ 1 +\frac{3}{\beta}\left( \frac{\Omega_{Dlog}}{c^2 \gamma_{log}} - \alpha   \right) \right].
\end{eqnarray}
For the non interacting case, we can obtain the following expressions for $\Omega''_{Dpl}$ and $\Omega''_{Dlog}$:
\begin{eqnarray}
\Omega_{Dpl}'' &=& \Omega_{Dpl}'\left\{ \frac{2}{\beta}\left[  \frac{1-\Omega_{\phi} -\Omega_{Dpl}}{c^2\gamma_{pl}} - \left(\frac{\Omega_{Dpl}}{c^2\gamma_{pl}} +\beta - \alpha \right)  \right]  +u_{pl}\left(2n+1   \right) \right\}\nonumber\\
&& + u_{pl}'\Omega_{Dpl} \left(2n+1   \right),  \label{lgo19-2} \\
\Omega_{Dlog}'' &=& \Omega_{Dlog}'\left\{ \frac{2}{\beta}\left[  \frac{1-\Omega_{\phi} -\Omega_{Dlog}}{c^2\gamma_{log}} - \left(\frac{\Omega_{Dlog}}{c^2\gamma_{log}} +\beta - \alpha \right)  \right]  +u_{log}\left(2n+1   \right) \right\} \nonumber\\
&&+ u_{log}'\Omega_{Dlog} \left(2n+1   \right). \label{lgo19log-2}
\end{eqnarray}
We have already derived the expressions $\Omega'_{Dpl}$, $\Omega'_{Dlog}$ for the non interacting case, we now need to find the expression of $u_{pl}'$ and $u_{log}'$. \\
Differentiating with respect to $x$ the general expression of $u$, i.e. $u=\rho_m/\rho_D$, we find the following relation:
\begin{eqnarray}
u'=\frac{\rho'_m}{\rho_D} - \frac{\rho'_D \rho_m}{\rho_D^2} = \frac{\rho'_m}{\rho_D} - u\left(\frac{\rho'_D}{\rho_D}\right).\label{uprime}
\end{eqnarray}
From the continuity equations for DE and DM given in Eqs. (\ref{46}), (\ref{46log}) and (\ref{47}), we can find the following results:
\begin{eqnarray}
\rho'_{Dpl} &=& -3\rho_{Dpl}\left( 1+\omega_{Dpl}  \right), \label{rhodprimevs}  \\
\rho'_{Dlog} &=& -3\rho_{Dlog}\left( 1+\omega_{Dlog}  \right), \label{rhodprimevs2}  \\
\rho'_m &=&  -3\rho_m = -3u_{pl,log}\rho_{D(pl,log)}.\label{rhomprimevs3}
\end{eqnarray}
We must underline that in Eq. (\ref{rhomprimevs3}) we used the general definition of $u$, i.e. $u=\rho_m/\rho_D$.\\
Inserting in Eq. (\ref{uprime}) the results of Eqs. (\ref{rhodprimevs}), (\ref{rhodprimevs2}) and (\ref{rhomprimevs3}), we obtain, after some algebraic calculations, the following relations for $u'_{pl}$ and $u'_{log}$:
\begin{eqnarray}
u_{pl}' &=& 3u_{pl}\omega_{Dpl}, \label{uprimenonpl}\\
u_{log}' &=& 3u_{log}\omega_{Dlog}.\label{uprimenonlog}
\end{eqnarray}
Using the expressions of $\Omega'_{Dpl}$, $\Omega'_{Dlog}$, $u'_{pl}$ and $u'_{log}$ derived in Eqs. (\ref{52kissnat}), (\ref{52logkissnat}), (\ref{uprimenonpl}) and (\ref{uprimenonlog}), we obtain the following expressions for $\Omega_{Dpl}''$ and $\Omega_{Dlog}''$:
\begin{eqnarray}
\Omega_{Dpl}'' &=& \left[\frac{2}{\beta} \left(  \frac{\Omega_{Dpl}}{c^2\gamma_{pl}} +\beta  - \alpha \right)\left(1-\Omega_{\phi} - \Omega_{Dpl}   \right)    + u_{pl}\Omega_{Dpl}\left( 2n+1\right)\right]\times \nonumber \\
&&\left\{ \frac{2}{\beta}\left[  \frac{1-\Omega_{\phi} -\Omega_{Dpl}}{c^2\gamma_{pl}} - \left(\frac{\Omega_{Dpl}}{c^2\gamma_{pl}} +\beta - \alpha \right)  \right]  +u_{pl}\left(2n+1   \right) \right\}\nonumber\\
&& + 3u_{pl} \omega_{Dpl}\Omega_{Dpl} \left(2n+1   \right),  \label{lgo19} \\
\Omega_{Dlog}'' &=& \left[  \frac{2}{\beta} \left(  \frac{\Omega_{Dlog}}{c^2\gamma_{log}} +\beta  - \alpha \right)\left(1-\Omega_{\phi} - \Omega_{Dlog}  \right)    +u_{log}\Omega_{Dlog}\left( 2n+1\right)  \right]\times \nonumber \\
&&\left\{ \frac{2}{\beta}\left[  \frac{1-\Omega_{\phi} -\Omega_{Dlog}}{c^2\gamma_{log}} - \left(\frac{\Omega_{Dlog}}{c^2\gamma_{log}} +\beta - \alpha \right)  \right]  +u_{log}\left(2n+1   \right) \right\} \nonumber\\
&&+ 3u_{log} \omega_{Dlog}\Omega_{Dlog} \left(2n+1   \right). \label{lgo19log}
\end{eqnarray}
Then, using the expressions of $\Omega'_{Dpl}$, $\Omega'_{Dlog}$, $\Omega''_{Dpl}$, and $\Omega''_{Dlog}$ derived in Eqs. (\ref{52kissnat}), (\ref{52logkissnat}), (\ref{lgo19}) and (\ref{lgo19log}),  we have that the expressions of $s_{pl,cosmo}$ and $s_{log,cosmo}$ are given, respectively,  by:
\begin{eqnarray}
s_{pl,cosmo} &=& \frac{1}{\beta c^2\gamma_{pl}}\times \left[\frac{2}{\beta} \left(  \frac{\Omega_{Dpl}}{c^2\gamma_{pl}} +\beta  - \alpha \right)\left(1-\Omega_{\phi} - \Omega_{Dpl}   \right)    + u_{pl}\Omega_{Dpl}\left( 2n+1\right)\right]\times \nonumber \\
&&\left\{ \frac{2}{\beta}\left[  \frac{1-\Omega_{\phi} -\Omega_{Dpl}}{c^2\gamma_{pl}} - \left(\frac{\Omega_{Dpl}}{c^2\gamma_{pl}} +\beta - \alpha \right)  \right]  +u_{pl}\left(2n+1   \right) \right\}\nonumber\\
&& + \frac{3u_{pl} \omega_{Dpl}\Omega_{Dpl} \left(2n+1   \right)}{\beta c^2\gamma_{pl}}  \nonumber \\
&&+\left[ \frac{2}{\beta c^2\gamma_{pl}} \left(  \frac{\Omega_{Dpl}}{c^2\gamma_{pl}} +\beta  - \alpha \right)\left(1-\Omega_{\phi} - \Omega_{Dpl}   \right)    + \frac{u_{pl}\Omega_{Dpl}}{c^2\gamma_{pl}}\left( 2n+1\right)   \right]\times \nonumber \\
&&\left[\frac{4}{\beta} +  \frac{7}{\beta^2}\left(\frac{\Omega_{Dpl}}{c^2\gamma_{pl}}  -\alpha\right) \right] \nonumber \\
&&+\left\{1 +  \frac{3}{\beta}\left(\frac{\Omega_{Dpl}}{c^2\gamma_{pl}}  -\alpha\right) +\frac{2}{\beta^2}\left[\left(\frac{\Omega_{Dpl}}{c^2\gamma_{pl}}  -\alpha\right)\right]^2 \right\} \times \nonumber \\
&& \left[ 1 +\frac{3}{\beta}\left( \frac{\Omega_{Dpl}}{c^2 \gamma_{pl}} - \alpha   \right) \right],  \label{scosmo1} \\
s_{log,cosmo} &=& \frac{1}{\beta c^2\gamma_{log}}\times \left[\frac{2}{\beta} \left(  \frac{\Omega_{Dlog}}{c^2\gamma_{log}} +\beta  - \alpha \right)\left(1-\Omega_{\phi} - \Omega_{Dlog}   \right)    + u_{log}\Omega_{Dlog}\left( 2n+1\right)\right]\times \nonumber \\
&&\left\{ \frac{2}{\beta}\left[  \frac{1-\Omega_{\phi} -\Omega_{Dlog}}{c^2\gamma_{log}} - \left(\frac{\Omega_{Dlog}}{c^2\gamma_{log}} +\beta - \alpha \right)  \right]  +u_{log}\left(2n+1   \right) \right\}\nonumber\\
&& + \frac{3u_{log} \omega_{Dlog}\Omega_{Dlog} \left(2n+1   \right)}{\beta c^2\gamma_{log}}  \nonumber \\
&&\left[ \frac{2}{\beta c^2\gamma_{log}} \left(  \frac{\Omega_{Dlog}}{c^2\gamma_{log}} +\beta  - \alpha \right)\left(1-\Omega_{\phi} - \Omega_{Dlog}   \right)    + \frac{u_{log}\Omega_{Dlog}}{c^2\gamma_{log}}\left( 2n+1\right)   \right]\times \nonumber \\
&&\left[\frac{4}{\beta} +  \frac{7}{\beta^2}\left(\frac{\Omega_{Dlog}}{c^2\gamma_{log}}  -\alpha\right) \right] \nonumber \\
&&+\left\{1 +  \frac{3}{\beta}\left(\frac{\Omega_{Dlog}}{c^2\gamma_{log}}  -\alpha\right) +\frac{2}{\beta^2}\left[\left(\frac{\Omega_{Dlog}}{c^2\gamma_{log}}  -\alpha\right)\right]^2 \right\} \times \nonumber \\
 &&\left[ 1 +\frac{3}{\beta}\left( \frac{\Omega_{Dlog}}{c^2 \gamma_{log}} - \alpha   \right) \right]. \label{scosmo2}
\end{eqnarray}
Inserting in Eqs. (\ref{scosmo1}) and (\ref{scosmo2}) the values of the parameters involved, we obtain, for the non flat case, that:
\begin{eqnarray}
s_{pl,cosmo}  = s_{log} &=& 1.79942. \label{scosmores}
\end{eqnarray}
Moreover, considering the flat Universe case, i.e. for $\alpha  = 0.8502$ and $\beta = 0.4817$, we obtain:
\begin{eqnarray}
s_{pl,cosmo}  = s_{log} &=& 2.0227. \label{scosmoresflat}
\end{eqnarray}
Instead, at Ricci scale, i.e. for $\alpha =2$ and $\beta =1$, we obtain:
\begin{eqnarray}
s_{pl,cosmo}  = s_{log} &=& -0.88631 \label{scosmores3}.
\end{eqnarray}
We find that the values of $s$ obtained for the three different limiting cases are between the errors of the values obtained in Capozziello $\&$ Izzo \cite{values0} and John \cite{values0-1}.

\subsection{Interacting Case}
We now consider the three interacting cases chosen in this paper.\\
We start calculating the expression of $u'$ since it will be useful for the following calculations.\\
We follow the same procedure of the non interacting case, we then consider the following general expression:
\begin{eqnarray}
u'=\frac{\rho'_m}{\rho_D} - \frac{\rho'_D \rho_m}{\rho_D^2} = \frac{\rho'_m}{\rho_D} - u \frac{\rho'_D}{\rho_D}. \label{uprimeQ}
\end{eqnarray}
From the continuity equations for DE and DM given in Eqs. (\ref{46}), (\ref{46log})  and (\ref{47}), we find the following relations for $\rho'_D$ and $\rho'_m$ for the interacting case:
\begin{eqnarray}
\rho'_{D,pl} &=& -3\rho_{D,pl}\left( 1+\omega_{D,pl}  \right) - \frac{Q}{H}, \label{darkQ} \\
\rho'_{D,log} &=& -3\rho_{D,log}\left( 1+\omega_{D,log}  \right) - \frac{Q}{H}, \label{darkQlog} \\
\rho'_m &=&  -3\rho_m + \frac{Q}{H}.\label{darkQm}
\end{eqnarray}
Inserting in Eq. (\ref{uprimeQ}) the expressions of $\rho'_D$ and $\rho'_m$ obtained in Eqs. (\ref{darkQ}), (\ref{darkQlog}) and (\ref{darkQm}) along with the definitions of $Q_1$, $Q_2$ and $Q_3$, we find the following expressions for $u_{pl}'$ and $u_{log}'$ for the three different interacting case considered:
\begin{eqnarray}
u_{pl}' &=& 3u_{pl}\omega_{D,pl} + 3b^2 u{pl} \left(  1+u_{pl}  \right), \label{uprimeint3}\\
u_{log}' &=& 3u_{log}\omega_{D,log} + 3b^2 u_{log} \left(  1+u_{log}  \right), \label{uprimeint3-1}\\
u_{pl}' &=& 3u_{pl}\omega_{D,pl} + 3b^2 \left(  1+u_{pl}  \right), \label{uprimeint3-2}\\
u_{log}' &=& 3u_{log}\omega_{D,log} + 3b^2 \left(  1+u_{log}  \right), \label{uprimeint3-3}\\
u_{pl}' &=& 3u_{pl}\omega_{D,pl} + 3b^2 \left(  1+u_{pl}  \right)^2, \label{uprimeint3-4}\\
u_{log}' &=& 3u_{log}\omega_{D,log} + 3b^2 \left(  1+u_{log}  \right)^2. \label{uprimeint3-5}
\end{eqnarray}
For the case corresponding to $Q_1$, using the definitions of $\Omega_{Dpl}'$ and $\Omega_{Dlog}'$ obtained for this case, we derive:
\begin{eqnarray}
\Omega_{Dpl}'' &=&  \Omega_{Dpl}'\left\{ \frac{2}{\beta}\left[  \frac{1-\Omega_{\phi} -\Omega_{Dpl}}{c^2\gamma_{pl}} - \left(\frac{\Omega_{Dpl}}{c^2\gamma_{pl}} +\beta - \alpha \right)  \right]  +u_{pl}\left(2n+1 -b^2\right) \right\}\nonumber \\
&& +\Omega_{Dpl} \left[u_{pl}'\left(2n+1-b^2\right) \right], \label{lgo19} \\
\Omega_{Dlog}'' &=&  \Omega_{Dlog}'\left\{ \frac{2}{\beta}\left[  \frac{1-\Omega_{\phi} -\Omega_{Dlog}}{c^2\gamma_{log}} - \left(\frac{\Omega_{Dlog}}{c^2\gamma_{log}} +\beta - \alpha \right)  \right]  +u_{log}\left(2n+1 -b^2\right) \right\}\nonumber \\
&& +\Omega_{Dlog} \left[u_{log}'\left(2n+1-b^2\right) \right]. \label{lgo19log}
\end{eqnarray}
Therefore, using the expressions of $\Omega'_{Dpl}$, $\Omega'_{Dlog}$, $u'_{pl}$ and $u'_{log}$  obtained in Eqs. (\ref{52}), (\ref{52log}), (\ref{uprimeint3}) and (\ref{uprimeint3-1}),  we have then that the expressions of $\Omega''_{Dpl}$ and $\Omega''_{Dlog}$ are given, respectively,  by:
\begin{eqnarray}
\Omega_{Dpl}'' &=&  \left[  \frac{2}{\beta} \left(  \frac{\Omega_{Dpl}}{c^2\gamma_{pl}} +\beta  - \alpha \right)\left(1-\Omega_{\phi} - \Omega_{Dpl}   \right)    + u_{pl}\Omega_{Dpl}\left(  2n+1-b^2\right)  \right] \times \nonumber \\
&&\left\{ \frac{2}{\beta}\left[  \frac{1-\Omega_{\phi} -\Omega_{Dpl}}{c^2\gamma_{pl}} - \left(\frac{\Omega_{Dpl}}{c^2\gamma_{pl}} +\beta - \alpha \right)  \right]  +u_{pl}\left(2n+1 -b^2\right) \right\}\nonumber \\
&& +\Omega_{Dpl}\left(2n+1-b^2\right) \left[ 3u_{pl} \omega_{Dpl} +3b^2 u_{pl}\left(  1+ u_{pl}\right)  \right] , \label{lgo19} \\
\Omega_{Dlog}'' &=&  \left[ \frac{2}{\beta} \left(  \frac{\Omega_{Dlog}}{c^2\gamma_{log}} +\beta  - \alpha \right)\left(1-\Omega_{\phi} - \Omega_{Dlog}  \right)    + u_{log}\Omega_{Dlog}\left(  2n+1-b^2\right)   \right] \times \nonumber \\
&&\left\{ \frac{2}{\beta}\left[  \frac{1-\Omega_{\phi} -\Omega_{Dlog}}{c^2\gamma_{log}} - \left(\frac{\Omega_{Dlog}}{c^2\gamma_{log}} +\beta - \alpha \right)  \right]  +u_{log}\left(2n+1 -b^2\right) \right\}\nonumber \\
&& +\Omega_{Dlog}\left(2n+1-b^2\right)\left[ 3u_{log} \omega_{Dlog} +3b^2 u_{log}\left(  1+ u_{log}\right)  \right] . \label{lgo19log}
\end{eqnarray}
Using the expressions of $\Omega'_{Dpl}$, $\Omega'_{Dlog}$, $\Omega''_{Dpl}$, $\Omega''_{Dlog}$,  we have then that the  expressions of $s_{pl}$ and $s_{log}$ are given, respectively,  by:
\begin{eqnarray}
s_{pl} &=& \frac{1}{\beta c^2\gamma_{pl}} \times \left[  \frac{2}{\beta} \left(  \frac{\Omega_{Dpl}}{c^2\gamma_{pl}} +\beta  - \alpha \right)\left(1-\Omega_{\phi} - \Omega_{Dpl}   \right)    + u_{pl}\Omega_{Dpl}\left(  2n+1-b^2\right)  \right] \times \nonumber \\
&&\left\{ \frac{2}{\beta}\left[  \frac{1-\Omega_{\phi} -\Omega_{Dpl}}{c^2\gamma_{pl}} - \left(\frac{\Omega_{Dpl}}{c^2\gamma_{pl}} +\beta - \alpha \right)  \right]  +u_{pl}\left(2n+1 -b^2\right) \right\}\nonumber \\
&& +\frac{\Omega_{Dpl}\left(2n+1-b^2\right)}{\beta c^2\gamma_{pl}} \left[ 3u_{pl} \omega_{Dpl} +3b^2 u_{pl}\left(  1+ u_{pl}\right)  \right]\nonumber \\
&&\left[ \frac{2}{\beta c^2\gamma_{pl}} \left(  \frac{\Omega_{Dpl}}{c^2\gamma_{pl}} +\beta  - \alpha \right)\left(1-\Omega_{\phi} - \Omega_{Dpl}   \right)    + \frac{u_{pl}\Omega_{Dpl}}{c^2\gamma_{pl}}\left(  2n+1-b^2\right)  \right]\times \nonumber \\
&&\left[\frac{4}{\beta} +  \frac{7}{\beta^2}\left(\frac{\Omega_{Dpl}}{c^2\gamma_{pl}}  -\alpha\right) \right] \nonumber \\
&&+\left\{1 +  \frac{3}{\beta}\left(\frac{\Omega_{Dpl}}{c^2\gamma_{pl}}  -\alpha\right) +\frac{2}{\beta^2}\left[\left(\frac{\Omega_{Dpl}}{c^2\gamma_{pl}}  -\alpha\right)\right]^2 \right\} \times \nonumber \\
 &&\left[ 1 +\frac{3}{\beta}\left( \frac{\Omega_{Dpl}}{c^2 \gamma_{pl}} - \alpha   \right) \right],  \label{arpia1}\\
s_{log} &=& \frac{1}{\beta c^2\gamma_{log}} \times \left[  \frac{2}{\beta} \left(  \frac{\Omega_{Dlog}}{c^2\gamma_{log}} +\beta  - \alpha \right)\left(1-\Omega_{\phi} - \Omega_{Dlog}   \right)    + u_{log}\Omega_{Dlog}\left(  2n+1-b^2\right)  \right] \times \nonumber \\
&&\left\{ \frac{2}{\beta}\left[  \frac{1-\Omega_{\phi} -\Omega_{Dlog}}{c^2\gamma_{log}} - \left(\frac{\Omega_{Dlog}}{c^2\gamma_{log}} +\beta - \alpha \right)  \right]  +u_{log}\left(2n+1 -b^2\right) \right\}\nonumber \\
&& +\frac{\Omega_{Dlog}\left(2n+1-b^2\right)}{\beta c^2\gamma_{log}} \left[ 3u_{log} \omega_{Dlog} +3b^2 u_{log}\left(  1+ u_{log}\right)  \right]\nonumber \\
&&\left[ \frac{2}{\beta c^2\gamma_{log}} \left(  \frac{\Omega_{Dlog}}{c^2\gamma_{log}} +\beta  - \alpha \right)\left(1-\Omega_{\phi} - \Omega_{Dlog}   \right)    + \frac{u_{log}\Omega_{Dlog}}{c^2\gamma_{log}}\left(  2n+1-b^2\right)  \right]\times \nonumber \\
&&\left[\frac{4}{\beta} +  \frac{7}{\beta^2}\left(\frac{\Omega_{Dlog}}{c^2\gamma_{log}}  -\alpha\right) \right] \nonumber \\
&&+\left\{1 +  \frac{3}{\beta}\left(\frac{\Omega_{Dlog}}{c^2\gamma_{log}}  -\alpha\right) +\frac{2}{\beta^2}\left[\left(\frac{\Omega_{Dlog}}{c^2\gamma_{log}}  -\alpha\right)\right]^2 \right\} \times \nonumber \\
 &&\left[ 1 +\frac{3}{\beta}\left( \frac{\Omega_{Dlog}}{c^2 \gamma_{log}} - \alpha   \right) \right]. \label{arpia2}
\end{eqnarray}
Inserting in Eqs. (\ref{arpia1}) and (\ref{arpia2}) the values of the parameters involved, we obtain, for the non flat case, that:
\begin{eqnarray}
s_{pl,cosmo}  = s_{log,cosmo}   &=& 1.79942 +\left( 1.81536 - 2.09948  b^2\right)b^2  \label{scosmoresint1}.
\end{eqnarray}
Instead, for a flat Universe,  i.e. for $\alpha  = 0.8502$ and $\beta = 0.4817$, we obtain the following value:
\begin{eqnarray}
s_{pl,cosmo}  = s_{log,cosmo}   &=& 2.0227 + \left(1.70528 -2.09683  b^2 \right)b^2 \label{scosmoresint1flat}.
\end{eqnarray}
Finally, at Ricci scale, i.e. for $\alpha =2 $ and $\beta =1$, we obtain the following value:
\begin{eqnarray}
s_{pl,cosmo}  = s_{log,cosmo}   &=& -0.88631 +\left( 3.95059 -2.77329b^2   \right)b^2 \label{scosmoresint1ricci}.
\end{eqnarray}

Following the same procedure of the first interacting case, for the case corresponding to $Q_2$ we have:
\begin{eqnarray}
\Omega_{Dpl}'' &=&  \Omega_{Dpl}'\left\{ \frac{2}{\beta}\left[  \frac{1-\Omega_{\phi} -\Omega_{Dpl}}{c^2\gamma_{pl}} - \left(\frac{\Omega_{Dpl}}{c^2\gamma_{pl}} +\beta - \alpha \right)  \right]  +u_{pl}\left(2n+1 \right)-b^2 \right\}\nonumber \\
&& +\Omega_{Dpl} \left[u_{pl}'\left(2n+1\right)   \right], \label{lgo19} \\
\Omega_{Dlog}'' &=&  \Omega_{Dlog}'\left\{ \frac{2}{\beta}\left[  \frac{1-\Omega_{\phi} -\Omega_{Dlog}}{c^2\gamma_{log}} - \left(\frac{\Omega_{Dlog}}{c^2\gamma_{log}} +\beta - \alpha \right)  \right]  +u_{log}\left(2n+1 \right)-b^2 \right\}\nonumber \\
&& +\Omega_{Dlog} \left[u_{log}'\left(2n+1\right)  \right]. \label{lgo19log}
\end{eqnarray}
Using the expressions of $\Omega'_{Dpl}$, $\Omega'_{Dlog}$, $u'_{pl}$ and $u'_{log}$ obtained in Eqs. (\ref{52caro}), (\ref{52logcaro}), (\ref{uprimeint3-2}) and (\ref{uprimeint3-3}),  we have then that the  expressions of $\Omega''_{Dpl}$ and $\Omega''_{Dlog}$ are given, respectively,  by:
\begin{eqnarray}
\Omega_{Dpl}'' &=&  \left\{ \frac{2}{\beta} \left(  \frac{\Omega_{Dpl}}{c^2\gamma_{pl}} +\beta  - \alpha \right)\left(1-\Omega_{\phi} - \Omega_{Dpl}   \right)    + \Omega_{Dpl}\left[u_{pl}\left(  2n+1\right) -b^2\right]  \right\} \times \nonumber \\
&&\left\{ \frac{2}{\beta}\left[  \frac{1-\Omega_{\phi} -\Omega_{Dpl}}{c^2\gamma_{pl}} - \left(\frac{\Omega_{Dpl}}{c^2\gamma_{pl}} +\beta - \alpha \right)  \right]  +u_{pl}\left(2n+1 \right)-b^2 \right\}\nonumber \\
&& +\Omega_{Dpl}\left(2n+1\right) \left[ 3u_{pl} \omega_{Dpl} +3b^2\left(  1+ u_{pl}\right) \right] , \label{lgo19} \\
\Omega_{Dlog}'' &=&  \left\{ \frac{2}{\beta} \left(  \frac{\Omega_{Dlog}}{c^2\gamma_{log}} +\beta  - \alpha \right)\left(1-\Omega_{\phi} - \Omega_{Dlog}  \right)    + \Omega_{Dlog}\left[ u_{log}\left(  2n+1\right) - b^2\right]  \right\} \times \nonumber \\
&&\left\{ \frac{2}{\beta}\left[  \frac{1-\Omega_{\phi} -\Omega_{Dlog}}{c^2\gamma_{log}} - \left(\frac{\Omega_{Dlog}}{c^2\gamma_{log}} +\beta - \alpha \right)  \right]  +u_{log}\left(2n+1 \right)-b^2 \right\}\nonumber \\
&& +\Omega_{Dlog} \left(2n+1\right) \left[ 3u_{log} \omega_{Dlog} +3b^2\left(  1+ u_{log}\right) \right]. \label{lgo19log}
\end{eqnarray}
We finally have then that expressions of $s_{pl}$ and $s_{log,cosmo}  $ are given, respectively,  by:
\begin{eqnarray}
s_{pl} &=& \frac{1}{\beta c^2\gamma_{pl}} \times \left\{ \frac{2}{\beta} \left(  \frac{\Omega_{Dpl}}{c^2\gamma_{pl}} +\beta  - \alpha \right)\left(1-\Omega_{\phi} - \Omega_{Dpl}   \right)    + \Omega_{Dpl}\left[u_{pl}\left(  2n+1\right) -b^2\right]  \right\} \times \nonumber \\
&&\left\{ \frac{2}{\beta}\left[  \frac{1-\Omega_{\phi} -\Omega_{Dpl}}{c^2\gamma_{pl}} - \left(\frac{\Omega_{Dpl}}{c^2\gamma_{pl}} +\beta - \alpha \right)  \right]  +u_{pl}\left(2n+1 \right)-b^2 \right\}\nonumber \\
&& +\frac{\Omega_{Dpl}\left(2n+1\right)}{\beta c^2\gamma_{pl}}\left[ 3u_{pl} \omega_{Dpl} +3b^2\left(  1+ u_{pl}\right) \right]\nonumber \\
&&+\left\{ \frac{2}{\beta c^2\gamma_{pl}} \left(  \frac{\Omega_{Dpl}}{c^2\gamma_{pl}} +\beta  - \alpha \right)\left(1-\Omega_{\phi} - \Omega_{Dpl}   \right)    + \frac{\Omega_{Dpl}}{c^2\gamma_{pl}}\left[ u_{pl}\left(2n+1\right)-b^2\right]  \right\}\times \nonumber \\
&&\left[\frac{4}{\beta} +  \frac{7}{\beta^2}\left(\frac{\Omega_{Dpl}}{c^2\gamma_{pl}}  -\alpha\right) \right] \nonumber \\
&&+\left\{1 +  \frac{3}{\beta}\left(\frac{\Omega_{Dpl}}{c^2\gamma_{pl}}  -\alpha\right) +\frac{2}{\beta^2}\left[\left(\frac{\Omega_{Dpl}}{c^2\gamma_{pl}}  -\alpha\right)\right]^2 \right\} \times \nonumber \\
 &&\left[ 1 +\frac{3}{\beta}\left( \frac{\Omega_{Dpl}}{c^2 \gamma_{pl}} - \alpha   \right) \right] , \label{rituccia1}\\
s_{log,cosmo}   &=&  \frac{1}{\beta c^2\gamma_{log}} \times \left\{ \frac{2}{\beta} \left(  \frac{\Omega_{Dlog}}{c^2\gamma_{log}} +\beta  - \alpha \right)\left(1-\Omega_{\phi} - \Omega_{Dlog}   \right)    + \Omega_{Dlog}\left[u_{log}\left(  2n+1\right) -b^2\right]  \right\} \times \nonumber \\
&&\left\{ \frac{2}{\beta}\left[  \frac{1-\Omega_{\phi} -\Omega_{Dlog}}{c^2\gamma_{log}} - \left(\frac{\Omega_{Dlog}}{c^2\gamma_{log}} +\beta - \alpha \right)  \right]  +u_{log}\left(2n+1 \right)-b^2 \right\}\nonumber \\
&& +\frac{\Omega_{Dlog}\left(2n+1\right)}{\beta c^2\gamma_{log}}\left[ 3u_{log} \omega_{Dlog} +3b^2\left(  1+ u_{log}\right) \right]\nonumber \\
&&+\left\{ \frac{2}{\beta c^2\gamma_{log}} \left(  \frac{\Omega_{Dlog}}{c^2\gamma_{log}} +\beta  - \alpha \right)\left(1-\Omega_{\phi} - \Omega_{Dlog}   \right)    + \frac{\Omega_{Dlog}}{c^2\gamma_{log}}\left[ u_{log}\left(2n+1\right)-b^2\right]  \right\}\times \nonumber \\
&&\left[\frac{4}{\beta} +  \frac{7}{\beta^2}\left(\frac{\Omega_{Dlog}}{c^2\gamma_{log}}  -\alpha\right) \right] \nonumber \\
&&+\left\{1 +  \frac{3}{\beta}\left(\frac{\Omega_{Dlog}}{c^2\gamma_{log}}  -\alpha\right) +\frac{2}{\beta^2}\left[\left(\frac{\Omega_{Dlog}}{c^2\gamma_{log}}  -\alpha\right)\right]^2 \right\} \times \nonumber \\
 &&\left[ 1 +\frac{3}{\beta}\left( \frac{\Omega_{Dlog}}{c^2 \gamma_{log}} - \alpha   \right) \right]. \label{rituccia2}
\end{eqnarray}
Inserting in Eqs. (\ref{rituccia1}) and (\ref{rituccia2}) the values of the parameters involved, we obtain, for the non flat case, that:
\begin{eqnarray}
s_{pl,cosmo}  = s_{log,cosmo}   &=&  1.79942 +\left(0.174543 +2.07428   b^2\right)b^2. \label{scosmoresint2}
\end{eqnarray}
Instead, for a flat Universe,  i.e. for $\alpha  = 0.8502$ and $\beta = 0.4817$, we obtain that:
\begin{eqnarray}
s_{pl,cosmo}  = s_{log,cosmo}   &=& 2.0227 + \left(-0.0892113 + 2.14416 b^2 \right)b^2 \label{scosmoresint2flat}.
\end{eqnarray}
Finally, at Ricci scale, i.e. for $\alpha =2$ and $\beta =1$, we obtain the following result:
\begin{eqnarray}
s_{pl,cosmo}  = s_{log,cosmo}   &=& -0.88631 +\left(4.22919 + 1.04046   b^2   \right)b^2 \label{scosmoresint2ricci}.
\end{eqnarray}

Finally, for the case corresponding to $Q_3$, we obtain:
\begin{eqnarray}
\Omega_{Dpl}'' &=&  \Omega_{Dpl}'\left\{ \frac{2}{\beta}\left[  \frac{1-\Omega_{\phi} -\Omega_{Dpl}}{c^2\gamma_{pl}} - \left(\frac{\Omega_{Dpl}}{c^2\gamma_{pl}} +\beta - \alpha \right)  \right]  +u_{pl}\left(2n+1 -b^2\right)-b^2 \right\}\nonumber \\
&& +\Omega_{Dpl} \left[u_{pl}'\left(2n+1-b^2\right) \right], \label{lgo19} \\
\Omega_{Dlog}'' &=&  \Omega_{Dlog}'\left\{ \frac{2}{\beta}\left[  \frac{1-\Omega_{\phi} -\Omega_{Dlog}}{c^2\gamma_{log}} - \left(\frac{\Omega_{Dlog}}{c^2\gamma_{log}} +\beta - \alpha \right)  \right]  +u_{log}\left(2n+1 -b^2\right) -b^2\right\}\nonumber \\
&& +\Omega_{Dlog} \left[u_{log}'\left(2n+1-b^2\right)   \right]. \label{lgo19log}
\end{eqnarray}
Using the expressions of $\Omega'_{Dpl}$, $\Omega'_{Dlog}$, $u'_{pl}$ and $u'_{log}$ obtained in Eqs. (\ref{52caro2}), (\ref{52logcaro2}), (\ref{uprimeint3-4}) and (\ref{uprimeint3-5}),  we have then that the  expressions of $\Omega''_{Dpl}$ and $\Omega''_{Dlog}$ are given, respectively,  by:
\begin{eqnarray}
\Omega_{Dpl}'' &=&  \left\{ \frac{2}{\beta} \left(  \frac{\Omega_{Dpl}}{c^2\gamma_{pl}} +\beta  - \alpha \right)\left(1-\Omega_{\phi} - \Omega_{Dpl}   \right)  + \Omega_{pl}\left[u_{pl}\left(2n+1 -b^2  \right)  -b^2 \right]  \right\}\times \nonumber \\
&&\left\{ \frac{2}{\beta}\left[  \frac{1-\Omega_{\phi} -\Omega_{Dpl}}{c^2\gamma_{pl}} - \left(\frac{\Omega_{Dpl}}{c^2\gamma_{pl}} +\beta - \alpha \right)  \right]  +u_{pl}\left(2n+1 -b^2\right)-b^2 \right\}\nonumber \\
&& +\Omega_{Dpl}\left(2n+1-b^2\right) \left[  3u_{pl} \omega_{Dpl} +3b^2\left(  1+ u_{pl}\right)^2  \right], \label{lgo19} \\
\Omega_{Dlog}'' &=&  \left\{  \frac{2}{\beta} \left(  \frac{\Omega_{Dlog}}{c^2\gamma_{log}} +\beta  - \alpha \right)\left(1-\Omega_{\phi} - \Omega_{Dlog}  \right)  + \Omega_{log}\left[u_{log}\left(2n+1 -b^2  \right)  -b^2 \right] \right\}\times \nonumber \\
&&\left\{ \frac{2}{\beta}\left[  \frac{1-\Omega_{\phi} -\Omega_{Dlog}}{c^2\gamma_{log}} - \left(\frac{\Omega_{Dlog}}{c^2\gamma_{log}} +\beta - \alpha \right)  \right]  +u_{log}\left(2n+1 -b^2\right) -b^2\right\}\nonumber \\
&& +\Omega_{Dlog} \left(2n+1-b^2\right) \left[ 3u_{log} \omega_{Dlog}  +3b^2\left(  1+ u_{log}\right)^2  \right]. \label{lgo19log}
\end{eqnarray}
We finally obtain that the expressions of $s_{pl}$ and $s_{log,cosmo}  $ are given, respectively,  by:
\begin{eqnarray}
s_{pl} &=& \frac{1}{\beta c^2\gamma_{pl}} \times \left\{ \frac{2}{\beta} \left(  \frac{\Omega_{Dpl}}{c^2\gamma_{pl}} +\beta  - \alpha \right)\left(1-\Omega_{\phi} - \Omega_{Dpl}   \right)  + \Omega_{pl}\left[u_{pl}\left(2n+1 -b^2  \right)  -b^2 \right]  \right\}\times \nonumber \\
&&\left\{ \frac{2}{\beta}\left[  \frac{1-\Omega_{\phi} -\Omega_{Dpl}}{c^2\gamma_{pl}} - \left(\frac{\Omega_{Dpl}}{c^2\gamma_{pl}} +\beta - \alpha \right)  \right]  +u_{pl}\left(2n+1 -b^2\right)-b^2 \right\}\nonumber \\
&& +\frac{\Omega_{Dpl}\left(2n+1-b^2\right)}{\beta c^2\gamma_{pl}} \left[  3u_{pl} \omega_{Dpl} +3b^2\left(  1+ u_{pl}\right)^2  \right]\nonumber \\
&&+\left\{ \frac{2}{\beta c^2\gamma_{pl}} \left(  \frac{\Omega_{Dpl}}{c^2\gamma_{pl}} +\beta  - \alpha \right)\left(1-\Omega_{\phi} - \Omega_{Dpl}   \right)    + \frac{\Omega_{Dpl}}{c^2\gamma_{pl}}\left[ u_{pl}\left(2n+1-b^2\right)-b^2\right]  \right\}\times \nonumber \\
&&\left[\frac{4}{\beta} +  \frac{7}{\beta^2}\left(\frac{\Omega_{Dpl}}{c^2\gamma_{pl}}  -\alpha\right) \right] \nonumber \\
&&+\left\{1 +  \frac{3}{\beta}\left(\frac{\Omega_{Dpl}}{c^2\gamma_{pl}}  -\alpha\right) +\frac{2}{\beta^2}\left[\left(\frac{\Omega_{Dpl}}{c^2\gamma_{pl}}  -\alpha\right)\right]^2 \right\} \times \nonumber \\
 &&\left[ 1 +\frac{3}{\beta}\left( \frac{\Omega_{Dpl}}{c^2 \gamma_{pl}} - \alpha   \right) \right], \label{rituccia10} \\
s_{log,cosmo}   &=& \frac{1}{\beta c^2\gamma_{log}} \times \left\{ \frac{2}{\beta} \left(  \frac{\Omega_{Dlog}}{c^2\gamma_{log}} +\beta  - \alpha \right)\left(1-\Omega_{\phi} - \Omega_{Dlog}   \right)  + \Omega_{log}\left[u_{log}\left(2n+1 -b^2  \right)  -b^2 \right]  \right\}\times \nonumber \\
&&\left\{ \frac{2}{\beta}\left[  \frac{1-\Omega_{\phi} -\Omega_{Dlog}}{c^2\gamma_{log}} - \left(\frac{\Omega_{Dlog}}{c^2\gamma_{log}} +\beta - \alpha \right)  \right]  +u_{log}\left(2n+1 -b^2\right)-b^2 \right\}\nonumber \\
&& +\frac{\Omega_{Dlog}\left(2n+1-b^2\right)}{\beta c^2\gamma_{log}} \left[  3u_{log} \omega_{Dlog} +3b^2\left(  1+ u_{log}\right)^2  \right]\nonumber \\
&&+\left\{ \frac{2}{\beta c^2\gamma_{log}} \left(  \frac{\Omega_{Dlog}}{c^2\gamma_{log}} +\beta  - \alpha \right)\left(1-\Omega_{\phi} - \Omega_{Dlog}   \right)    + \frac{\Omega_{Dlog}}{c^2\gamma_{log}}\left[ u_{log}\left(2n+1\right)-b^2\right]  \right\}\times \nonumber \\
&&\left[\frac{4}{\beta} +  \frac{7}{\beta^2}\left(\frac{\Omega_{Dlog}}{c^2\gamma_{log}}  -\alpha\right) \right] \nonumber \\
&&+\left\{1 +  \frac{3}{\beta}\left(\frac{\Omega_{Dlog}}{c^2\gamma_{log}}  -\alpha\right) +\frac{2}{\beta^2}\left[\left(\frac{\Omega_{Dlog}}{c^2\gamma_{log}}  -\alpha\right)\right]^2 \right\} \times \nonumber \\
&& \left[ 1 +\frac{3}{\beta}\left( \frac{\Omega_{Dlog}}{c^2 \gamma_{log}} - \alpha   \right) \right]. \label{rituccia11}
\end{eqnarray}
Inserting in Eqs. (\ref{rituccia10}) and (\ref{rituccia11}) the values of the parameters involved, we obtain, for the non flat case:
\begin{eqnarray}
s_{pl,cosmo}  = s_{log,cosmo}   &=&  1.79942 +\left( 1.9899 - 3.79521  b^2\right)b^2. \label{scosmoresint3}
\end{eqnarray}
For a flat Universe, i.e. for $\alpha  = 0.8502$ and $\beta = 0.4817$, we obtain the following result:
\begin{eqnarray}
s_{pl,cosmo}  = s_{log,cosmo}   &=& 2.0227 + \left( 1.61607 - 3.68551 b^2 \right)b^2 \label{scosmoresint3flat}.
\end{eqnarray}
At Ricci scale, i.e. for $\alpha =2$ and $\beta =1$, we obtain the following result:
\begin{eqnarray}
s_{pl,cosmo}  = s_{log,cosmo}   &=&  -0.88631 +\left( 8.17977 -7.47243  b^2   \right)b^2\label{scosmoresint3ricci}.
\end{eqnarray}

We observe that, in the limiting case of $b^2=0$, i.e. in absence of interaction, the results of the three interacting cases lead to the same results of the non interacting case for all the three limiting cases taken into account. Moreover, for the range of values of $b^2=\left[0,0.025   \right]$, we obtain that the values of $s_{cosmo}$ still stay between the range of the errors of the values derived in Capozziello $\&$ Izzo \cite{values0} and John \cite{values0-1}.

\section{Concluding Remarks}
In this work, we considered the Power Law Entropy Corrected Holographic Dark Energy (PLECHDE) and the Logarithmic Entropy Corrected Holographic Dark Energy (LECHDE) models in the framework of Brans-Dicke cosmology and we considered as infrared cut-off of the system the Granda-Oliveros scale $L_{GO}$, which contains a term proportional to the first time derivative of the Hubble parameter $H$ (i.e. $\dot{H}$) and one term proportional to the Hubble parameter squared (i.e. $H^2$). The GO cut-off is also characterized by two free parameters indicated with $\alpha$ and $\beta$.  Both the power law and the logarithmic corrections to the entropy are well motivated from the Loop Quantum Gravity (LQG).

We considered a non-flat FLRW background in Brans-Dicke cosmology, which involves a scalar field $\phi$ accounting for a dynamical gravitational constant. We assumed a particular ansatz, wherein the BD scalar field $\phi$ has an evolution depending on the expansion of the Universe. Therefore, the correspondence between the field and the PLECHDE and LECHDE models with GO scale as  IR cut-off is established, in order to study their dynamics, which is governed by some dynamical parameters, like the Equation of State (EoS) parameter $\omega_D$, the evolutionary form of energy density parameter of DE $\Omega_D'$ and the deceleration parameter $q$. We calculated them in the non-flat Universe for both cases corresponding to absence of interaction and presence of interaction between DE and DM. We have chosen three different interaction terms  between DE and DM, indicated with $Q_1$, $Q_2$ and $Q_3$, which expressions are given, respectively, by: $Q_1=3b^2H\rho_m$, $Q_2=3b^2H\rho_D$ and $Q_3=3b^2H\left(\rho_m + \rho_D \right) $.  Moreover, we calculated the limiting cases corresponding to $\lambda = 0$ (i.e., $\gamma_{pl} =1$), to $\varrho = \epsilon =0$ (i.e. $ \gamma_{log} = 1$), to Einstein's gravity, to $\lambda = 0$ and Einstein's gravity together and to $\varrho = \epsilon =0$ and Einstein's gravity together as well. Furthermore, we considered the case corresponding to the Ricci scale, which is recovered for $\alpha = 2$ and $\beta = 1$.\\
We studied the statefinder diagnostic for the models we considered. We obtained  that, for the  non flat  and the flat Universe,  the models we are studying lead to  two points which are slightly far from the point $\left\{r,s\right\} = \left\{1, 0 \right\}$ (corresponding to the $\Lambda$CDM model), with the non flat case closer to the point corresponding to the $\Lambda$CDM model. Moreover, since we obtained $s<0$ for both cases, we can conclude that the models considered have a phantom-like behavior in the case of absence of interacting between Dark Sectors for both flat and non flat cases.\\
Instead, for the case corresponding to the Ricci scale, i.e. for $\alpha =2$ and $\beta =1$, the models we are studying lead to point  in the $\left\{r,s\right\}$ plane which is closer to $\left\{r,s\right\} = \left\{1, 0 \right\}$ if compared with the other two models. However, since for the Ricci scale we obtained $s>0$, it means that for this particular limiting case we deal with a quintessence-like model. \\
For the interacting case, we can observe that, in the limiting case of $b^2=0$, i.e. in absence of interaction, we recover the same results of the non interacting case for all the three interacting cases and for all the three limiting cases considered.\\
Considering $b^2$ in the range $[0,0.025]$ for the cases corresponding to the non flat and to the flat cases, we can observe that, for all the three interacting cases considered, we find a point in the $\left\{r,s\right\}$ plane which is closer to the one corresponding to the $\Lambda$CDM model if compared with the non interacting case. This fact is more evident for the third interacting case, while it is slightly less for the second and the first interacting cases. Moreover, for all the three cases considered, we have $s<0$, which indicates a phantom-like model. For the limiting case corresponding to the Ricci scale, we obtain that for all the three interacting case, there is a departure from the point $\left\{ r,s  \right\} =\left\{ 1,0  \right\}$, corresponding to the $\Lambda$CDM model.\\
We finally made some considerations about the expressions for the cosmographic parameters we obtained for the models studied in this work. We found that the value of the cosmographic parameter $s$ is in agreement with some results obtained in other recent  works.

\section*{Acknowledgements}
S. Chattopadhyay acknowledges financial support from SERB/DST, Govt of India under project grant no. SR/FTP/PS-167/2011.


\begin{thebibliography}{99}
\bibitem{1} P. de Bernardis  et al.,  Nature \textbf{404}, 955 (2000)
\bibitem{1a} S. Perlmutter  et al.,  Astrophys. J.   \textbf{517}, 565 (1999)
\bibitem{1b} A.G. Riess  G et al.,   Astron. J.   \textbf{116}, 1009 (1998)
\bibitem{1c} U. Seljak  et al.,   Phys. Rev. D   \textbf{71}, 103515 (2005)
\bibitem{1d} P. Astier  et al.,    Astron. Astrophys.   \textbf{447}, 31 (2006)
\bibitem{cmb3} E. Komatsu et al.,     Astrophys. J. Suppl.   \textbf{180}, 330 (2009)
\bibitem{planck} Planck Collaboration, P.A.R. Ade, Aghanim N et al., 2013, [arXiv:1303.5076]
\bibitem{sds1} M. Tegmark et al.,    Phys. Rev. D   \textbf{69}, 103501 (2004)
\bibitem{sds2} K. Abazajian  et al.,   Astron. J.   \textbf{128}, 502 (2004)
\bibitem{xray} S.W. Allen   et al.,   Mon. Not. Roy. Astron. Soc.   \textbf{353}, 457 (2004)
\bibitem{copeland-2006} E.J. Copeland, M. Sami, S. Tsujikawa,    Int. J. Mod. Phys. D   \textbf{15}, 1753 (2006)
\bibitem{delcampo} S. del Campo,  R. Herrera, D. Pavon,    J. Cosmol. Astropart. Phys.   \textbf{0901}, 020 (2009)
\bibitem{delcampoa} G. Leon, E.N. Saridakis,    Phys. Lett. B   \textbf{693}, 1 (2010)
\bibitem{delcampoc} M.S. Berger, H. Shojae,   Phys. Rev. D   \textbf{73}, 083528 (2006)
\bibitem{delcampod} X. Zhang,     Mod. Phys. Lett. A   \textbf{20}, 2575 (2005)
\bibitem{delcampoe} K. Griest,    Phys. Rev. D   \textbf{66}, 123501 (2002)
\bibitem{delcampoi} M. Jamil, M.U. Farooq,    J. Cosmol. Astropart. Phys.   \textbf{03}, 001 (2010)
\bibitem{delcampol} M. Jamil, A. Sheykhi, M.U. Farooq,     Int. J. Mod. Phys. D   \textbf{19}, 1831 (2010)
\bibitem{twothirds} H.V. Peiris et al.,    Astrophys. J. Suppl. Ser.   \textbf{148}, 213 (2003)
\bibitem{dil1} N. Arkani-Hamed, P. Creminelli, S. Mukohyama, M. Zaldarriaga,     J. Cosmol. Astropart. Phys.   \textbf{4}, 1 (2004)
\bibitem{dil2} M. Gasperini, F. Piazza, G. Veneziano,     Phys. Rev. D   \textbf{65}, 023508 (2001)
\bibitem{dil2-1} E. Elizalde et al.,     Eur. Phys. J. C   \textbf{53}, 447 (2008)
\bibitem{kess3} T. Chiba, T. Okabe, M. Yamaguchi,    Phys. Rev. D   \textbf{62}, 023511 (2000)
\bibitem{kess4} C. Armendariz-Picon, T. Damour, V. Mukhanov,     Phys. Lett. B   \textbf{458}, 209 (1999)
\bibitem{quint1} B. Ratra, P.J.E. Peebles,    Phys. Rev. D   \textbf{37}, 3406 (1988)
\bibitem{quint2} C. Wetterich,     Nucl. Phys. B   \textbf{302}, 668 (1988)
\bibitem{quint3} I. Zlatev, L. Wang, P.J. Steinhardt,    Phys. Rev.  Lett.   \textbf{82}, 896 (1999)
\bibitem{quint4} P.J.E. Peebles, B. Ratra,      Astrophys. J. Lett.   \textbf{325}, L17 (1988)
\bibitem{tac1} A. Sen,     Journal of High Energy Physics   \textbf{4}, 48 (2002)
\bibitem{tac3} T. Padmanabhan, T.R. Choudhury,     Phys. Rev. D   \textbf{66}, 081301 (2002)
\bibitem{tac1-2} Y. Shao, Y.X. Gui, W. Wang,     Mod. Phys. Lett. A   \textbf{22}, 1175 (2007)
\bibitem{tac1-3} G. Calcagni, A.R. Liddle,    Phys. Rev. D   \textbf{74}, 043528 (2006)
\bibitem{tac2-1} A. Sen,   Journal of High Energy Physics   \textbf{10}, 8 (1999)
\bibitem{pha1} R.R. Caldwell,    Phys. Lett. B   \textbf{545}, 23 (2002)
\bibitem{pha2} S. Nojiri, S.D. Odintsov,      Phys.Lett.B   \textbf{565}, 1 (2003)
\bibitem{pha5} L.P. Chimento, R. Lazkor,     Phys. Rev. Lett.   \textbf{91}, 211301 (2003)
\bibitem{qui2} E. Elizalde E, S. Nojiri, S.D. Odintsov,    Phys. Rev. D   \textbf{70}, 043539 (2004)
\bibitem{qui3} S. Nojiri, S.D. Odintsov, S. Tsujikawa,     Phys. Rev. D   \textbf{71}, 063004 (2005)
\bibitem{qui4} B. Feng, X.L. Wang, X.M. Zhang,     Phys. Lett. B   \textbf{607}, 35 (2005)
\bibitem{2a} Y.F. Cai, E.N. Saridakis, M.R. Setare,  J.Q. Xia,     Phys. Rep.   \textbf{493}, 1 (2010)
\bibitem{cgas1} A. Kamenshchik, U. Moschella, V. Pasquier,    Phys. Lett. B   \textbf{511}, 265 (2001)
\bibitem{cgas3} M.R. Setare,     European Phys. J. C   \textbf{52}, 689 (2007)
\bibitem{cgas41} A.E. Bernardini, O. Bertolami,     Phys. Rev. D   {\bf 77}, 083506 (2008)
\bibitem{ade1} H. Wei, R.G. Cai,     Phys. Lett. B   \textbf{660}, 113 (2008)
\bibitem{ade2} R.G. Cai,    Phys. Lett. B   \textbf{657}, 228 (2007)
\bibitem{3} M. Li,   Phys. Lett. B   \textbf{603}, 1 (2004)
\bibitem{3b} Y.S. Myung,  M.G. Seo,    Phys. Lett. B   \textbf{671}, 435 (2009)
\bibitem{4} Q.G. Huang, M. Li,    J. Cosmol. Astropart. Phys.   \textbf{8}, 13 (2004)
\bibitem{5} G. 't Hooft,    Int. J. Mod. Phys. D   \textbf{15}, 1587 (2006)
\bibitem{5a} L. Susskind,   J. Math. Phys.   \textbf{36}, 6377 (1995)
\bibitem{7} A.G. Cohen, D.B. Kaplan, A.E. Nelson,    Phys. Rev. Lett.   \textbf{82}, 4971 (1999)
\bibitem{n2primo} M. Li, X. D. Li, S.Wang, Y.Wang, X. Zhang,   J. Cosmol. Astropart. Phys.   \textbf{0912}, 014 (2009).
\bibitem{n2secondo} M. Li, X. D. Li, S. Wang, X. Zhang,   J. Cosmol. Astropart. Phys.   \textbf{0906}, 036 (2009).

\bibitem{8} B. Guberina, R. Horvat, H. Nikolic,     J. Cosmol. Astropart. Phys.   \textbf{1}, 12 (2007)
\bibitem{9} J. D. Bekenstein,    Phys. Rev. D   \textbf{7}, 2333 (1973)
\bibitem{9a}  S.W. Hawking,   Phys. Rev. D   \textbf{13}, 191 (1976)
\bibitem{10} B. Chen, M. Li, Y. Wang,     Nucl. Phys. B   \textbf{774}, 256 (2007)
\bibitem{11} M. Jamil, E.N. Saridakis, M.R.  Setare,   Phys. Lett. B   \textbf{679}, 172 (2009)
\bibitem{12} L. Xu,    J. Cosmol. Astropart. Phys.   \textbf{9}, 16 (2009)
\bibitem{12a} H.M. Sadjadi,  M. Jamil,    Gen. Rel. Grav.   \textbf{43}, 1759 (2011)
\bibitem{12c} K. Karami, J.  Fehri,      Int. J. Theor. Phys.   \textbf{49}, 1118 (2010)
\bibitem{12d} M. Jamil,  M.U. Farooq,     Int. J. Theor. Phys.   \textbf{49}, 42 (2010)
\bibitem{12e} M. Jamil, M.U. Farooq, , M.A. Rashid,    European Phys. J. C   \textbf{61}, 471 (2009)
\bibitem{13} B. Wang, Y. Gong , E. Abdalla,     Phys. Lett. B   \textbf{624}, 141 (2005)
\bibitem{13c} A. Sheykhi,     Class.  Quant. Grav.   \textbf{27}, 025007 (2010)
\bibitem{14} S. Chattopadhyay,  U. Debnath,   Astrophys. Space Sci.   \textbf{319}, 183 (2009)
\bibitem{14a} K. Karami,  J. Fehri,    Phys. Lett. B   \textbf{684}, 61 (2010)
\bibitem{14b} K. Karami,  M.S. Khaledian,  M. Jamil,    Phys. Scr.   \textbf{83}, 025901 (2011)
\bibitem{16} C. Feng, B. Wang, Y. Gong, R.K. Su,     J. Cosmol. Astropart. Phys.   \textbf{9}, 5 (2007)
\bibitem{16a} B. Wang, J. Zang, C.Y. Lin, E. Abdalla, S. Micheletti,     Nucl. Phys. B   \textbf{778}, 69 (2007)
\bibitem{16b} Q. Wu, Y. Gong, A. Wang, J.S. Alcaniz,     Phys. Lett. B   \textbf{659}, 34 (2008)
\bibitem{16c} M. Li, X.D. Li, S. Wang, Y. Wang, X. Zhang,    J. Cosmol. Astropart. Phys.   \textbf{12}, 14 (2009)
\bibitem{nojod} S. Nojiri, S.D. Odintsov,     Int. J. Geom. Meth. Mod. Phys.   \textbf{4}, 115 (2007)
\bibitem{15} C.J. Fen, X.Z. Li,    Phys. Lett. B   \textbf{679}, 151 (2009)
\bibitem{15b} H. Wei,    Nucl. Phys. B   \textbf{819}, 210 (2009)
\bibitem{15c} Y. Bisabr,    Gen. Rel.  Grav.   \textbf{41}, 305 (2009)
\bibitem{15d} K. Nozari, N. Rashidi,     Int. J. Mod. Phys. D   \textbf{19}, 219 (2010)
\bibitem{15g} K. Karami, M.S. Khaledian,    Journal of High Energy Physics   \textbf{3}, 86 (2011)
\bibitem{15i} M.R. Setare,  M. Jamil,      Europhys. Lett.   \textbf{92}, 49003 (2010)
\bibitem{mio} A. Pasqua, I. Khomenko,        Int. J. Theor. Phys.   \textbf{52}, 3981 (2013)
\bibitem{bra1} C. Deffayet, G. Dvali , G. Gabadadze,    Phys. Rev. D   \textbf{65}, 044023 (2002)
\bibitem{bra2} V. Sahni V, Y. Shtanov,     J. Cosmol. Astropart. Phys.   \textbf{11}, 14 (2003)
\bibitem{mioft1} A. Pasqua,  S. Chattopadhyay,     Can. J. Phys.   \textbf{91}, 351 (2013)
\bibitem{mioft2} S. Chattopadhyay,  A. Pasqua,     Astrophys. Space Sci.   \textbf{344}, 269 (2013)
\bibitem{ft2} E.V. Linder,     Phys. Rev. D   \textbf{81} 127301 [Erratum-ibid. D \textbf{82}, 109902 (2010)]
\bibitem{ft4}  B. Li, T.P. Sotiriou, J.D. Barrow,    Phys. Rev. D   \textbf{83}, 104017 (2011)
\bibitem{ft8} K. Bamba, C.Q. Geng, C.C. Lee,  L.W. Luo,    J. Cosmol. Astropart. Phys.   \textbf{1101}, 021 (2011)
\bibitem{fr1} S. Capozziello,   Int. J. Mod. Phys. D   \textbf{11}, 483 (2002)
\bibitem{fr2} T.P. Sotiriou, V. Faraoni,     Rev. Mod. Phys.   \textbf{82}, 451 (2010)
\bibitem{miofr} A. Jawad, S. Chattopadhyay,  A. Pasqua,     Astrophys. Space Sci.   \textbf{346}, 273 (2013)
\bibitem{fr9} S. Capozziello, S. Carloni , A. Troisi,    Recent Res. Dev. Astron. Astrophys.   \textbf{1}, 625 (2003)
\bibitem{fr8} N. Arkani-Hamed, H.C. Cheng, M.A. Luty, S. Mukohyama,     Journal of High Energy Physics     \textbf{05}, 043528 (2004)
\bibitem{fr10} S. Nojiri, S.D. Odintsov,     Phys. Rev. D   \textbf{68}, 123512 (203)
\bibitem{fr15} A. Aghmohammadi, K. Saaidi , M.R. Abolhassani,     Int. J. Theor Phys.   \textbf{49}, 709 (2010)
\bibitem{fr12} S. Capozziello, S. Nojiri, S.D. Odintsov, A. Troisi,    Phys. Lett. B   \textbf{639}, 135 (2006)
\bibitem{fr14} D.A. Easson,    Int. Mod. Phys. A   \textbf{19}, 5343 (2004)
\bibitem{miofg1} A. Jawad, S. Chattopadhyay, A. Pasqua,    European Phys. J. Plus   \textbf{128}, 88 (2013)
\bibitem{frt2} R. Myrzakulov,    European Phys. J. C   \textbf{72}, 2203 (2012)
\bibitem{frt4} A. Pasqua, S. Chattopadhyay, I,. Khomenko,          Canad. J. Phys.   \textbf{91}, 632 (2013)
\bibitem{dgp1} G. Dvali, G. Gabadadze, M. Porrati,    Phys. Lett. B   \textbf{485}, 208 (2000)
\bibitem{miodbi} S. Chattopadhyay, A. Pasqua,        Int. J. Theor. Phys.   \textbf{52}, 3945 (2013)
\bibitem{miohl} A. Pasqua, S. Chattopadhyay,        Astrophys. Space Sci.   \textbf{348}, 541 (2013)
\bibitem{miobd2} A. Pasqua,  S. Chattopadhyay,     Astrophys. Space Sci.   \textbf{348}, 283 (2013)
\bibitem{23} Y. Gong,    Phys. Rev. D   \textbf{70}, 064029 (2004)
\bibitem{23a} H. Kim, H.W. Lee, Y.S. Myung,     Phys. Lett. B   \textbf{628}, 11 (2005)
\bibitem{24a} Y. Gong,  Phys. Rev. D   \textbf{61}, 043505 (2000)
\bibitem{24b} B. Nayak, L.P. Singh,   Mod. Phys. Lett. A   \textbf{24}, 1785 (2009)
\bibitem{24c} L. Xu, W. Li,  J. Lu,    European Phys. J. C   \textbf{60}, 135 (2009)
\bibitem{24d} A. Sheykhi,    Phys. Rev. D   \textbf{81}, 023525 (2010)
\bibitem{24e} A. Sheykhi,  M. Jamil,     Phys. Lett. B   \textbf{694}, 284 (2011)
\bibitem{25} A. Sheykhi,    Phys. Lett. B   \textbf{681}, 205 (2009)
\bibitem{das18} S. Das, S. Shankaranarayanan, S. Sur,     Phys. Rev. D   \textbf{77}, 064013 (2008)
\bibitem{das18a} N. Radicella , D. Pavon,     Phys. Lett. B   \textbf{691} 121 (2010)
\bibitem{das18d} A. Sheykhi, M. Jamil,     Gen. Relativ. Grav.   \textbf{43} 2661 (2011)
\bibitem{hei17} H. Wei,    Commun. Theor. Phys.   \textbf{52}, 743 (2009)
\bibitem{grandaoliveros} L.N. Granda ,  A. Oliveros,    Phys. Lett. B   \textbf{671}, 199 (2009)
\bibitem{grandaoliverosa} L.N. Granda ,  A. Oliveros,    Phys. Lett. B   \textbf{669}, 275 (2008)
\bibitem{wangalfa} Y. Wang, L. Xu,       Phys. Rev. D   \textbf{81}, 083523 (2010)
\bibitem{26-1} M. Arik, M.C. {\c C}alik,    Mod. Phys. Lett. A   \textbf{21}, 1241 (2006)
\bibitem{28bane} N. Banerjee,  D. Pav{\'o}n,     Phys. Lett. B   \textbf{647}, 477 (2007)
\bibitem{29} L. Xu, J.  Lu, W.  Li,    Mod. Phys. Lett. A   {\bf 25}, 1441 (2010)
\bibitem{30} V. Acquaviva, L. Verde,     J. Cosmol. Astropart. Phys.   \textbf{12}  1 (2007)
\bibitem{30-1} B. Bertotti, L. Iess, P. Tortora,     Nature   \textbf{425}, 374 (2003)
\bibitem{jgo} M. Jamil, K. Karami,  A. Sheykhi, E. Kazemi, Z. Azarmi,    Int. J. Theor. Phys.   \textbf{51}, 604 (2012)
\bibitem{kho8} A. Khodam-Mohammadi, A. Pasqua, M. Malekjani, I. Khomenko, M. Monshizadeh,     Astrophys. Space Sci.   \textbf{345}, 415 (2013)
\bibitem{q2} O. Bertolami, F. Gil Pedro, M. Le Delliou,    Phys. Lett. B   \textbf{654}, 165 (2007)
\bibitem{q2-2} M. Jamil, M.A. Rashid,     European Phys. J. C   \textbf{58}, 111 (2008)
\bibitem{q3} C. Feng, B. Wang,  Y. Gong, R.K.  Su,    J. Cosmol. Astropart. Phys.   \textbf{9}, 5 (2007)
\bibitem{1abd}  E. Abdalla, L.R. Abramo, L. Sodr{\'e}, B. Wang,   Phys. Lett. B   \textbf{673}, 107 (2009)
\bibitem{10abd} O. Bertolami, F. Gil Pedro, M. Le Delliou,     Gen. Rel. Grav.   \textbf{41}, 2839 (2009)
\bibitem{A173} Z.K. Guo, N. Ohta, S. Tsujikawa,    Phys. Rev. D   \textbf{76}, 023508 (2007)
\bibitem{22abd} J.H. He, B. Wang, P.  Zhang P,    Phys. Rev. D   \textbf{80}, 063530 (2009)
\bibitem{q1} L. Amendola , D. Tocchini-Valentini,     Phys. Rev. D   \textbf{64}, 043509 (2001)
\bibitem{q1-4} A. Sheykhi,  M. Jamil,     Phys. Lett. B   \textbf{694}, 284 (2011)
\bibitem{q1-8} W. Zimdahl, D. Pavon,     Gen. Rel. Grav.   \textbf{35}, 413 (2003)
\bibitem{q1-9} M.R. Setare,  M. Jamil,    J. Cosmol. Astropart. Phys.   \textbf{02}, 010 (2010)
\bibitem{feng08} C. Feng et al.,     Phys. Lett. B   \textbf{665}, 111 (2008)
\bibitem{q4} K. Ichiki  et al.,     J. Cosmol. Astropart. Phys.   \textbf{06}, 005 (2008)
\bibitem{zhang-02-2006} H. Zhang , Z.H. Zhu,    Phys. Rev. D   \textbf{73}, 043518 (2006)

\bibitem{sah} V. Sahni, T.D. Saini, A.A. Starobinsky, U. Alam,    Soviet Journal of Experimental, Theoretical Physics Letters   \textbf{77}, 201 (2003)
\bibitem{alam} U. Alam, V. Sahni, T. Deep Saini , A.A. Starobinsky,      Mon. Not. R. Astron. Soc., \textbf{344}, 1057 (2003)
\bibitem{huang} Z.G. Huang,  X.M. Song, H.Q. Lu, W. Fang,       Astrophys. Space Sci.   \textbf{315}, 175 (2008)
\bibitem{wu1} P. Wu, H. Yu,      Physics Letters B   \textbf{693}, 415 (2010)
\bibitem{wang} F.Y. Wang, Z.G. Dai , S. Qi,     Astron. Astrophys.   \textbf{507}, 53 (2009)
\bibitem{kho1} A. Khodam-Mohammadi, M. Malekjani,       Communications in Theoretical Physics, \textbf{55}, 942 (2011)
\bibitem{kho5} M. Malekjani, A. Khodam-Mohammadi,       Astrophys. Space Sci.   \textbf{343}, 451 (2013)
\bibitem{kho7} M. Malekjani, A. Khodam-Mohammadi, N. Nazari-Pooya,     Astrophys. Space Sci.   \textbf{332}, 515 (2011)




\bibitem{cosmo1} S. Weinberg,     Gravitation, cosmology, Wiley  , New York 1972.
\bibitem{cosmo2} M. Visser,   Class. Quant. Grav.   \textbf{21}, 2603 (2004)
\bibitem{arab} M. Arabsalmani, V. Sahni,      Phys. Rev. D   \textbf{83}, 043501 (2011)
\bibitem{dabro} M.P. D{\c a}browski,      Physics Letters B \textbf{625}, 184 (2005)
\bibitem{duna} M. Dunajski,, G. Gibbons,      Classical, Quantum Gravity \textbf{25}, 235012 (2008)

\bibitem{values0} S. Capozziello, L. Izzo,     Astronomy and Astrophysics \textbf{490}, 31 (2008)
\bibitem{values0-1} M.V. John      The Astrophysical Journal \textbf{630}, 667 (2005)

\end{thebibliography}
\end{document}